\documentclass[]{aa}

\usepackage[varg]{txfonts}
\usepackage{natbib}
\def\apjl{ApJ}%
\def\apjs{ApJS}%
\def\ao{Appl.~Opt.}%
\def\aap{A\&A}%
\bibpunct{(}{)}{;}{a}{}{,} % to follow the A&A style

\begin{document}

\title{Rotation periods for cool stars in the open cluster Ruprecht\,147~(NGC\,6774)%
    \thanks{Table\,\ref{tab_sample_periods} and the processed light curves of the sample stars as plotted in Appendix\,\ref{apx_lightcurves} are available in electronic form at the CDS via anonymous ftp to cdsarc.u-strasbg.fr (130.79.128.5) or via http://cdsweb.u-strasbg.fr/cgi-bin/qcat?J/A+A/ }%
    }

\subtitle{Implications for gyrochronology}

\author{D. Gruner\inst{1,2} \and S. A. Barnes\inst{1,3}}

\institute{
    Leibniz-Institute for Astrophysics Potsdam (AIP), An der Sternwarte 16, 14482, Potsdam, Germany
        \and 
    Institut für Physik und Astronomie, Universit\"at Potsdam, Karl-Liebknecht-Str. 24/25, 14476 Potsdam, Germany
        \and
    Space Science Institute, Boulder, CO, USA
    }

\date{Received date / Accepted date}

\abstract 
  {Gyrochronology allows the derivation of ages for cool main sequence stars based on their observed rotation periods and masses, or a suitable proxy thereof. It is increasingly well-explored for FGK stars, but requires further measurements for older ages and K\,--\,M-type stars.
  }
  {We study the nearby, 3\,Gyr-old open cluster Ruprecht\,147 to compare it with the previously-studied, but far more distant, NGC\,6819 cluster, and especially to measure cooler stars than was previously possible there. }
  {We constructed an inclusive list of 102 cluster members from prior work, including \emph{Gaia}\,DR2, and for which light curves were also obtained during Campaign~7 of the \emph{Kepler/K2} space mission.
  We placed them in the cluster color-magnitude diagram and checked the related information against appropriate isochrones. 
  The light curves were then corrected for data systematics using Principal Component Analysis on all observed K2\,C07 stars and subsequently subjected to periodicity analysis.}
  {Periodic signals are found for 32 stars, 21 of which are considered to be both highly reliable and to represent single, or effectively single, Ru\,147 stars. 
  These stars cover the spectral types from late-F to mid-M stars, and they have periods ranging from 6\,d\,--\,33\,d, allowing for a comparison of Ruprecht\,147 to both other open clusters and to models of rotational spindown.
  The derived rotation periods connect reasonably to, overlap with, and extend to lower masses the known rotation period distribution of the 2.5\,Gyr-old cluster NGC\,6819.}
  {The data confirm that cool stars lie on a single surface in rotation~period-mass-age space, and they simultaneously challenge its commonly assumed shape.
  The shape at the low mass region of the color-period diagram at the age of Ru\,147 favors a recently-proposed model, which requires a third mass-dependent timescale in addition to the two timescales required by a former model, suggesting that a third physical process is required to model rotating stars effectively.}

\titlerunning{Rotation periods for cool stars in the open cluster Ruprecht\,147}

\authorrunning{Gruner \& Barnes}

\keywords{Stars: rotation, Stars: late-type, (Stars:) starspots,
        Open clusters and associations: individual: \object{Ruprecht 147}, \object{NGC 6774} }

\maketitle

\section{Introduction} \label{sec:introduction}

 Studies that require coeval groups of stars older than $\sim$1\,Gyr are often hindered by their distance. The younger Hyades (46\,pc; $\sim$600\,Myr) and Pleiades (130\,pc; $\sim$150\,Myr) are the nearest open clusters and, consequently, have been extensively studied, including with respect to the rotation periods of their cool stars \citep{1987ApJ...321..459R,1987A&AS...67..483V, 2016AJ....152..113R, 2019ApJ...879..100D}. The closest open cluster of near-solar age ($\sim$4\,Gyr) is M67 at a distance of roughly 900\,pc \citep[][see also \cite{1955ApJ...121..616J}]{2005A&A...438.1163K}; a fortuitous proximity that provides valuable samples of solar analogs and many other cluster stars of non-solar mass, enabling detailed studies \citep[e.g.,][]{1957ApJ...126..326S,1971ApJ...168..393R,1992AJ....103..151D} including of its rotational properties \citep{ApJ...823..2016.16B}.

 At intermediate ages, say 2\,--\,3\,Gyr for instance, the closest open cluster that has been well-studied with respect to stellar rotation is the 2.3\,kpc-distant cluster NGC\,6819. This object fortuitously was in the field observed by the Kepler satellite, permitting a careful rotational study despite its relative distance by using data acquired over the 4yr Kepler observational baseline \citep{2015Natur.517..589M}\footnote{This rotational study was itself built upon extensive prior work on the cluster in the literature, including a near-decade-long radial velocity survey for cluster membership and multiplicity, and also a ground-based proper-motion study \citep{2013AJ....146...43P}.}. The study of another cluster of a similar age would permit the independent verification of the NGC\,6819 rotation results (if the results were similar); and additionally, if that cluster were substantially closer than NGC\,6819, this would also allow the derivation of rotation periods for lower mass cool stars than was possible in NGC\,6819. Observations of the nearby ($305$\,pc), $\sim3$~Gyr-old open cluster \object{Ruprecht 147} ( = NGC\,6774; Ru147 hereafter) with the \emph{K2} reincarnation of the \emph{Kepler} satellite permit exactly this type of work, as described in this paper. 
 
 A key motivation for our work is to examine whether Ru147 can be used as an additional benchmark for ``gyrochronology,'' the technique for deriving the age of a main sequence star from its (measured) rotation period and mass, or a suitable mass proxy such as color \citep[e.g.,][]{2003ApJ...586..464B, 2007ApJ...669.1167B, 2008ApJ...687.1264M, 2010ApJ...722..222B, 2020A&A...636A..76S}. The spindown of stars of solar mass was famously described by \cite{1972ApJ...171..565S}\footnote{A power law was fitted to the averaged $v \sin i$ values of solar mass stars in a limited number of open clusters.} and is now well-known as originating in angular momentum loss caused by magnetized stellar winds \citep{1958ApJ...128..664P,1967ApJ...148..217W,1988ApJ...333..236K}. However, its generality and applicability to stars of non-solar mass, the basis of gyrochronology, are by no means assured.

 \citet[][Fr20 hereafter]{Fr2020} have recently shown that the measured rotation period distributions of the well-studied Zero Age Main Sequence (ZAMS) open clusters Pleiades, M\,35, M\,50, Blanco\,1, and NGC\,2516 are indistinguishable \cite[with data from][respectively]{2016AJ....152..113R, 2009ApJ...695..679M, 2009MNRAS.392.1456I, 2014ApJ...782...29C,Fr2020}. This fact suggests that the ZAMS cool star rotational distribution is indeed identical in otherwise identical clusters, that such a distribution is a natural outcome of pre-main sequence evolution, and perhaps of the star formation process itself. However the paucity of suitable cluster data at older ages has not allowed such a corresponding check to date for older stars. Ru147 allows such a comparison to be made for 3\,Gyr-old stars by comparison with the similarly-old \object{NGC 6819} cluster, previously studied by \cite{2015Natur.517..589M}.

 Data for late-F to mid-K-type stars, in a series of clusters of increasing age; for Hyades \citep[625\,Myr;][]{1987ApJ...321..459R,2019ApJ...879..100D}, NGC\,6811 \citep[1\,Gyr;][]{2011ApJ...733L...9M}, NGC\,6819 \citep[2.5\,Gyr;][]{2015Natur.517..589M}, and M\,67 \citep[4\,Gyr;][]{ApJ...823..2016.16B} show that the spindown for those stars follows the generalized Skumanich relationship $P(m) \propto \sqrt{t}$, where $P, m,$ and $t$ represent a cool star's rotation period, mass, and age respectively. Such models are called ``separable'' because the dependence of $P$ on $m$ and $t$ is factorized into separate functions $f(m)$ and $g(t)$, of stellar mass and age respectively. 

 However, data for lower-mass stars in the Praesepe \citep{2011ApJ...740..110A} and NGC\,6811 open clusters \citep{2019ApJ...879...49C} indicate deviations from the simple $P(m) \propto \sqrt{t}$ spindown relationship. Certain deviations are expected because, as has been clearly explained in \citet[][Ba10 hereafter]{2010ApJ...722..222B} and \citet[][see also \citet{2016AN....337..810B}, BSW16 hereafter]{2015ApJ...799L..23M}, second-generation (i.e., non-separable) gyrochronology models (e.g., Ba10) only require that $P(m) \propto \sqrt{t}$ hold in the asymptotic limit of large Rossby Number, $Ro$. The rotational evolution at small $Ro$ is both intrinsically different and also modulated by the initial rotational distribution, resulting in different predicted shapes for rotation period distributions as a function of stellar mass and age\footnote{Observed rotation period distributions are of course not completely homologous and display patterns that are a combination of intrinsic differences and also observational sensitivity.}. Regardless, non-separable models are also believed to have deficiencies and the data mentioned above have prompted \cite{2020A&A...636A..76S} to develop a model with one additional degree of freedom as compared with Ba10\footnote{These models have a pedigree that dates back to \cite{1991ApJ...376..204M} in mathematical form.}. This enters via the parameter $p$, which specifies the power law in the mass dependence of the internal coupling in their two-zone rotational model, which otherwise follows the Ba10 spindown formulation. The \cite{2015ApJ...799L..23M}, \cite{2018ApJ...862...90G}, and \cite{2019A&A...631A..77A} models allow several more degrees of freedom, with varying success in describing the observations. (A detailed summary comparison of these models in connection with Zero Age Main Sequence (ZAMS) open clusters can be found in Fr20.)

 From a field star viewpoint (as opposed to the cluster viewpoint above), \citet{2016Natur.529..181V} and \citet{2019ApJ...871...39M} have used asteroseismic ages for field stars to claim significant deviations of theoretical gyrochronology models from observations, originating in a drastic decrease of angular momentum loss when stars reach middle age, initially broadly interpreted as the main sequence career beyond 2\,Gyr; and more recently as the point where stars reach a rotation period such that $Ro \approx 2$. However, as noted by, for example, \cite{2013ApJ...771L..31D,2014ApJ...790L..23D}, BSW16, and \cite{2016A&A...589A..27B}, there appear to be a number of problems and disagreements regarding the ages, metallicities, and binary status of many of these field star samples, where determination of a star's evolutionary status and stellar parameters is inherently far more challenging than that in open cluster member stars. Indeed, \cite{2020MNRAS.tmpL..51L} have recently published a secure 35\,d rotation period determination for the 8\,Gyr-old solar twin star HD\,197027  \citep[=\,HIP\,102152, see also][]{2019MNRAS.485L..68L,2020AN....341..497S}. These results appear to refute the proposal that stars stop spinning down in middle age. Ongoing large scale surveys like \emph{Gaia} and TESS provide increasingly large samples of field stars for gyrochronology \citep[e.g.][respectively]{2018A&A...616A..16L,2020arXiv200703079C} and future studies, such as PLATO \citep{2014ExA....38..249R}, will further expand the amount of available data.

 Wide binaries bridge the gap between field stars and open clusters; to a certain extent, they could be considered the smallest open clusters. Rotational studies of such systems in the Kepler field \citep[e.g.][]{2017ApJ...835...75J,2017AAS...22924026O} have also provided some intriguing evidence for deviations. Approximately $60$\% of the systems in \cite{2017ApJ...835...75J} appear to agree with rotational isochrones calculated using the Ba10 models. However, the remaining systems display partial-to-significant disagreements, with the secondary star rotation periods largely located below the rotational isochrone for the primary component. This result modulates the original result from \citet[][Ba07 hereafter]{2007ApJ...669.1167B}, where the three wide binary systems with measured rotation periods for both components then known ($\alpha$\,Cen\,A/B, 16\,Cyg\,A/B, and $\xi$\,Boo\,A/B) all agreed within their uncertainties with their respective rotational isochrones. The discrepant systems have not to date been investigated carefully for tertiary components or other pathologies\footnote{The faintness of the systems in the \emph{Kepler} field is an obstacle to detailed spectroscopic investigation.}. 

 With the present study on Ru\,147, we approach the above mentioned problems from the open cluster perspective. Ru\,147 (also known as NGC\,6774) was originally discovered by \cite{1833RSPT..123..359H}, who designated it as GC\,481 \citep{1863RSPS...13....1H}, and has been mentioned occasionally since then in various catalogs \citep[e.g.][]{1888MmRAS..49....1D,1958csca.book.....A,1966BAICz..17...33R,1987A&A...188...35L}. However, it has recently attracted significant interest because of the combination of its relative proximity ($\sim 300\,$pc) and age. In fact, Ru147 is the oldest nearby open cluster with 2\,--\,3\,Gyr age \citep[][Cu13 hereafter]{2013AJ....145..134C}. Several other recent studies have identified member stars and derived cluster properties using a variety of techniques including photometry, astrometry, and radial velocities \citep[e.g.][]{2017A&A...600A.106C,2018A&A...618A..93C,2018A&A...619A.176B,2018A&A...616A..10G,2019A&A...625A.115O}. The combination of all of the above-mentioned information with additional results from astrometric surveys such as Gaia\,DR2 \citep[][see also \cite{2016A&A...595A...1G}]{2018A&A...616A...1G} provides extensive information about the cluster's membership, stellar multiplicity, and other fundamental properties.

 Additional studies have focused on individual objects within the cluster, such as eclipsing binaries \citep{2018ApJ...866...67T,2019ApJ...887..109T,2020arXiv200413032T}, brown dwarfs \citep{2017AJ....153..131N}, and exoplanets \citep{2018AJ....155..173C}. 
 Finally, \cite{2019AJ....157..115Y} have suggested that Ru147 is imminently likely to dissolve into the galactic disk. 

 We take advantage of all relevant prior work and combine it with detailed analysis of high-precision time series photometry acquired using the Kepler/K2 mission to measure the rotation periods of cool stars in Ru147. Unfortunately, the Kepler/K2 data for Ru147 both have an abbreviated observing baseline as compared with the original Kepler data for NGC\,6819 and are of significantly lower photometric quality. These observational realities will require special efforts to overcome, as described below. In short, we use Principal Component Analysis (PCA), reprising a technique that our group used successfully in our analysis of similar Kepler/K2 data for the 4\,Gyr-old open cluster M\,67 \citep{ApJ...823..2016.16B}. Ru147 also presents a peculiar difficulty. Because of its proximity and perhaps its imminent dissolution, it is spread out over a large area on the night sky, making it operationally difficult to obtain the detailed membership analysis required to distinguish the cluster stars from non-members. Fortunately, the cluster has offsets with respect to the surrounding field stars in both radial velocity and proper motion, allowing member identification when such data are actually available. Gaia DR2 is particularly helpful in this regard. We rely on a combination of prior work from the literature for this membership and other basic cluster information. An overview of the adopted parameters of Ru147 is provided in Table\,\ref{tab_ru147}.

\begin{table}
    \centering
    \caption{Astrometric and physical parameters adopted for Ru\,147.}
    \label{tab_ru147}
    \begin{tabular}{lcrc}
        \hline\hline
        Parameter         & Unit     & Value   & Reference \\
        \hline
        Ra                & deg           & 289.087   & 1 \\
        Dec               & deg           & -16.333   & 1  \\
        $\mu_\text{Ra}$   & mas/yr        & -0.939	& 1  \\
        $\mu_\text{Dec}$  & mas/yr        & -26.576   & 1  \\
        parallax $\pi$    & mas           & 3.250	    & 1  \\
        $v_\text{rad}$   & km\,s$^{-1}$  & 41.79     & 2 \\
        distance $d$      & pc            & 305.0     & 3 \\
        $[$Fe/H$]$        &               & +0.08     & 4 \\
        $[$Fe/H$]$        &               & +0.12     & 5 \\
        Age               & Gyr           & 2.7       & 6  \\
        $E_{G_{BR}-G_{RP}}$& mag          & 0.1       & 7 \\
        $A_G$             & mag           & 0.2       & 7 \\
        \hline
    \end{tabular}
    \tablebib{
        (1) \citet{2018A&A...618A..93C};
        (2) \citet{2018A&A...616A..10G};
        (3) \citet{2017A&A...600A.106C};
        (4) \citet{2018A&A...619A.176B};
        (5) \citet{2020AJ....159..199D};
        (6) \citet{2019ApJ...887..109T};
        (7) this work.
    }
\end{table}

 This paper is structured as follows. In Sect.\,\ref{sec_observations}, we describe the construction of the sample of stars for detailed study, including the construction of Color-Magnitude Diagrams (CMDs) in multiple relevant colors. The issues with K2 lightcurves and our treatment of those using Principal Component Analysis, followed by period analysis, are outlined in Sect.\,\ref{lightcurves}, while the resulting periods are discussed in Sect.\,\ref{sec_results}. We compare our results to data from other comparable clusters in Sect.\,\ref{sec_empiricalcomparison}, and to widely-used stellar spindown relations in Sect.\,\ref{sec_model_comparison}. Section\,7 outlines our conclusions, and there are also  four Appendices containing ancillary information, together with all relevant light curves.

\section{K2 coverage, cluster membership, and CMD} \label{sec_observations}

 We now describe the sample selection based on the archival data, the K2 coverage, and the construction of the cluster color-magnitude diagram. For our analysis of Ru147, we use light curves obtained during Campaign\,7 (C07) of the Kepler K2 mission, during which a part of Ru147 was monitored over the 82.5\,d interval from Dec 26th, 2015 to Apr 20th, 2016. 15085 lightcurves were recorded during C07 of K2. Of those, 13483 correspond to individual sources listed in the EPIC catalog \citep{2017yCat.4034....0H}. These are used as the starting point for our study, hereafter called the ``full sample''\footnote{The others correspond to special targets that require a different pixel mask for each cadence. Those can be identified by their EPIC IDs (2000\#\#\#\#\#) and correspond to Pluto, and Trojan and Hilda asteroids.}.

\subsection{Source catalogs}

 The EPIC catalog uses the 2MASS\footnote{CDS: II/246/out} \citep{2003yCat.2246....0C} and the UCAC4\footnote{CDS: I/322A/out} \citep{2012yCat.1322....0Z} catalogs as inputs and, therefore, contains identifiers from those two catalogs for a large number of targets. Consequently, it conveniently lists $J, H, K_s, g,$ and $r$ magnitudes for most stars. The cross-match by  \citet{2019A&A...621A.144M} of the Gaia DR2 catalog with other large scale surveys, among them the 2MASS point source catalog \citep[PSC,][]{2006AJ....131.1163S}, is also helpful to us and facilitates identification. 
 
 The identification of members, their evolutionary status and possible multiplicity is crucial to our analysis and interpretation of results. Therefore, we adopt the \emph{Gaia} photometry \citep[$G$, $B_P$, $R_P$;][Ev18 hereafter]{2018A&A...616A...4E} and parallax \citep[$\pi$,][]{2018A&A...616A...2L}. We initially use the extinction and reddening parameters \cite[$A_G$, $E_{G_{BP}-G_{RP}}$][]{2018A&A...616A...8A} from the Gaia DR2 catalog, before coming up with an alternative.

\subsection{Cluster membership} \label{sec_member}

\begin{figure}
    \centering
    \includegraphics[width=\linewidth]{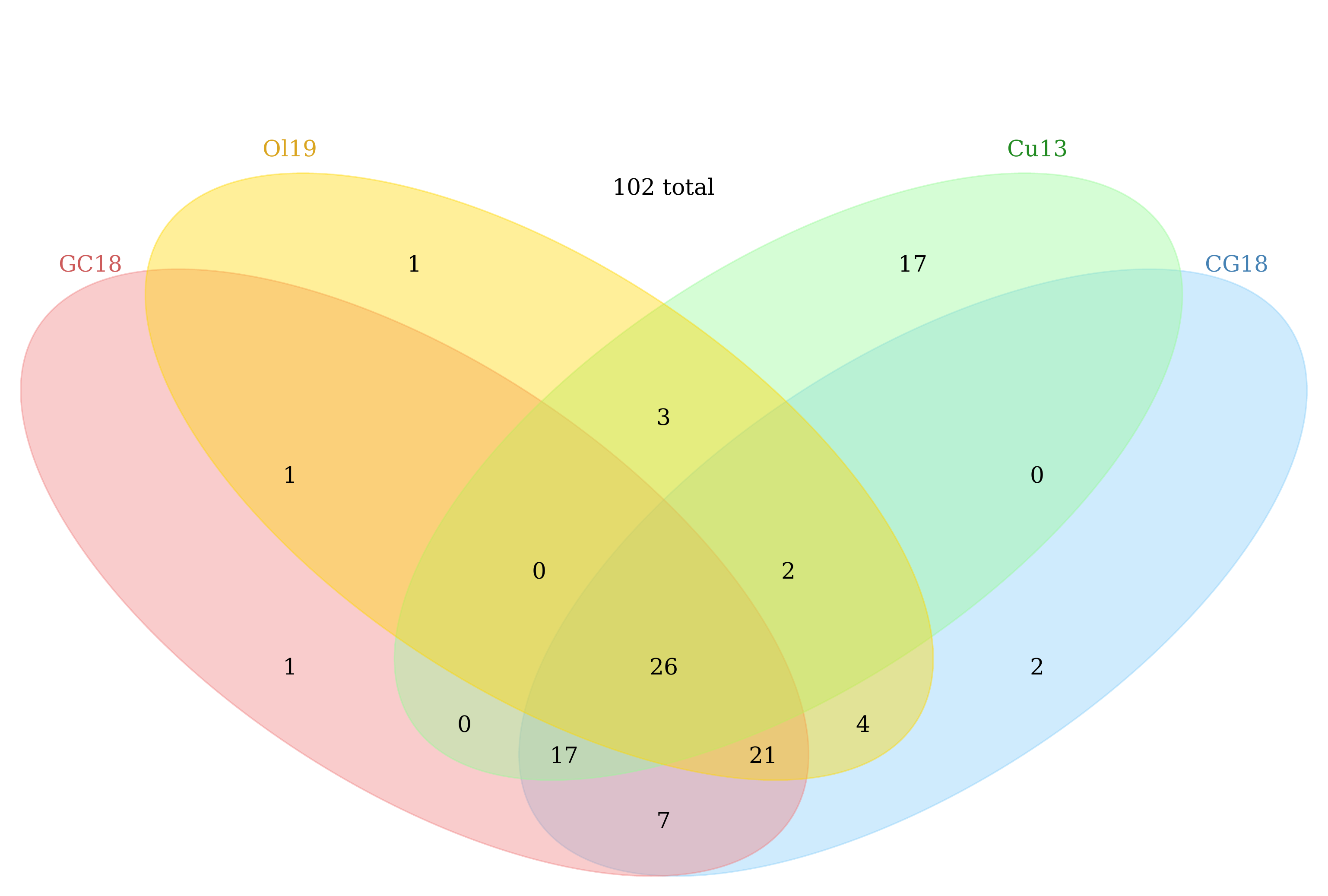}
    \caption{
        Venn diagram of the 102 cluster members observed by Kepler/K2. The numbers indicate the quantity of stars in each corresponding subset. Stars included from the individual studies are: all those listed by \citet[][GC18]{2018A&A...616A..10G}, those labeled $P$ or $Y$ by \citet[][Cu13]{2013AJ....145..134C}, and those listed in \citet[][CG18]{2018A&A...618A..93C} or \citet[][Ol19]{2019A&A...625A.115O} with membership probability greater than 0.5. 
    }
    \label{member_venn}
\end{figure}

 Fortunately, several studies of Ru147's cluster membership have been carried out over the years. Notable ones are the membership analysis of \citet[][Cu13 hereafter]{2013AJ....145..134C},  based on pre-\emph{Gaia} astrometry and spectroscopic data, and \emph{Gaia}-related work by \citet[][GC18 hereafter]{2018A&A...616A..10G} and \citet[][CG18 hereafter]{2018A&A...618A..93C}, both using Gaia astrometry to identify cluster members. The most recent census of Ru147 was performed by \citet[][Ol19 hereafter]{2019A&A...625A.115O}  using all information then available in the literature. Consequently we need not carry out our own membership analysis and can simply adopt the results of these four prior studies as inputs. It is important to note that these studies have by no means identified the same set of stars as members. However, there is a large degree of overlap between the candidates found; see Fig\,\ref{member_venn}. We begin by adopting all stars that are identified as members in at least one of the studies for our sample, that is to say we work with the union of the prior data sets. We will review the membership and multiplicity information again, after the rotation period work in our study has been completed. For a summary of the details regarding the differences between the four membership studies see Ol19.

\begin{figure}
    \centering
    \includegraphics[width=\linewidth]{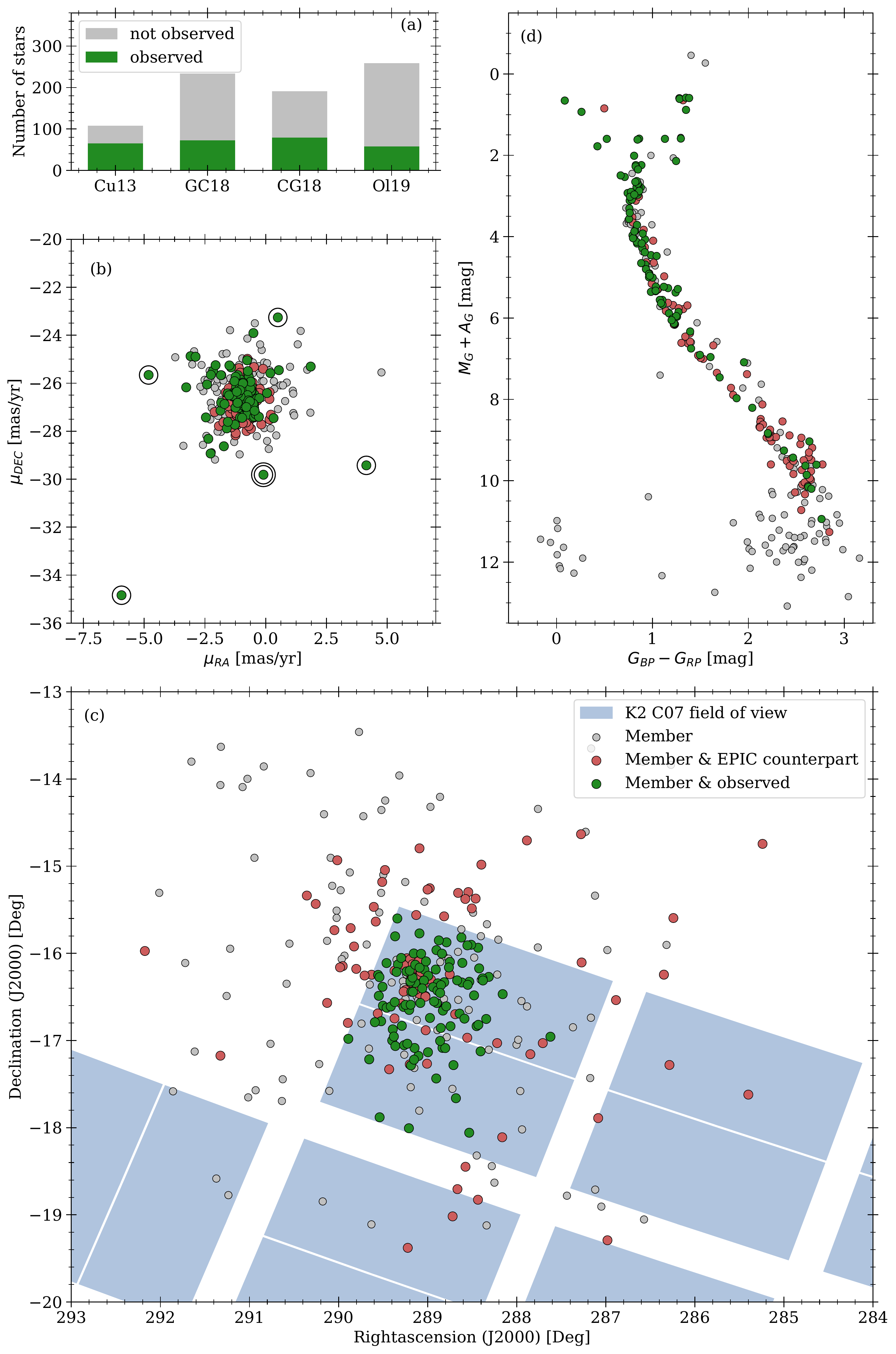}
    \caption{
        Overview of the identified Ru147 cluster members in K2. The histogram in panel (a) shows the fraction of stars in each membership catalog that was observed during K2 as compared with the total number of identified stars. Panel (b) shows the \emph{Gaia} vector point diagram for our compiled member list. We note that certain observed stars (encircled: EPIC\,219665632, 219515762, 219560884, 219437560, and 219855372) are apparent outliers. However, with the exception of EPIC\,219515762 (encircled twice), none are in our final sample, and 219515762 itself is discarded from the interpretation of the results since it is clearly past the cluster turnoff. Panel (c) shows a map of the sky centered on Ru147, with all member stars from our merged sample, assimilated from Cu13, CG18, GC18, and Ol19 (gray). Red symbols indicate those stars which have a counterpart in EPIC while the green ones are those that were actually observed. The shaded regions indicate the approximate layout of the CCDs in the Kepler/K2 field of view. Panel (d) shows a distance-corrected (but not reddening-corrected) CMD of Ru147 cluster members with the color coding as in panel (b).
    }
    \label{member_map}
\end{figure}

 An operational difficulty is that of these four studies, only Ol19 and CG18 list an actual membership probability. Cu13 assigns stars to one of the three categories: ``non-member'', ``very-likely-member'', and ``member'', while GC18 only list members according to their own analysis. We match stars to Cu13 based on their 2MASS IDs; to GC18 and CG18 based on their Gaia IDs; and to Ol19 based on their EPIC and Gaia IDs. For our analysis we adopt a star as a possible member if it is identified as such in at least one of the four above-mentioned catalogs. From Cu13, we take all stars labeled as ``very-likely-member'' and ``member''. We also include all stars listed by GC18, and all stars from CG18 and Ol19 with $P\geq0.5$. We emphasize that this selection includes stars that are listed as members in one study, but that are labeled as non-members in another. Whenever this occurs, we break the impasse by prioritizing the four studies in the order 
 \begin{equation}
    \textrm{Ol19} > \textrm{CG18} > \textrm{GC18} > \textrm{Cu13},
 \end{equation}
 and where two studies of the set \{ GC18, CG18, Ol19 \} may overrule the third in case of disagreement. This procedure enabled us to identify 310 unique cluster members, of which 102 were also observed by Kepler. Figure\,\ref{member_map} provides an overview of the coverage of Kepler/K2, including a comparison with numbers of stars represented in each member list. While the cluster center does indeed lie within the region covered by C07 of K2 (cf. Fig.\,\ref{member_map} panel (b)), a large fraction of the Ru\,147 stars is located outside the K2 field of view. In fact, as the numbers above show, fewer than a third of the identified members from the four studies, as provided by our procedure above, were actually observed as part of the K2 C07 target sample. 

 In Fig.\,\ref{member_map}, we also display a (distance-corrected) color-magnitude diagram (CMD) in \emph{Gaia} $G_{BP}-G_{RP}$ color for context. This CMD shows that despite minor issues with Gaia DR2 photometry at the faint end, there is good consensus between the membership and photometry. Correspondingly, we see a well-defined cluster main sequence, turnoff, giant branch, red clump, blue stragglers, and even several white dwarfs.

 We note that the crossmatch between Gaia and 2MASS does not cover all targets in K2 C07. However, all targets relevant to our rotation period work are covered. Furthermore, we also independently cross-matched the EPIC and Gaia DR2 catalogs based on astrometry and magnitudes, finding the same matches as in \citet{2019A&A...621A.144M} for the Ru147 stars. In summary, our procedure has identified 102 cluster members from the four membership studies discussed above that have been observed by Kepler/K2.

\subsection{Cluster reddening, extinction, and color transformation} \label{sec_color_transformation}

\begin{figure}
    \centering
    \includegraphics[width=\linewidth]{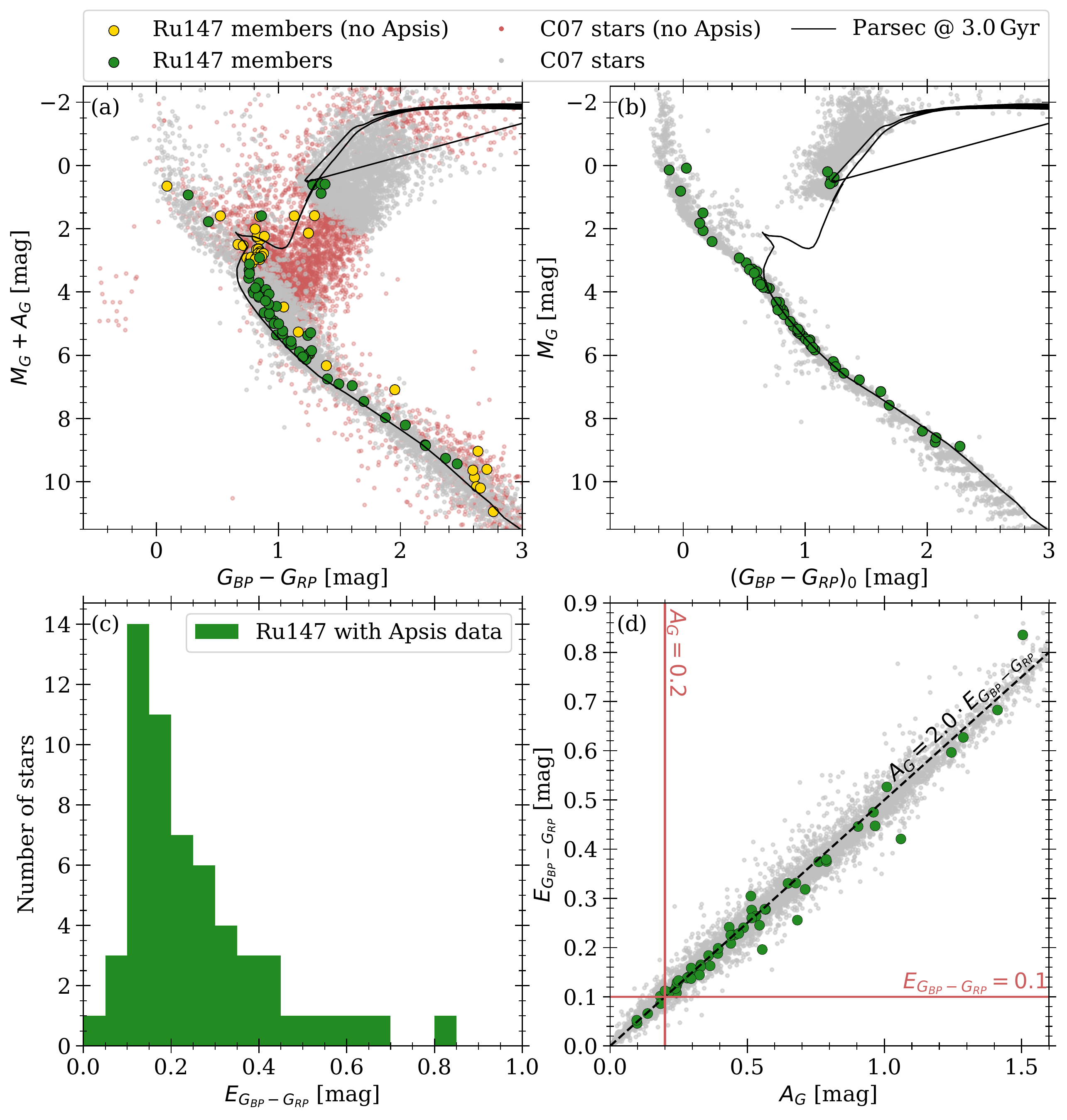}
    \caption{
        Reddening-related aspects of Ru147. \emph{Top:} Color-magnitude-diagrams for all K2 C07 objects (pale symbols); (a) only distance corrected according to their parallax and (b) additionally corrected for both distance and reddening according to Gaia DR2. The gray symbols refer to the subset of stars with reddening parameters provided and red to those without. Ru147 members (green for stars with reddening parameter, yellow for those without) are superimposed in both panels (a) and (b), the latter corrected for reddening $E_{G_{BP}-G_{RP}}$ according to Gaia DR2. A PARSEC isochrone for 3\,Gyr is overplotted. \emph{Bottom:} The panels show that while the reddening values are greatly divergent (c), there is a strong (and unsurprising) correlation between reddening and extinction (d). The ratio between $E_{G_{BP}-G_{RP}}$ and $A_G$ allows us to set $A_G$ for Ru147 based on our $E_{G_{BP}-G_{RP}}$ estimate. This is indicated by the red lines in panel (d).  
    }
    \label{reddening_effects}
\end{figure}

 To verify this combined membership information and also the age of Ru147 in light of it, we have plotted a number of color magnitude diagrams of the member stars and the field, including distance-calibrated and dereddened ones. While so doing, we noticed a suspicious trend with the reddening and extinction parameters provided in Gaia DR2. As can be seen in Fig.\,\ref{reddening_effects}, panels (a) and (b), various problems become evident when photometry is dereddened using the reddening and extinction parameters provided by \emph{Gaia} DR2. Firstly, suspicious horizontal structures are introduced in the CMD for the late-type dwarfs, secondly, Ru147 loses definition near the cluster turn-off, and, thirdly, barely any stars are located above the zero age main sequence.

 We infer that the machine learning approach (Apsis) advocated and described by \cite{2018A&A...616A...8A} in dealing with the Gaia DR2 data is biased toward stars in highly populated regions such as the Main Sequence and the Red Clump. This approach appears to simply \emph{de-redden every star} in low-stellar-density regions of the CMD back onto the main sequence. The strongly varying (and sometimes very large) extinction values between the individual cluster stars (cf. Fig.\,\ref{reddening_effects} panel (c)) are improbable and thus another telling indicator. Yet another suspicious trend is that the majority of stars with reddening estimates (gray in Fig.3) are already situated close to densely populated regions in the CMD. While these issues could potentially be resolved with improved spectrophotometry in the future Gaia DR3 data release, we dismiss these reddening and extinction parameters as unreliable for the purposes of this work.

 Instead, we see (e.g., panel (a) in Fig.\,\ref{reddening_effects}), that only a small uniform reddening for all Ru147 stars is required to bring a 3.0\,Gyr isochrone into agreement with the observed colors (Fig.\,\ref{reddening_effects} panel (b)). Consequently, guided by the linear relationship between Apsis \citep{2018A&A...616A...8A} reddening and extinction estimates, that is, the ratio of selective to total extinction (cf. Fig.\,\ref{reddening_effects} panel (d)), the values $E_{G_{BP}-G_{RP}} = 0.1$ and $A_G=0.2$ were adopted\footnote{These values are not intended to provide a definitive estimate for the cluster reddening and are only a consistiency check, motivated by our interest in the rotational properties of Ru147 cluster members observed with Kepler/K2.}. A small change in the adopted reddening parameters does not impact the results of our main study in any significant way.

 The availability and quality of parameters for the individual cluster stars varies strongly across the sample. Optical and IR photometry are not available for all stars, with especially uncertain $B$ and $JHK$ magnitudes for the red, faint stars in the sample. Fortunately, most stars have extensive (and relatively well-constrained) magnitudes from \emph{Gaia}. However, the relation between Gaia and Johnson colors is non-trivial. The relationship between magnitudes and colors provided by \cite{2018A&A...616A...4E}, itself calibrated on standard stars, fails for late-type stars. The region with $B-V > 1.4$ is especially problematical. Therefore, we create our own empirical color transformation based on photoelectric photometry of Hyades and Pleiades stars in the literature and those of \citet[][and continuously updated afterwards, PM13 hereafter]{2013ApJS..208....9P}\footnote{``A Modern Mean Dwarf Stellar Color and Effective Temperature Sequence'', Version 2019.3.22, \url{pas.rochester.edu/~emamajek/EEM_dwarf_UBVIJHK_colors_Teff.txt}.}. We note that the PM13 results do not list individual stars but averaged results for various intrinsic stellar parameters of local dwarfs ($\leq 30$\,kpc) as a function of the spectral type. We are gratified that both approaches provide similar results, thereby verifying one another's results. The derivation of the relationships is described in detail in Appendix\,\ref{sec_color_trafo}. However, readers are cautioned that this relation is only valid on the main sequence and generally fails for giants. (A related disagreement is highlighted in the Appendix figure.)

\subsection{Color Magnitude Diagram}

\begin{figure*}
    \centering
    \includegraphics[width=\linewidth]{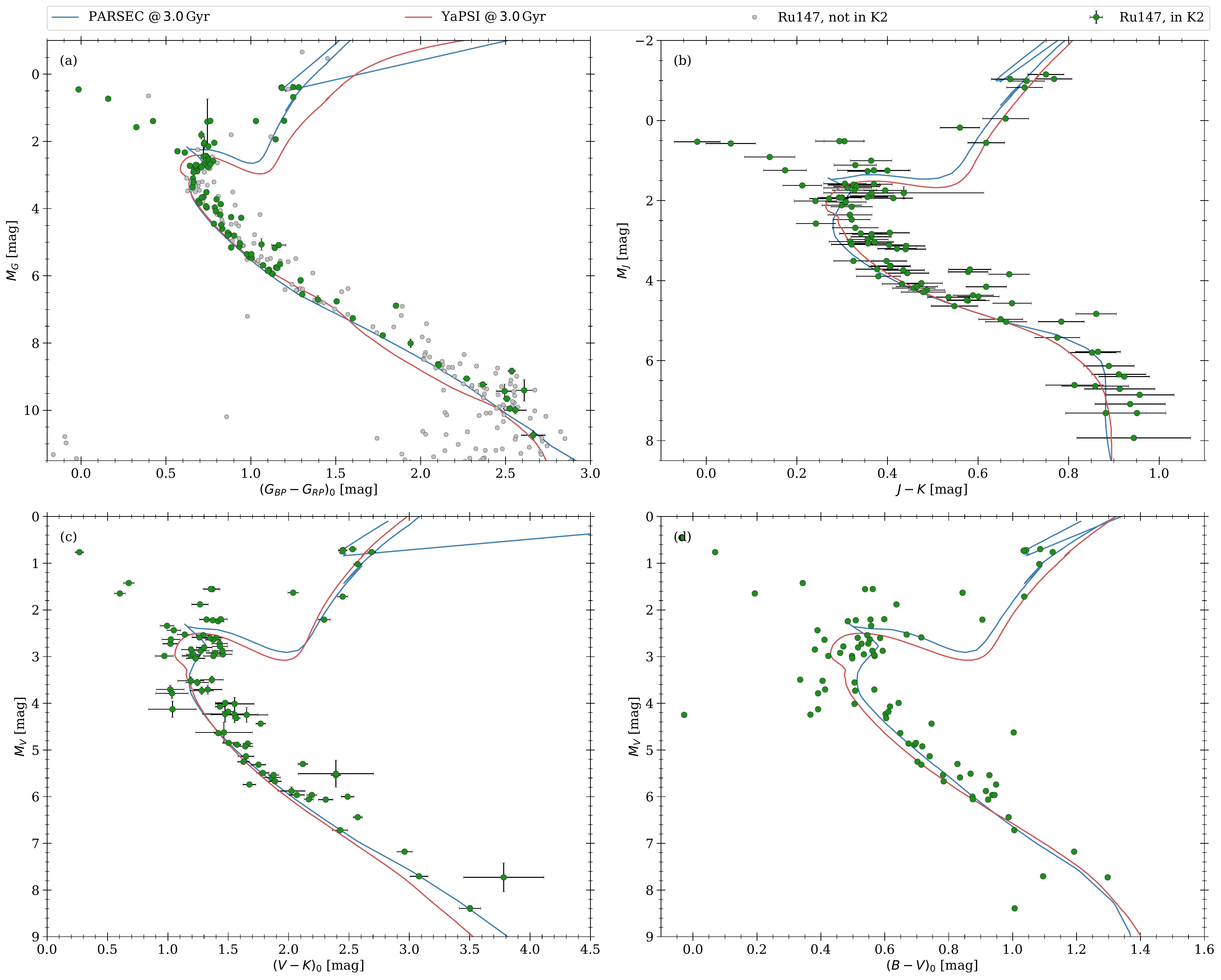}
    \caption{
        Color-magnitude diagrams in \emph{Gaia}, 2MASS, and Johnson photometry for the Ru147 member stars, both observed in K2 (green) and not (gray). Main sequence stars, photometric binaries, the turnoff, giant branch, and also blue stragglers are clearly distinguishable. PARSEC and YaPSI isochrones for 3.0\,Gyr, are overplotted for comparison. Agreement is largely better than satisfactory, especially in the (most comprehensive) \emph{Gaia} CMD. The stellar samples differ in each panel because of varying levels of completeness in the member photometry for the relevant color. The uncertainties in $B-V$ photometry are suppressed for visibility reasons. The individual stellar positions in the \emph{Gaia} CMD are distance-corrected using the Gaia parallaxes; reddening and extinction-corrected uniformly with $E_{G_{BP}-G_{RP}}=0.1$\,mag and $A_G = 0.2$\,mag, in preference to the \emph{Gaia} reddening and extinction values. (For additional corrections, see text.) The PARSEC isochrones use the \emph{Gaia} colors provided within the isochrones; for the YaPSI isochrones we use our empirical color transformation from $B-V$. This causes the redward displacement of the giant branch in the YaPSI isochrones, because our transformation is not valid on the giant branch and tends to predict too red \emph{Gaia} colors (cf. the encircled region in Fig.\,\ref{color_trafo}).
    }
    \label{fig_member_cmd}
\end{figure*}

 The final CMD is shown in Fig.\,\ref{fig_member_cmd} in both \emph{Gaia} and other commonly used colors. Stars with available K2 light curves are highlighted with colored symbols, while the remaining cluster members are displayed in the background. We see a significantly more realistic cluster sequence, as compared with the versions in Fig.\,\ref{reddening_effects}. In particular, we now see a well-defined cluster turnoff, a tight blue hook region, and additional stars populating the giant branch, the red clump, and even the blue straggler regions. A number of photometric binaries are also clearly present here, in contrast with the CMD that uses the Gaia DR2 extinction and reddening values. This CMD is more compelling than the earlier versions, in our opinion, and provides the confidence needed to place stars effectively in the color-period diagram later.

 Reddening and extinction in Johnson colors are calculated using the mean extinction coefficient from \cite{2018MNRAS.479L.102C} as $E_{B-V} = E_{G_{BP}-G_{RP}}/1.339=0.075\,\text{mag}$ and $A_V = 3.1 \cdot E_{B-V} = 0.23$\,mag. From the latter we calculate $E_{V-K}=0.21$ according to \citet[][their Table\,12]{1968nim..book..167J}. We note that despite the fact that all extinctions are calculated rather than fitted, they agree very well with the observations. No dereddening is applied to the 2MASS ($J-K$) CMD.

 We also display two sets of 3.0\,Gyr isochrones in Fig.\ref{fig_member_cmd}, in both \emph{Gaia} and other colors. We show isochrones from the PAdova and TRieste Stellar Evolution Code\footnote{\url{stev.oapd.inaf.it/cgi-bin/cmd}} \citep[PARSEC:][]{2012MNRAS.427..127B,2014MNRAS.445.4287T,2017ApJ...835...77M,2019MNRAS.485.5666P} and the Yale-Potsdam-Isochrones\footnote{\url{astro.yale.edu/yapsi/}} \citep[YaPSI: ][]{2017ApJ...838..161S}. We note that the YaPSI isochrones were transformed into \emph{Gaia} colors using the transformations we derived, as described above, and detailed in the Appendix. The PARSEC isochrones provide \emph{Gaia} colors based on the revised passbands from \cite{2018A&A...617A.138W}. We observe a generally satisfactory agreement between the finally-selected cluster members and the isochrones. 

 During this comparison, we found that our relationship between $G_{BP}-G_{RP}$ and $B-V$ and that constructed from the colors provided in the PARSEC isochrones are incompatible with each other for stars redder than $B-V>1.5$. The PARSEC isochrones fail to reproduce the observed color distribution of the Hyades and Pleiades (see Appendix Fig.\,\ref{color_trafo}). We are unable to explain this difference in the colors and proceed as follows: For the CMD we always display both YaPSI with our transformed colors and PARSEC with their provided colors. Whenever we need to transform \emph{Gaia} colors of the cluster stars to $B-V$ (or vice versa), we use our derived relation. We see later that all stars for which we find rotation periods have $G_{BP}-G_{RP}<2.3$ ($B-V\approx 1.4$) and for these stars the difference is small, posing no problem for this study.

 The metallicity of Ru147 is generally reported to be $\text{[Fe/H]}\approx0.1$, with values ranging from $\text{[Fe/H]}=+0.08 \pm 0.07$  \citep[][$Z\approx0.017$]{2018A&A...619A.176B} to $\text{[Fe/H]}=+0.12 \pm 0.03$  \citep{2020AJ....159..199D}. We adopt $\text{[M/H]}=0.08$ for the PARSEC isochrone, corresponding to $Y=0.28$ and $Z=0.0175$, and $\text{[Fe/H]}= 0.0$ and $Y = 0.28$ (corresponding to solar metallicity and $Z=0.0162$)\footnote{The next higher metallicity available is $\text{[Fe/H]}= 0.3$} for YaPSI, aiming to be as close to Ru147 as is feasible.

 Although we display only 3.0\,Gyr isochrones, we find that both the 2.5 and 3.0\,Gyr PARSEC and YaPSI isochrones provide plausible fits to the cluster data, as commonly suggested in the literature, all the way from the lowest-mass stars to the clump stars on the giant branch. Neither isochrone is completely satisfactory in the blue hook region past the turnoff, where the convective core appears. We tend to favor the higher age because it requires lower values of extinction and reddening when matching the cluster data with the isochrone in Gaia colors. This is reasonable for a cluster as close as Ru147 is. \cite{2019ApJ...887..109T} have suggested an isochrone age of $2.7\pm 0.61 $\,Gyr, based on a PARSEC model fit to eclipsing binary systems in the cluster. We have no objection to this result, noting that both 2.5\,Gyr and 3.0\,Gyr are well within the uncertainties.

\section{Analysis of the K2 lightcurves} \label{lightcurves}

 We determine the rotation periods of stars by measuring the modulation of the stellar flux caused by the carriage of surface inhomogeneities such as star spots or plage across the stellar disk as the star rotates. When the orientation of the stellar rotation axis is sufficiently favorable, and the asymmetries are large enough and stable enough, periodicity can be observed and measured, even visually in the best cases, by counting the number of pattern repetitions over the time baseline available.  

\subsection{Basic K2 lightcurve information}

 The 82.5\,d observational baseline, while long by the standards of most ground-based campaigns, still limits the detectability of long periods, potentially problematic for the late type stars in Ru147. To identify a period reliably based on spot motion, we typically need to recognize three occurrences of the spot. Therefore, the observational baseline needs to be longer than double the period. It is, however, possible to identify periods with a shorter baseline when more than one spot is visible in the light curve,  so that their individual signatures can be assigned unambiguously and yield similar periods. Differential rotation and spot evolution often further complicate the period analysis. The reduced amplitude of smaller spots in old stars, combined with the limited (82\,d) window of observation, makes the detection of long periods in old stars a matter of good fortune. (Ground-based studies \citep[e.g.][]{1994A&A...282..535S, 2013ApJ...768..155H, 2018ApJ...855...75R,2018A&A...614A..35M} often monitor stars for multiple years, and occasionally even decades, before they are able to detect a large-enough spot group to derive the rotation period.)

\begin{figure}
    \centering
    \includegraphics[width=\linewidth]{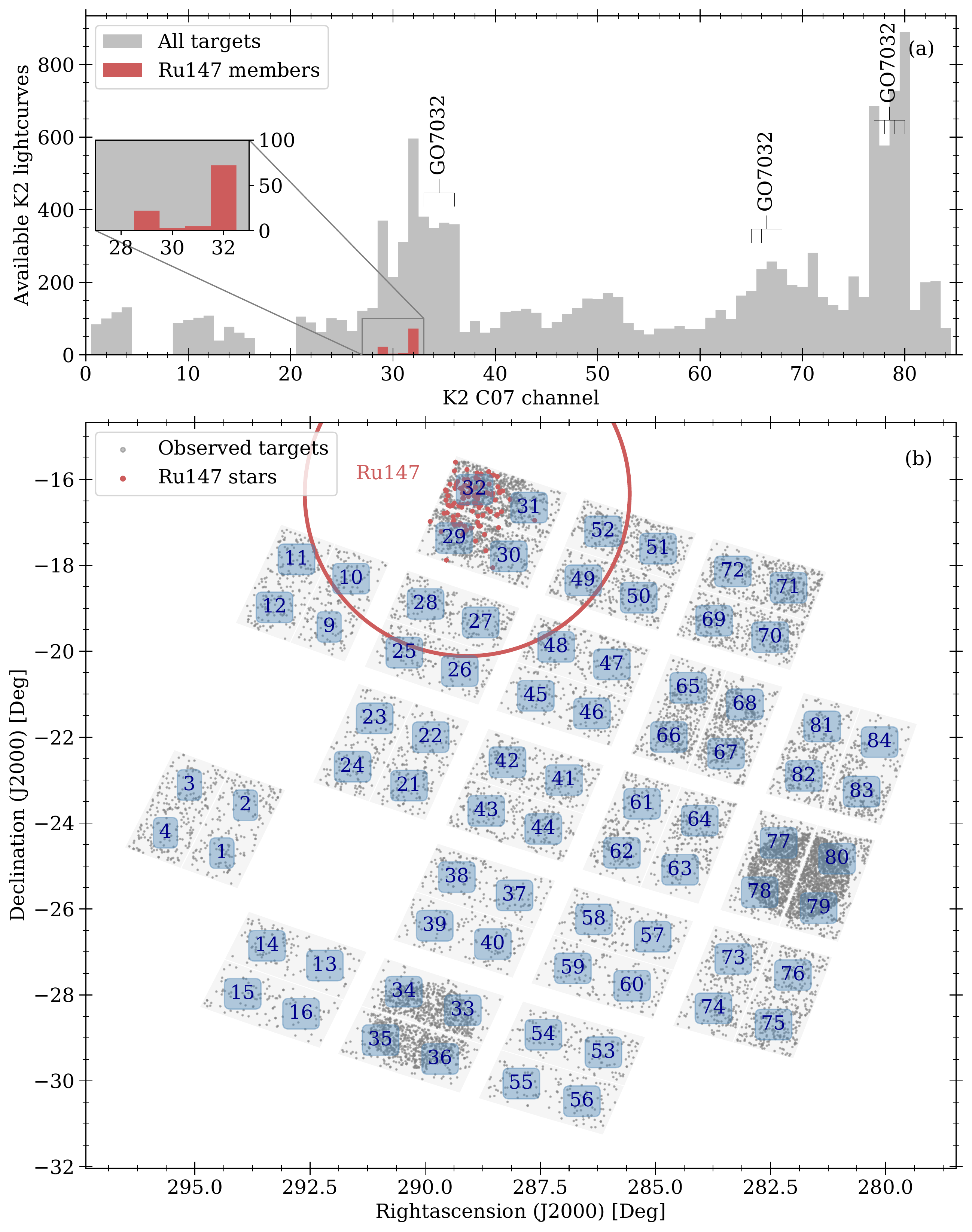}
    \caption{
        Overview of K2 campaign 07. Panel (a): Histogram displaying the C07 target distribution over the channels (enumerated as displayed) that constitute the K2 CCD array. Ru147 targets (red) are mainly in channels 29, 31, and 32, while the others (gray) are everywhere.             Panel (b): Sky map displaying the spatial coverage of all K2 C07 targets. Ru147 (encircled) is located at the northern edge. The footprints of a galactic archaeology campaign (GO7032) are prominent in both panels.
    }
    \label{fig_k2_c07}
\end{figure}

 The K2 field of C07 would initially have missed the cluster were it not for a community-driven effort that led to a change in the telescope pointing to include the center of the cluster at the edge of the field of view (cf. lower panel in Fig.\,\ref{fig_k2_c07}). The related K2 observing program ``K2 survey of Ruprecht 147 - the oldest nearby star cluster'' (GO7035\footnote{\url{keplerscience.arc.nasa.gov/data/k2-programs/GO7035.txt}}) added about 1000 stars of interest to the K2 target list distributed over 60 target pixel files (TPFs). As described above, we find that 102 member stars of Ru147 (by our determination, integrating prior membership determinations) actually have K2 light curves. Because of data transmission limitations during the K2 mission, only target pixel files containing predefined pixel masks for selected stars are available, that is, no full frame images (FFIs) are available.

 Figure\,\ref{fig_k2_c07} (upper panel) shows the channel distribution for all C07 targets over the K2 field. Ru147 is located in the northern portion (channels 29, 31, and 32). Although our scientific interests are here confined to Ru147, we also make use of data from other channels for light curve corrections, as discussed further below. Notably, a galactic archaeology campaign (GO7032) observed $>4000$ targets whose footprints are also visible in the lower panel. (GO7012, which observed Pluto and GO7025, which observed Trojan and Hilda asteroids, is already removed in this overview.) Two Kepler modules became dysfunctional early in the Kepler mission and account for the blank spaces in both panels of Fig.\,\ref{fig_k2_c07}.

\begin{figure*}
    \centering
    \includegraphics[width=\linewidth]{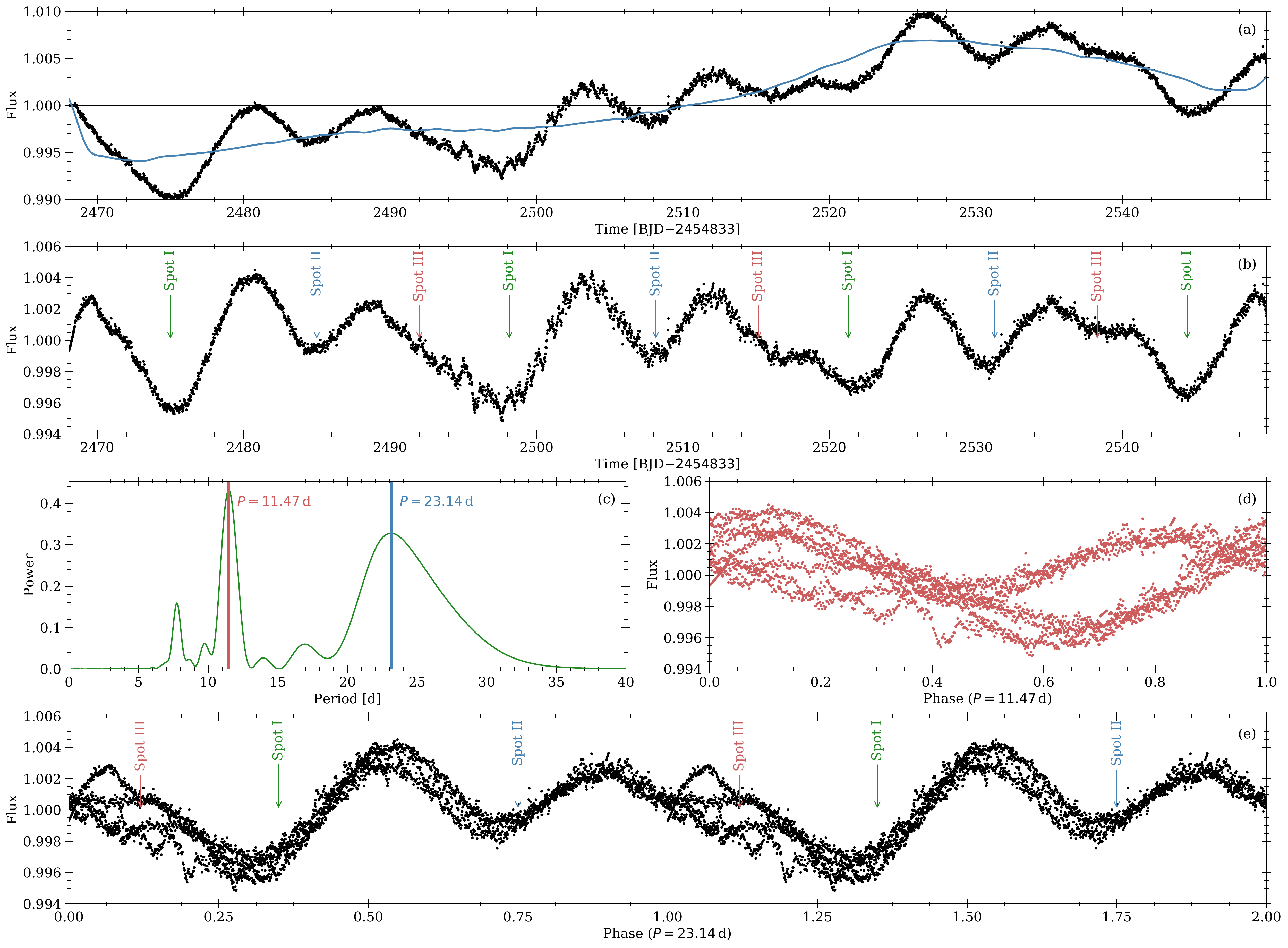}
    \caption{
        Light curve processing and period analysis for EPIC\,219297228. The light curve exhibits a complex structure from multiple spot groups and spot evolution; panel (a) shows the \texttt{EVEREST} lightcurve (black) and the reconstruction from the PCA (blue). Panel (b) displays the PCA-corrected lightcurve (black) with large spot features marked. Panel (c) shows the result of a Lomb-Scargle analysis with the maximum in the power spectrum indicated in red and the period determined by manual inspection of the light curve in blue. Panels (d) and (e) show the light curve phase folded with the periods obtained from the Lomb-Scargle and the manual inspection, respectively. For this particular light curve, the Lomb-Scargle analysis preferentially picks out half the true period despite the clearly different shapes of the recognizable spot features.  
    }    
    \label{fig_example_lc}
\end{figure*}

 For the acquired TPFs, various attempts have been carried out to extract de-trended light curves that are free of systematics in the observations and to correct, for example, the image drift during the observation. We do not attempt to perform this very-specialized data extraction process ourselves; instead, we use the \verb+EVEREST+ light curves \citep{2016AJ....152..100L} as our basic input. These were extracted using a pixel level de-correlation function \citep[PLD,][]{2015ApJ...805..132D}. Other methods, for example, \verb+k2sc+ \citep{2016MNRAS.459.2408A}, are useful primarily for exoplanet search and use purely mathematical approaches on K2 lightcurves extracted using simple aperture photometry (SAP) to correct for common trends in the data. These latter methods generally overfit the lightcurves and eliminate all traces of intrinsic, long-term, stellar variability. As a result, such lightcurves are unsuitable for our purposes; hence our usage of the \verb+EVEREST+ lightcurves.

\subsection{Light curve detrending}\label{sec_pca_1}
  
 Despite the sophistication of the method employed by \cite{2016AJ....152..100L} to extract the \verb+EVEREST+ light curves, various trends and correlations are still apparent. We are unable to determine the origin of such trending conclusively, but it could plausibly be attributed to instrumental systematics which appear to become more pronounced in the lightcurves after the extraction by \cite{2016AJ....152..100L}. In any case, the lightcurves, as provided, are unsuitable for analysis of periodicity, and must be reworked. To ameliorate the trending in the curves, we perform a Principal Component Analysis (PCA) on sets of lightcurves. 
  
 The technical and procedural details of the PCA are described fully in Appendix\,\ref{sec_pca}. We also display the 32 individual light curves and their PCA corrections in the Appendix (Figs. \ref{fig_lc_1}\,--\,\ref{fig_lc_32}). The upper panel (a) in each figure shows the \verb+EVEREST+ light curve and the reconstruction used, while the lower panels show the corrected and phased light curves. We note that our correction is superior to that with, for instance, a simple, higher order polynomial. While both approaches involve some level of subjectivity in the fitting process, that subjectivity is strongly reduced for the PCA. This is achieved via the dominant role played by the common trends in the PCA components, in opposition to a polynomial fit which only acts on the individual light curve, ignoring prior knowledge of shared systematics. Consequently, a polynomial fit is prone to overfit stellar signal with a long baseline and to ignore short baseline systematics. This can, in principle, be overcome by a manually fine-tuned fit of a higher order polynomial ($\ge$5) but this only replaces the identification of systematics using the PCA  with a more subjective one that varies from star-to-star. The PCA correction provides us with detrended light curves for subsequent periodicity analysis. We believe that these light curves  (See Fig.\,\ref{fig_example_lc} and Appendix\,\ref{apx_lightcurves}) are far more representative of the underlying astrophysical reality than are the \verb+EVEREST+ light curves.

\subsection{Analysis of periodicity}\label{sec_analysis}

 Our principal targets from the point of view of rotation are the cooler stars among the K2 targets that are on the cluster main sequence. Nevertheless, we have inspected all 102 member stars in K2 for possible periodicity. We eliminated most stars in evolutionary states past the turnoff (TO) from our sample. Stars at the turnoff (i.e., Spectral type G0; $(G_{BP}-G_{RP}) = 0.7$) and even somewhat cooler ones display no discernible periodicity. However, we retained some bluer stars near the core-convection hook in the isochrone, one star blueward of the giant branch, and a number of photometric binaries. (These are flagged accordingly below.) We also eliminated lightcurves that clearly show features from eclipsing binary/planetary systems and those that are essentially featureless. 

 As one can see from the example of EPIC\,291722781 in Fig.\,\ref{lc_correction} and from the light curves constructed by our PCA procedure for the final sample, and displayed in the Appendix (Figs.\,\ref{fig_lc_1}\,--\,\ref{fig_lc_32}), periodicity is visually recognizable for all the light curves retained\footnote{Such a choice could be considered overly conservative, and others might have chosen inclusivity, but we prefer to retain an exclusive sample in this work.}. One can typically read off the approximate rotation period of the star in question by inspection, with the proviso that in most cases, more than one spot group is present. The periodicity analysis discussed below simply serves to quantify the visually observed periodicity.
 
 The periodicity in the processed light curves is measured by subjecting each one to Lomb-Scargle analysis \citep{1976Ap&SS..39..447L,1982ApJ...263..835S}. The algorithm is run for periods in the range $0.2\,\text{d} \leq P \leq 40.0\,\text{d}$\footnote{The K2 baseline of 82.5\,d does not permit secure identification of any periodicity longer than this.}, with a logarithmic spacing of $\Delta\,\text{log\,}P = 0.001\,\text{dex}$. For a minority of lightcurves that show periodic variations over the complete duration of C07, this approach is able to identify the correct periods without further intervention. However, most of our lightcurves display spot evolution and/or multiple spot groups. This requires that we inspect all lightcurves manually to identify the correct period. The lightcurves are then phase-folded to match features. 
 
 This process is illustrated in Fig.\,\ref{fig_example_lc}, but for EPIC\,219297228 which exhibits signals of at least three clearly distinct spot groups. In fact, we observe multiple spot features for the majority of our stars. And with the exception of  EPIC\,219353203, it is only the fastest rotating stars in our sample that show only one (large) feature. This observation is consistent with the findings of \cite{2018ApJ...863..190B}, where the incidence of multiple spot groups was found to increase with rotation period.

 The identification of rotation periods from starspot features can be hindered by both differential rotation, which yields slightly different periods for each spot, and also spot evolution, which occasionally makes spots (dis)appear. We estimate the period error from the phase-folded light curve. We do this by examining the above-mentioned effects and the extent to which they allow period changes that still result in an acceptable phase folded light curve. If no such effects are present, the period error is found to be generally small $\sim$3\%, owing to the photometric precision and the short cadence of Kepler data. In the worst case, the error is on the order of $\sim$25\%.

\begin{figure}
    \centering
    \includegraphics[width=\linewidth]{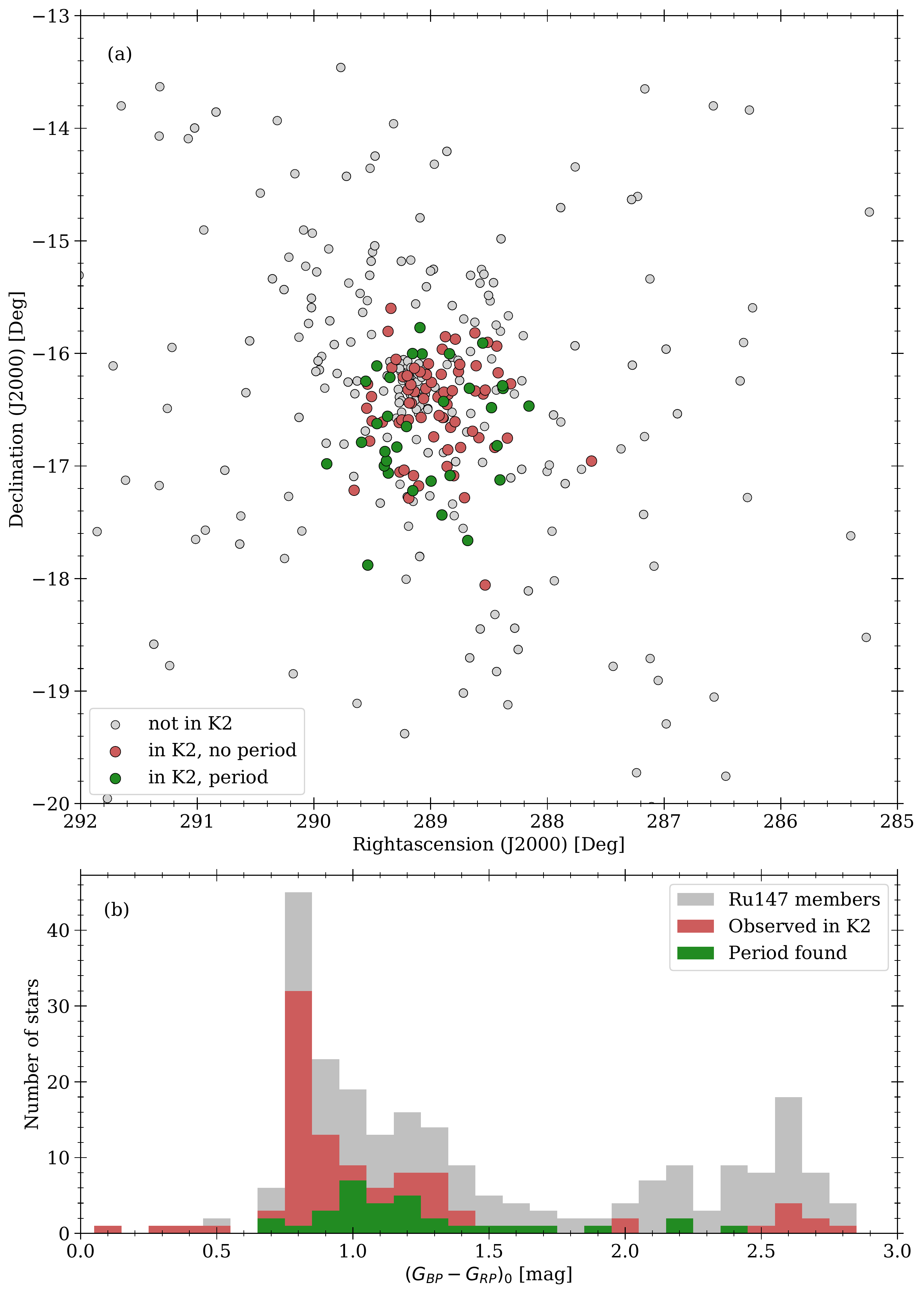}
    \caption{
        Fraction of stars with detected periods. Panel (a) shows the spatial distribution of the cluster stars, color coded to indicate both coverage in K2, and whether a periodic  signal was identified (green) or not (red). \label{fig_sample_map} Panel (b) shows a histogram of the stars identified as Ru147 members and its coverage in K2. Gray depicts all stars identified as Ru147 members in Sect.\,\ref{sec_member}. Stars for which we found periodic signals are displayed in green, while red symbols indicate the remaining stars observed by Kepler/K2. Red and green that denote stars that were observed during K2 C07 and green are the stars for which we found a periodic signal.
    }
    \label{fig_coverage_hist}
\end{figure}

\begin{table*}
    \centering
    \caption{ Periodicity and related information for the 32 sample stars. (This table is available in electronic form at the CDS via anonymous ftp to cdsarc.u-strasbg.fr (130.79.128.5) or via http://cdsweb.u-strasbg.fr/cgi-bin/qcat?J/A+A/)}
    \label{tab_sample_periods}
    \begin{tabular}{ccccccccc}
        \hline
        \hline
             &&\\[-0.7em]
        EPIC & $(G_{BP}-G_{RP})_0$ & $(B-V)_0$\tablefootmark{a} & $P$& $\Delta P$ & Components & Category & Flag\tablefootmark{b} & Final sample\tablefootmark{c} \\
             & [mag]& [mag] &[d]&[d]&\\
             &&\\[-0.7em]
        \hline &&\\[-0.7em]
        218933140 & 0.87 & 0.69 & 20.4 & 2.5 & 9 & 1 & MS & yes \\
        219037489 & 0.99 & 0.80 & 22.8 & 1.5 & 6 & 1 & MS& yes \\
        219141523 & 2.27 & 1.50 & 26.9 & 1.8 & 4 & 1 & MS& yes \\
        \it 219238231 & \it 0.78 & \it 0.61 & \it 28.1 & \it 0.9 & \it 3 & \it 2 & \it MS & \it no \\
        \hline &&\\[-0.7em]
        \it 219275512 & \it 0.88 & \it 0.70 & \it 20.4 & \it 0.5 & \it 6 & \it 2 & \it BIN & \it no \\
        219280168 & 1.15 & 0.95 & 23.0 & 0.8 & 5 & 1 & MS & yes \\
        219297228 & 1.13 & 0.93 & 23.1 & 0.4 & 6 & 1 & MS & yes\\
        219306354 & 0.98 & 0.79 & 22.8 & 0.8 & 5 & 1 & MS & yes\\
        \hline &&\\[-0.7em]
        \it 219333882 & \it 1.16 & \it 0.95 & \it 11.6\tablefootmark{f} & \it 0.5 & \it 11 & \it 2 & \it MS & \it no \\
        \it 219341906 & \it 0.75 & \it 0.58 & \it 1.6 & \it 0.1 & \it 14 & \it 1 & \it TO & \it no \\
        219353203 & 2.1 & 1.48 & 21.6 & 0.7 & 5 & 1 & MS & yes\\
        219388192 & 0.86 & 0.69 & 12.5 & 0.2 & 8 & 1 & MS\tablefootmark{d} & yes\\
        \hline &&\\[-0.7em]
        \it 219404735 & \it 0.79 & \it 0.62 & \it 24.4 & \it 1.1 & \it 5 & \it 1 & \it BIN & \it no \\
        \it 219409830 & \it 0.83 & \it 0.66 & \it 9.6 & \it 0.4 & \it 7 & \it 2 & \it MS & \it no \\
        219422386 & 1.01 & 0.81 & 22.6 & 1.3 & 5 & 1 & MS & yes\\
        219479319 & 1.6 & 1.31 & 20.1 & 0.7 & 4 & 1 & MS & yes\\
        \hline &&\\[-0.7em]
        219489683 & 1.78 & 1.40 & 19.0 & 0.5 & 5 & 1 & MS & yes\\
        \it 219515762 & \it 0.57 & \it 0.43 & \it 5.7 & \it 0.1 & \it 8 & \it 1 & \it TO\tablefootmark{e} & \it no \\
        219545563 & 0.93 & 0.75 & 22.2 & 1.6 & 4 & 1 & MS & yes\\
        219551103 & 1.0 & 0.81 & 22.0 & 1.8 & 5 & 1 & MS & yes\\
        \hline &&\\[-0.7em]
        219566703 & 1.51 & 1.25 & 23.2 & 1.5 & 5 & 1 & MS & yes\\
        \it 219610232 & \it 1.17 & \it 0.97 & \it 5.6 & \it 0.2 & \it 5 & \it 1 & \it BIN & \it no \\
        219610822 & 1.11 & 0.91 & 23.1 & 1.5 & 5 & 1 & MS & yes\\
        219619241 & 2.11 & 1.48 & 22.1 & 1.5 & 5 & 1 & MS & yes\\
        \hline &&\\[-0.7em]
        219634222 & 1.29 & 1.08 & 27.3 & 3.5 & 5 & 1 & MS & yes\\
        \it 219646472 & \it 0.61 & \it 0.47 & \it 22.1 & \it 2.0 & \it 3 & \it 1 & \it TO & \it no \\
        219683737 & 1.07 & 0.88 & 21.7 & 1.0 & 5 & 1 & MS & yes\\
        219721519 & 1.1 & 0.90 & 21.9 & 1.5 & 2 & 1 & MS & yes\\
        \hline &&\\[-0.7em]
        219722212 & 0.94 & 0.75 & 22.7 & 2.1 & 5 & 1 & MS & yes\\
        219722781 & 1.39 & 1.16 & 21.4 & 0.5 & 7 & 1 & MS & yes\\
        \it 219755108 & \it 0.94 & \it 0.76 & \it 29.4 & \it 0.5 & \it 4 & \it 1 & \it BIN & \it no \\
        \it 219800881 & \it 0.9 & \it 0.72 & \it 32.7 & \it 8.1\tablefootmark{f} & \it 5 & \it 2 & \it MS & \it no \\
        \hline
    \end{tabular}
    \tablefoot{
        \tablefoottext{a}{Calculated from $(G_{BP}-G_{RP})_0$ with our derived transformation.}\\  
        \tablefoottext{b}{MS\,=\,Main sequence, TO\,=\,Turn-off, BIN\,=\,(possible) binary?}\\ 
        \tablefoottext{c}{Star used for detailed comparison in Sect.\,\ref{sec_empiricalcomparison} and \ref{sec_model_comparison}}\\ 
        \tablefoottext{d}{Spectroscopic binary (G+M star) and eclipsing Brown Dwarf companion \citep[e.g.][]{2018AJ....156..168B}}\\ 
        \tablefoottext{e}{Suspicious proper motions, cf. Fig\,\ref{member_map}, and only mentioned in Cu13, cf. Table\,\ref{tab_sample_overview}, thus likely not a member}\\ 
        \tablefoottext{f}{Ambiguity in the matching of visible spot features}
    }
\end{table*}

 More difficulties arise in noisy data, or when the degree to which the PCA is performed results in ambiguities. Therefore, we assign each period found to one of two categories based on the reliability of the signal found. Category 1 denotes periods in which we have great confidence, while category 2 periods are those where doubts can reasonably be entertained. By this classification, we aim to reduce the impact of possible false-positives. Because we aim for the greatest confidence in the final sample of rotation periods, we have been relatively conservative in accepting light curves as periodic and more so when assigning Cat.\,1 to it. For the time being, and for the convenience of researchers interested in non-rotational variability, we retain evolved stars, binaries, etc., but we will mark or remove them in due course.  The results of the periodicity analysis are summarized in Table\,\ref{tab_sample_periods}.

 We note that our derived periods display a visually similar distribution to the one found by \cite{2018csss.confE..24C} (hitherto unpublished, but see below), with the exception of a handful of stars in the vicinity of $G_{BP} - G_{RP} \approx 0.6$, which is not present in our sample.   The lower (b) panel in Fig.\,\ref{fig_coverage_hist} displays an overview in \emph{Gaia} color that depicts the fraction of member stars by color for which a period was ultimately found in the K2 data. The upper (a) panel of Fig.\,\ref{fig_coverage_hist} shows the spatial locations of these stars. It is rather obvious in Fig.\,\ref{fig_coverage_hist} that late-type stars are vastly underrepresented. The total number of stars with $G_{BP}-G_{RP}>1.5$\,mag is probably much higher and a large number of stars is likely simply missing in our sample. Furthermore, K2 targets are clearly biased towards solar type stars. Aside from the obvious predominance of solar type stars, the faintness of M-dwarfs also contributes to this bias.

\begin{figure}
    \centering
    \includegraphics[width=\linewidth]{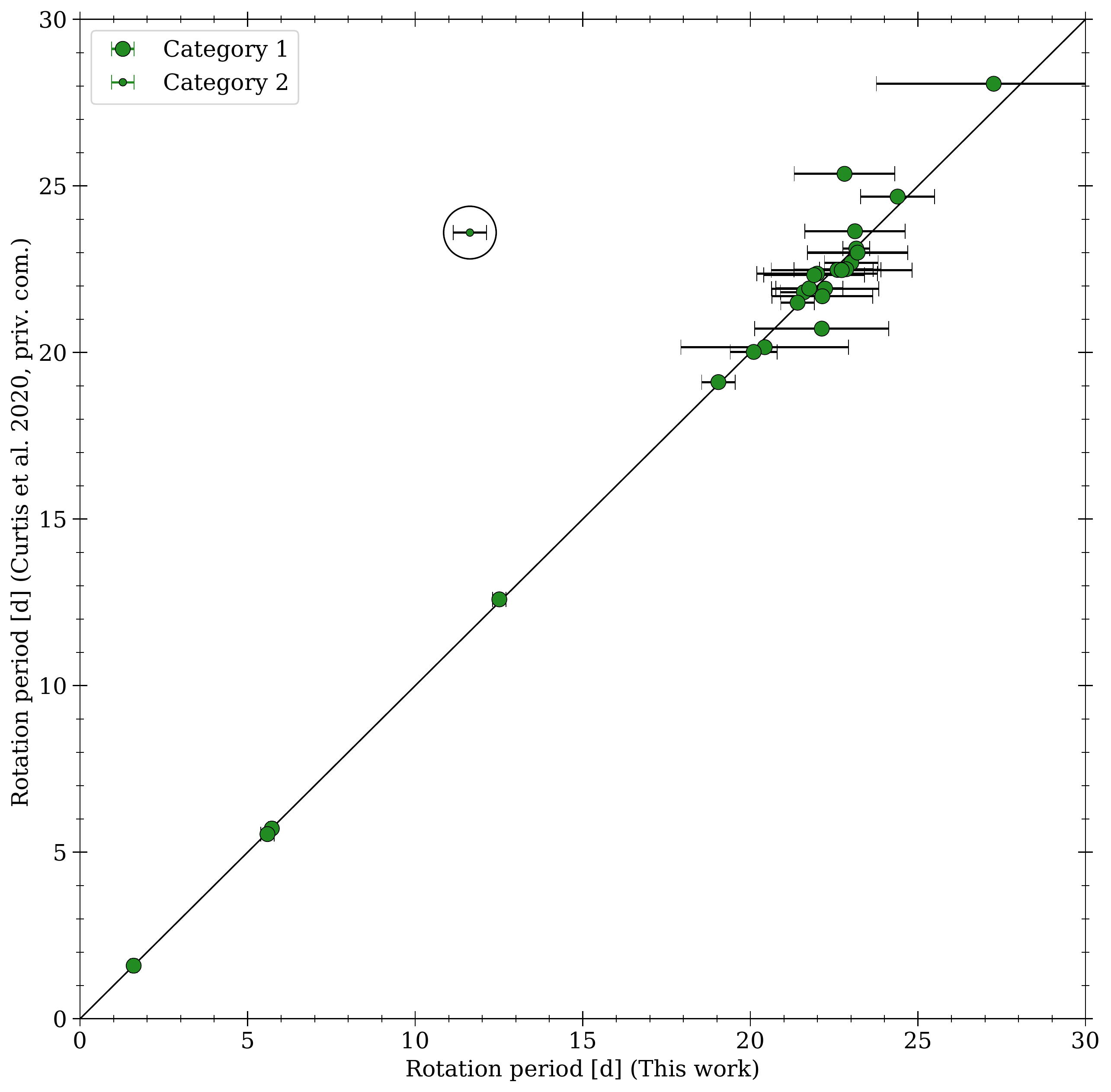}
    \caption{
        Comparison of our periods with those found by \citep[][ApJ subm., priv. comm.]{Curtis2020} for the 26 stars common to both samples.             The encircled outlier (EPIC\,219333882, which does not make it into our final 21-star sample) has an ambiguous light curve that permits multiples of the period listed.
    }
    \label{fig_period_comparison}
\end{figure}

 A parallel and independent study of this cluster has been carried out by \citet[][ApJ subm., priv. comm.]{Curtis2020}, with whom we have exchanged periodicity data (but no other information, to preserve independence) after both publications were essentially complete. This exchange allows us to compare the periods for all 26 stars in common to both studies, as shown in Fig.\,\ref{fig_period_comparison}. We are pleased to report very good agreement between their periods and ours, with the exception of one outlier (EPIC\,219333882, encircled in Fig.\,\ref{fig_period_comparison}).
 
 This outlier is assigned Cat.\,2 by us because of an ambiguity that, in principle, allows to double or even triple the associated period and still obtain a reasonable phased curve (cf. Fig.\,\ref{fig_lc_9}). Doubling our period would put it in good agreement with the distribution observed for the other stars and suggests that that we have likely identified half the true period. However, we have decided to list the star as is, because our light curve by itself evinces no preference for the longer period\footnote{In fact, we have taken some care in our work to compartmentalize each star and not to let the results of neighboring stars affect periodicity judgements. This makes our work comparable to field star studies, where the occasional multi-spotted star could potentially be assigned a submultiple of the true period and hence provide a significantly younger rotational age than the real one.}.  We also note that of our final 21 star sample (see below), 20 stars are common to both studies, 19 of which have periods that agree within $\leq 2$\,\%; only EPIC\,219037489 is more discrepant than the error bar \footnote{We infer from this that our uncertainties are likely reasonable.} ($\approx 10$\,\%).

\section{Rotation periods in the CMD and CPD}\label{sec_results}
 
 The near-final sample of periodic stars that constitute the result of our analysis contains 32 stars in which a periodic signal could be identified and plausibly attributed to stellar rotation. For the time being (and for the convenience of other researchers), we retain various objects unsuited to our main sequence rotation interests such as evolved stars, binaries, and stars as blue as spectral type F3V, the last clearly stars without surface convection zones. The cross-identifications and other basic properties of these stars are summarized in Table\,\ref{tab_sample_overview}.

\begin{figure}[ht!]
    \centering
    \includegraphics[width=1\linewidth]{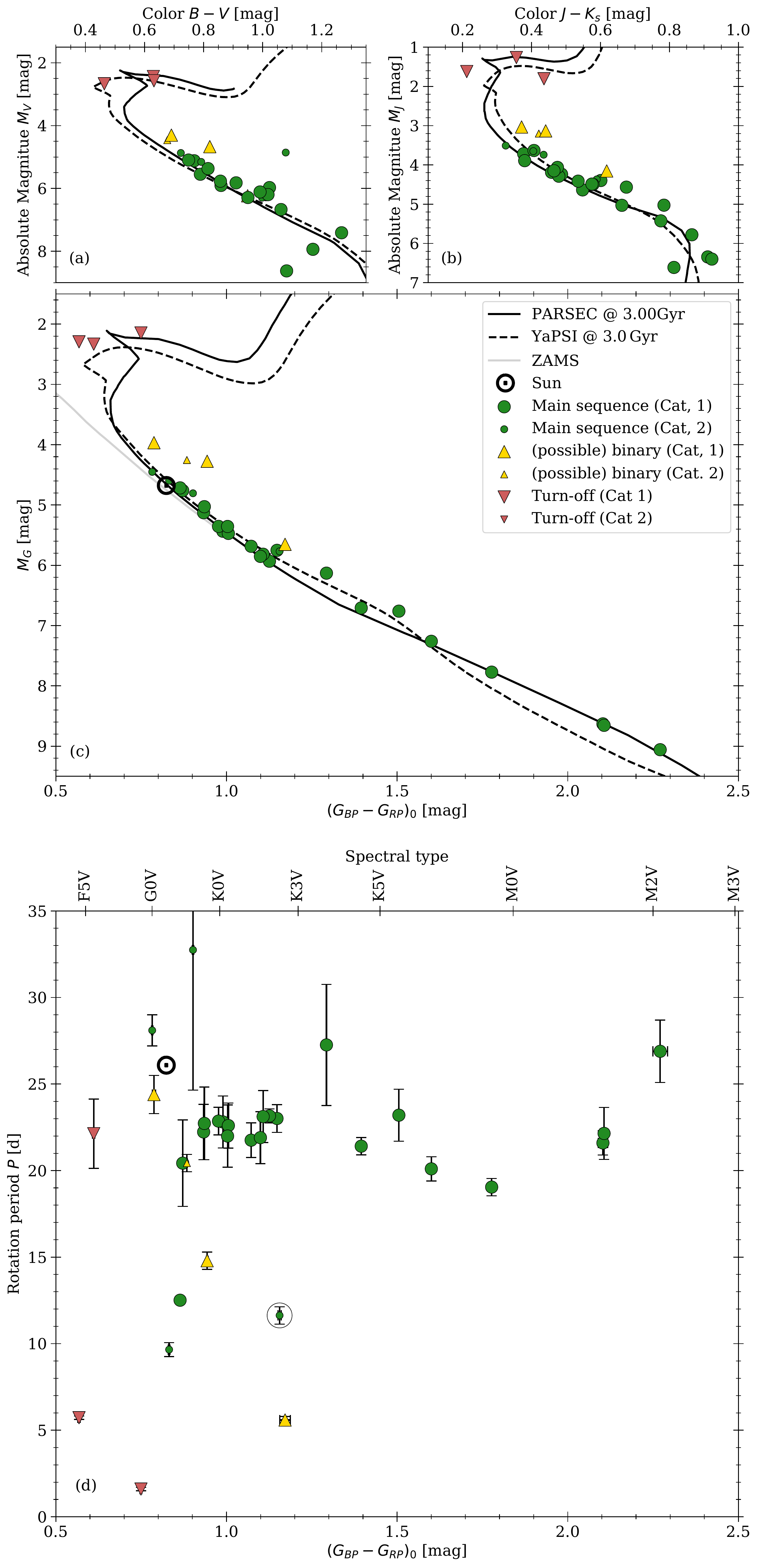}
    \caption{
        CMDs in various color systems (upper panels (a), (b), and (c)) for the 32 periodic Ru147 cluster members. The bottom panel (d) shows the corresponding color-period diagram (CPD) for the same stars. The 3 F-type stars past the cluster turnoff are marked with red symbols, single main sequence cluster members are in green, and yellow symbols indicate known binaries. The plot symbols also encode both object type and period category (large symbols for Cat.\,1 periods, small for Cat.\,2) as indicated in the legend. The sun is marked with its usual symbol in both the CMD and CPD and is displayed only as a reference point. The encircled star is EPIC\,219333882, for which multiples of our period are equally plausible (see text).
    }
    \label{fig_cmd_base}
\end{figure}

 In addition to the derived period and the number of components required, this table lists the assigned category of reliability as described before\footnote{For various reasons, we err on the side of caution, so it is quite possible that future work on the same data could yield a larger sample of acceptable periods.}. We also flag stars in Table\,\ref{tab_sample_periods} to indicate their evolutionary and binary status. The latter criterion is given when a star sits above the main sequence but is clearly redder then the turn-off (undetected multiplicity), or when the stellar environment suggests light contamination due to crowding.

 These periodic star results are displayed in color-magnitude diagrams (CMDs) in panels (a)\,--\,(c) of Fig.\,\ref{fig_cmd_base} and in a color-period diagram (CPD) in panel (d) of Fig.\,\ref{fig_cmd_base}. We see that three of the F-type stars (red triangles in the figures), indeed some of the bluest stars of our periodic sample, are clearly evolved past the turnoff and are in the vicinity of the blue hook. Two of their rotation periods are below $6$\,d, while one is far higher, at $\sim$22\,d. Four additional stars (yellow triangles) are photometric binaries which are located significantly above the single star sequence in the CMD. Their rotation periods also have a wide range, from $5.6$\,d to $29.4$\,d, with all but one being clear outliers also in the CPD.

 The remaining 25 periodic stars (green circles) are all plausibly on the cluster's single star main sequence in the CMDs. As can be seen in the corresponding CPD, these stars also display a wide range of rotation periods, ranging from under $10$\,d to almost $33$\,d. However, the majority (19 out of 25) of these GKM-type main sequence stars occupy a horizontal band between 19\,d and 27\,d periods across the GKM spectral range. These are all stars for which we have great confidence in the periods determined (Category 1).  The remaining 6 stars are outliers, based both on the measured distribution itself and prior expectations from studies of other open clusters. 

 We now trim our dataset down to those periodic stars that are on the main sequence, for which no contaminating flux is evident and for which we have a high degree of confidence (category 1) in the rotation periods. This leaves us with 21 stars, cf. the \emph{Final sample} column in Table\,\ref{tab_sample_periods}, which will be the only ones we use for the remainder of this paper.

\section{Comparison with other empirical cluster period work } \label{sec_empiricalcomparison}

\begin{figure*}
    \centering
    \includegraphics[width=\linewidth]{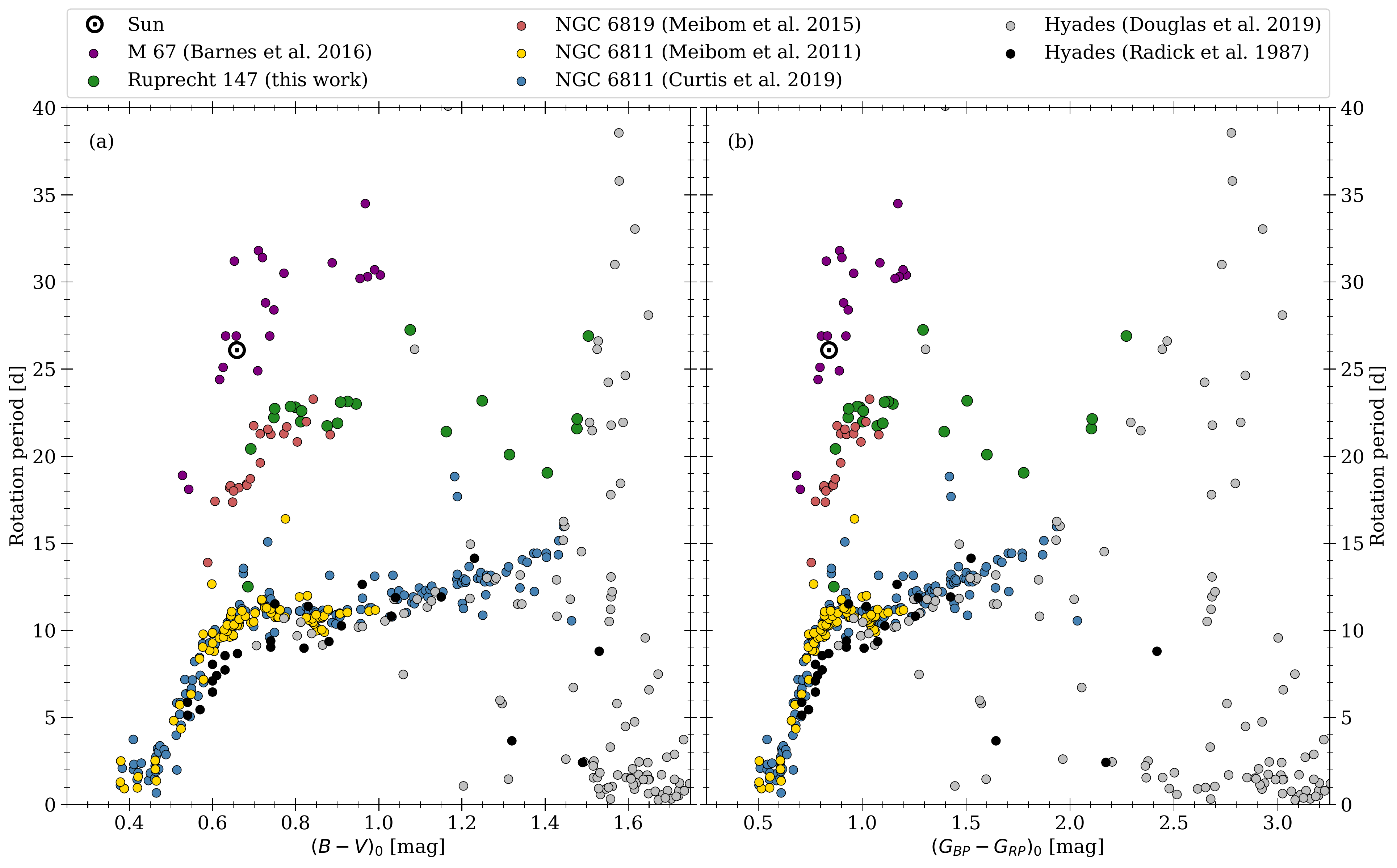}
    \caption{
        Color-Period-diagrams in Johnson $B-V$ and Gaia colors for Ru147 in relation to those for other relevant clusters. We see that Ru147 stars (green) are sandwiched between the 4\,Gyr-old cluster M67 (purple), and the 1\,Gyr-old cluster NGC\,6811 (yellow, blue), as expected. The Ru147 rotation periods also connect smoothly to the rotation period data for the 2.5\,Gyr cluster NGC\,6819 (red), overlapping well in the G-K spectral range. We also display the distribution(s) for the younger ($\sim$625\,Myr) Hyades open cluster (black, gray). (See legend and text for references.)
    }
    \label{fig_cpd_cluster}
\end{figure*}

 Before comparing our measured periods with models we wish to show the context of, and continuity with, other work in the literature. There are three other relevant open clusters for which rotation periods are available, all of which are based on work with Kepler or its K2 reincarnation. These are the 4\,Gyr-old open cluster M\,67 \citep{ApJ...823..2016.16B}, the 2.5\,Gyr-old open cluster NGC\,6819 \citep{2015Natur.517..589M}, and the 1\,Gyr-old cluster NGC\,6811 \citep{2011ApJ...733L...9M,2019ApJ...879...49C}. The measured rotational distributions for their cool stars are also displayed in Fig.\,\ref{fig_cpd_cluster}. In order to avoid any possible color-related inconsistencies, we associate the stars with measured rotation periods with their \emph{Gaia} colors (the most uniform currently available), which we subsequently deredden appropriately. The reddening parameters used are listed in Table\,\ref{tab_reddening_refs}. We also transform this $G_{BP} - G_{RP}$ color into $(B-V)_0$ color to display the same information in $B-V$ color in an additional panel for the reader's convenience.

 The M\,67 data display the greatest dispersion in rotation period, likely a consequence of the difficulty of determining rotation periods in this relatively old cluster, and the result of its having been observed for only one K2 quarter. The rotation periods are taken from the study of \cite{ApJ...823..2016.16B}, where the large rotation period uncertainties can be appreciated. Despite this dispersion, it is clear that all the M\,67 stars are located \emph{above} the Ru147 stars in a mass-dependent way, as expected. This fact informs us that Ru147 is younger than the 4\,Gyr-old M\,67 cluster.

\begin{table}
    \centering
    \caption{
        Reddenings used for the individual clusters in Fig.\,\ref{fig_cpd_cluster}. Calculated reddenings use the relationship $1.339\cdot E_{B-V} = E_{G_{BP} - G_{RP}}$ \citep{2018MNRAS.479L.102C} and the arrow indicates the direction of the calculation.
    }
    \label{tab_reddening_refs}
    \begin{tabular}{rcccc|cc}
        \hline\hline
        Cluster & $E_{B-V}$ &  & $E_{G_{BP} - G_{RP}}$ & Ref & [Fe/H] & Ref \\
        \hline
        NGC\,6819    & $0.15$  & $\Rightarrow$ & $0.201$ & 1 & 0.05  & 6 \\
        NGC\,6811    & $0.48$  & $\Rightarrow$ & $0.065$ & 2 & -0.05 & 6 \\
        Hyades       & $0.027$ & $\Rightarrow$ & $0.036$ & 3 & 0.13  & 7 \\
        M\,67        & $0.04$  & $\Rightarrow$ & $0.054$ & 4 & 0.03  & 8 \\
        Ru\,147      & $0.075$ & $\Leftarrow$  & $0.1$   & 5 & 0.12  & 6 \\
        \hline
    \end{tabular}
    \tablebib{(1)˜\cite{2015Natur.517..589M}; (2) \cite{2019ApJ...879...49C};
    (3) \cite{2006AJ....132..111J}; (4) \cite{ApJ...823..2016.16B}; (5) this work;
    (6) \cite{2020AJ....159..199D}; (7) \citet{2016A&A...585A.150N}; (8) \citet{2017MNRAS.470.4363C,2019MNRAS.490.1821C}.}
\end{table}

 Conversely, all the rotation periods measured in NGC\,6811  are located below those of Ru147, again in a mass-dependent way. This tells us that Ru147 is clearly older than the 1\,Gyr-old NGC\,6811 cluster. We display the rotation period determinations of both \cite{2011ApJ...733L...9M}, based on a single Kepler quarter, and those of \cite{2019ApJ...879...49C}, based on the entire 4 year Kepler dataset. We note the good agreement of the majority of the rotation periods between the two studies and especially the very well-defined sequence of NGC\,6811 in the CPD. The latter is likely the result of NGC\,6811's relative youth, which manifests itself in relatively large flux variations from starspots and the fact that NGC\,6811 was located in the Kepler field itself, allowing for it to be observed over the entire 4yr baseline.

 We also show the rotation periods measured by \cite{1987ApJ...321..459R} in the younger ($\sim$ 625\,Myr-old) Hyades open cluster. This sequence of rotation periods was the first to be measured, and provided the first significant clue to the mass dependence of stellar rotation in cool stars. As expected, they are located below the NGC\,6811 data, except in the mid-K spectral type region, where there is some overlap with the NGC\,6811 data. For completeness, we also display rotation periods from the recent Kepler/K2 study of \cite{2019ApJ...879..100D}, which seem to be marginally below the NGC\,6811 values.

 The comparison with the 2.5\,Gyr-old NGC\,6819 cluster is perhaps the most revealing. In the region of the spectral types G-K, the rotation periods of the two clusters overlap one another significantly enough that they could almost be merged. This fact confirms that Ru147 is roughly the same age as the 2.5\,Gyr-old NGC\,6819 cluster. Our Ru\,147 rotation periods also extend the empirical rotational isochrone for ($2.5-3$\,Gyr) towards much lower masses. These rotation periods for the lower mass stars are somewhat shorter than those of the G-K stars in the cluster. This is somewhat unexpected and will be discussed further in Sect.\,\ref{sec_model_comparison}. We have unfortunately been unable to derive rotation periods for early G-type stars in Ru147, to confirm any possible overlap with NGC\,6819 in this mass range. This could be the result of one of more of the following: (a) our study using overly strict requirements in accepting periodicity, (b) the relatively small amplitudes of spot variability for such 2.5\,Gyr-old stars, and (c) the relatively poor quality and shorter baseline of the K2 Ru\,147 light curves, as opposed to the 4yr baseline of the higher-quality Kepler NGC\,6819 data. The NGC\,6819 periods and $(B-V)_0$ colors are taken from Extended Data Table\,1 in \citet[][]{2015Natur.517..589M}.

 We conclude from this empirical comparison that all extant cool star rotation period data for open clusters between 1 and 4\,Gyr, including the current ones for Ru\,147, are compatible with all these data lying on a single surface in color-rotation period-age space, as originally proposed by \cite{2003ApJ...586..464B} and as emphasized by \cite{2015Natur.517..589M} in connection with rotation periods in the 2.5\,Gyr-old open cluster NGC\,6819. The period determinations for Ru147 herein extend this surface towards lower mass stars at this important intermediate age. However, the detailed shape of the surface proposed appears to require revision for lower-mass stars, as argued by \cite{2019ApJ...879...49C}, when they extended the NGC\,6811 (1\,Gyr) rotation period data of \cite{2011ApJ...733L...9M} to the low mass range.

\section{Comparison with models} \label{sec_model_comparison}

 Another aim of our study is to examine the predictions of stellar spindown models in a region of parameter space (lower mass, combined with higher age stars of well-defined age) than has not been possible thus far, as can be appreciated in Fig.\,\ref{fig_cpd_cluster}. The goal of such efforts is of course to construct an empirically-constrained model of stellar rotational evolution across the largest-possible parameter range. Such models could be used to derive stellar ages via gyrochronology if the relationship between the underlying variables is suitably well-behaved, and more generally, to understand the physics of magnetic braking.

 The first of such mass-dependent models was that proposed by \cite{1988ApJ...333..236K}, subsequently implemented in the Yale Rotational stellar Evolution Code \citep[YREC;][]{1989ApJ...338..424P}, following a method for computing rotational stellar models first explicated by \cite{1978ApJ...220..279E}. This method of modeling rotating stars has also been implemented in the Geneva code \citep[e.g.][]{2000ARA&A..38..143M} which, although it is generally used for modeling hot stars, has been updated for usage in particular cool star contexts \citep{2012A&A...539A..70E,2019A&A...631A..77A}. Rotational evolution in all extant stellar models is overlaid on non-rotating stellar models (so-called ``standard models''), using a number of additional parameters unique to the rotational aspect of the modeling \citep[see][]{1989ApJ...338..424P}, with various tradeoffs between their number and the fidelity of description of the data \citep[e.g.][]{2010ApJ...721..675B}. We note that relevant data prior to the mid-1990s typically consisted of measured $v \sin i$ values, with notable exceptions being the rotation period work of \citet[][Pleiades]{1987A&AS...67..483V} and \citet[][Hyades]{1987ApJ...321..459R}.

 The advent of large-format CCDs allowed increasingly large numbers of rotation periods to be measured for both pre-main sequence stars \citep[e.g.][]{1992ApJ...398L..61A,1993A&A...272..176B,1995A&A...299...89B} and for main sequence stars \citep[e.g.][]{1993PASP..105.1407P,2006MNRAS.370..954I,2010A&A...515A.100J}, with corresponding steady pressure on models. \cite{2003ApJ...586..464B} collected the open cluster rotation periods then available for cool main sequence stars and identified color- and age-dependent patterns in the rotation period data that could be described by a simple empirical relationship between rotation period, color, and age using only three fitted numerical constants. The possibility of deriving the age (otherwise hard to measure) from the measured periods and colors led to his proposing the neologism gyrochronology for the associated age-determination procedure. A subsequent publication \citep{2007ApJ...669.1167B} showed that the associated uncertainties in the derived stellar age ($\sim 15-20\%$) for cool main sequence stars were indeed small enough to be useful and similar empirical relationships have been been subsequently proposed by \citet{2008ApJ...687.1264M} and \cite{2020arXiv200509387A}, among others.

 The undesirability of constructing separate relationships with new fitted parameters for each relevant color prompted \cite{2010ApJ...721..675B} and \cite{2010ApJ...722..222B} to formulate an empirical spindown relationship that captures the fact that cluster stars appear to have a bimodal rotation period distribution of fast- and slow rotators at the ZAMS, that subsequently erodes into a unimodal slow rotator distribution in older clusters. The fast- and slow asymptotic rotation period behaviors were formulated mathematically symmetrically, using the convective turnover timescale, $\tau$, in stars as the mass variable, to describe the two mass-dependent timescales in the problem. The usage of $\tau$ allowed the model to be translated into any relevant observed color as necessary, and arguably more importantly, connected to stellar magnetic activity and dynamo theory, where the convective turnover timescale, or equivalently the Rossby Number $Ro = P/\tau$ (or its inverse, the Coriolis Number, $Co$) has long been recognized as an important variable  \citep{1978GApFD...9..241D,1993SoPh..145..207D,1984ApJ...279..763N,1996ApJS..106..489P,2018JPlPh..84d7304B}. This (Symmetric Empirical) Model\footnote{The name was coined by \cite{2014ApJ...789..101B}, who advocated a ``Metastable Dynamo Model'', where the shape of the slow rotator sequence does not change over time.} requires only two dimensionless constants, $k_C$ and $k_I$, to describe rotational evolution on the main sequence, and will also be shown below, unmodified from its original, now 10\,yr-old formulation. A key feature of the \cite{2010ApJ...722..222B} model is that the morphology of the predicted cluster rotation period distributions changes with cluster age (as seen in the observations) and in contrast to the  Ba07 model and a number of other subsequent ones.

 These proposed models all have additional degrees of freedom. They include the ones of \cite{2020A&A...636A..76S}, \cite{2015ApJ...799L..23M}, \cite{2018ApJ...862...90G}, and \cite{2019A&A...631A..77A}, in order of increasing number of degrees of freedom. (See Fr20 for an inter-comparison of these models in the context of an extensive rotation period dataset for the Pleiades-age southern open cluster NGC\,2516.)Of these models, we also display our Ru\,147 data against the model of \cite{2020A&A...636A..76S} because it appears to come closest to describing them with a minimum of parameters and to the \cite{2019A&A...631A..77A} models, which in principle have enough degrees of freedom to enable them to describe the observations with greater fidelity. For an alternative perspective on this subject, one that emphasizes the magnetic braking perspective and uses scaling relations and associated parameters liberally, see the recent work of \cite{2020A&A...635A.170A}.

\subsection{The \cite{2010ApJ...722..222B} model} 

 Our first detailed comparison is with the model of \citet[][Ba10 hereafter]{2010ApJ...722..222B}, which uses the relationship listed in \citet[][their Table\,1]{2010ApJ...721..675B} to convert between stellar mass, temperature, and $U,\,B,\,V,\,R,\,I,\,J,\,H,\,K$ colors. This fact permits us to use the dimensionless scaling constants $k_C$ and $k_I$ unchanged from that work. Equation (32) from Ba10, explicit for the age,
 \begin{align}
    t & = \frac{\tau}{k_c} \rm{ln}\,\left(\frac{P}{P_0}\right) + \frac{k_I}{2 \tau} \left(P^2 - P_0^2\right) \label{eq_barnes10}
 \end{align}
 also provides an implicit function for the rotation period $P$ for any given age, $t$, in terms of the convective turnover timescale, $\tau$ and the initial period $P_0$, adequately represented by the $1.1$\,d value for stars of sufficiently advanced ages\footnote{For young stars, the full range of possible ZAMS rotation periods ought to be considered.}.

 For computational convenience, we transform the above expression into an explicit one for the rotation period $P$. (For this particular use, we actually began with the explicit solution for $\tau$ in Eq.\,(22) from Ba10, which itself uses the fact that Eq.\,\eqref{eq_barnes10} above is quadratic in $\tau$, and hence, solvable.) Solving this equation yields
 \begin{align}
    P & = \sqrt{\frac{a\cdot\mathcal{L}(w,0)}{2b}} \label{eq_period} \\    
    \intertext{with}
    w &= \frac{2 \cdot \exp\left(2\cdot(b\cdot P_0^2 + t)/a\right)bP_0^2}{a},\notag\\
    a &= \frac{\tau}{k_c}, \textrm{ and } b = \frac{k_i}{2\tau}\notag
 \end{align}
 where $\mathcal{L}$ is the Lambert W function\footnote{see \url{docs.scipy.org/doc/scipy-0.14.0/reference/generated/scipy.special.lambertw.html} for the python implementation used.}, and $k_c=0.646$\,d\,Myr$^{-1}$, $k_i=452$\,Myr\,d$^{-1}$, exactly as in Ba10. (We note that for small $\tau$, solving Eq.\,\eqref{eq_period} can lead to numerical instabilities; this is only relevant for stars bluer than those considered here. In such cases, it may be necessary to solve the explicit function for $t$ from Ba10 and reproduced in Eq.\,\eqref{eq_barnes10}, numerically.)
 
\begin{figure}
    \centering
    \includegraphics[width=\linewidth]{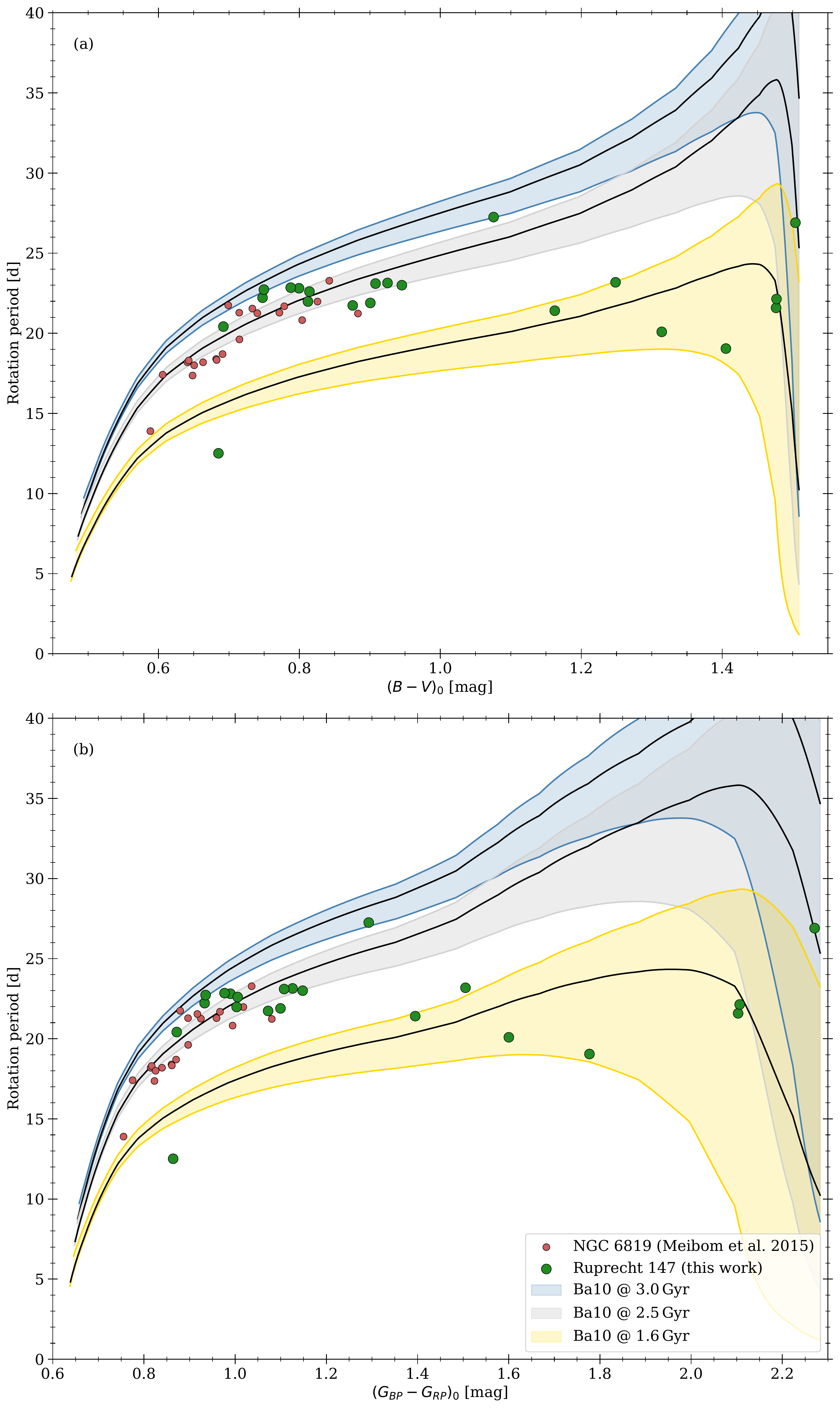}
    \caption{
        Color-Period-Diagrams in both Johnson $B-V$ (a) and Gaia (b) colors. Both show our Ru147 stars (green) and those from NGC\,6819 \citep[red;][]{2015Natur.517..589M}, compared with the Ba10 models. We only display stars that are neither evolved nor suspected binaries, and only those with periods classified as Category 1 (unambiguous). The rotational isochrones from the Ba10 model are also displayed for ages of 3.0, 2.5, and 1.6\,Gyr using blue, gray and yellow corridors respectively. (See text for details.) The central black lines within each corridor correspond to $P_0 = 1.1$\,d for the relevant age.
    }
    \label{fig_cpd_fit}
\end{figure}

 The convective turnover timescale, $\tau$, is obtained from \citet[][Table 1]{2010ApJ...721..675B}, which in turn relied on the $T_{\rm eff}$-color transformations of \cite{1997A&AS..125..229L,1998A&AS..130...65L}. Both Johnson and 2MASS colors are provided there as a function of convective turnover timescale, stellar mass, effective temperature, etc.  The conversion to $G_{BP}-G_{RP}$ is effected using the transformation from $B-V$ as described earlier, in Sect.\,\ref{sec_color_transformation}. The solution of Eq.\,\eqref{eq_period} returns a range of periods for a given age and color that is bounded by the spread permitted in the initial periods $P_0$. Cool stars span a range of periods at the ZAMS, from near-breakup rotation periods at 0.12\,d and up to 3.4\,d, the latter appearing to be longest rotation period found in very young clusters at the relevant mass range \citep{2010ApJ...722..222B}. Following Ba10, we use the intermediate $P_0 = 1.1$\,d as a representative reference value for solar-type stars in each range. The result can be seen in Fig.\,\ref{fig_cpd_fit}, where we show the Ba10 rotational isochrones for three different ages, 3.0, 2.5, and 1.6\,Gyr (top to bottom).

 In Fig.\,\ref{fig_cpd_fit}, we also display the most reliable of our rotation periods, defined as such if they are both in category 1 and also if the star is on the main sequence. These are all redward of $(G_{BP} - G_{RP})_0 = 0.86$ ($(B-V)_0 = 0.68$). Unfortunately, as the reader may see by glancing back at Fig.\,\ref{fig_coverage_hist}, very few stars bluer than solar color were observed in K2, and of these we have been able to determine rotation periods for no normal ones. Consequently, we supplement our Ru147 rotation periods with those in NGC\,6819 (also believed to be of similar age) that were determined by \cite{2015Natur.517..589M} to enable a full comparison. 

 The comparison between the data and the models shows that there is a reasonable match between the isochrone for 2.5\,Gyr for all the NGC\,6819 data \citep[as was also found by][]{2015Natur.517..589M} and the Ru147 rotation periods for stars warmer than spectral type K2V (i.e., $(G_{BP}-G_{RP})_0 \sim 1.2$; $(B-V)_0 \sim 0.95$).  However, while the models for all ages predict a steady rotation period increase with redder color at a given age up to early M stars, our rotation period data show a much more horizontal, even slightly declining trend for the cooler (mid-K and early-M) stars.  Consequently, for the 8 cooler stars with measured rotation periods, there appears to be a significant mismatch between the measurements and the Ba10 model, with the data points mostly accumulating in the region corresponding to the 1.6\,Gyr isochrone rather than that for 2.5\,Gyr.  This behavior was first pointed out by \cite{2018csss.confE..24C} where it was entitled the ``puzzle of K dwarf rotation''. 
 
 We also find a group of Ru147 stars located in the late-G and early-K region [$(G_{BP}-G_{RP})_0 \sim 0.85 - 1.1$). These appear to overlap well with their counterparts in NGC\,6819, and to be consistent with a rotational isochrone for 2.5\,Gyr, the age of Ru147 (e.g., Cu13; see also \cite{2018ApJ...866...67T}, which uses eclipsing binaries and Parsec models to propose a  2.7 Gyr age.). However, they are clearly inconsistent with an older 3.0\,Gyr rotational isochrone. In fact, we consider this region of the Ru147 color-period diagram to be populated well enough to have conclusive significance.
 
 We note that there is one outlier rotation period at 12.5\,d. We have been unable to convince ourselves that we have grossly underestimated its rotation period, or that we have only identified a period submultiple and that it should instead be recorded as a star with 25\,d period (cf. the light curve of EPIC~219388192 in Fig.\,\ref{fig_lc_6} in the Appendix). The star has been reported to be a (gravitationally bound) wide binary composed of G and M dwarfs, with the primary G\,star itself having an eclipsing brown dwarf companion with a $5.3$\,d orbital period \citep{2016csss.confE..95C,2017AJ....153..131N,2018AJ....156..168B}, whose transits are visible in our corrected lightcurve (c.f. appendix Fig.\,\ref{fig_lc_12}). This configuration could be responsible for the unexpected and discrepant rotational period. We exclude it from further consideration for this reason.
 
 Finally, of the 8 late-K and M-type stars that show a decidedly horizontal rotation period distribution, there is one star (EPIC 21963422, c.f. its light curve in Fig.\,\ref{fig_lc_25}) with a significantly longer period of 27\,d [($G_{Bp}-G_{Rp})_0 = 1.29$; SpT K4V] that appears to follow the model predictions for 2.5\,Gyr, but appears as a long-period outlier, as compared with neighboring data points. No peculiarities about this star are known at the time of this writing that may contribute to its atypical (compared the rest of Ru147) rotation. Given that this star is the only long period one in the present sample, we may not assign too much significance to its apparent agreement with the model predictions.

\subsection{Other models}
 
 We now compare the measured distribution of rotation periods in Ru\,147 and NGC\,6819 to other models of  stellar spin down proposed over the last decade. A careful inter-comparison between these has been published in Fr20, in connection with measured rotation periods in the ZAMS open cluster NGC\,2516. Consequently, the description here will be abbreviated.

\subsubsection{The \cite{2020A&A...636A..76S} model}
          
 The \citet[][SL20 hereafter]{2020A&A...636A..76S} isochrones incorporate a two zone model of internal stellar coupling in addition to implementing the (2-parameter) braking formulation of Ba10. The additional parameter is the index of the power-law describing the mass dependence of the coupling.  The angular momentum previously stored in the radiative core of the star is released to the surface convection zone on the related mass-dependent timescale, delaying the spindown of the star's surface, and potentially even spinning it up briefly\footnote{Models invoking decoupling (and subsequent recoupling) have been a steady presence in angular momentum evolution, most notably \cite{1991ApJ...376..204M}. We find this version to be more convincing than prior ones, partly because the time-scale for recoupling is transparently stated.}.  In principle, two additional parameters describe the initial period and effect of disk locking \citep{1991ApJ...370L..39K} during the pre-main sequence phase; both lose relevance as the star gets older. Finally, it should be noted that the SL20 models are technically formulated only for the slow rotators, that is, the fast rotators are not directly addressed.
 
\begin{figure}
    \centering
    \includegraphics[width=\linewidth]{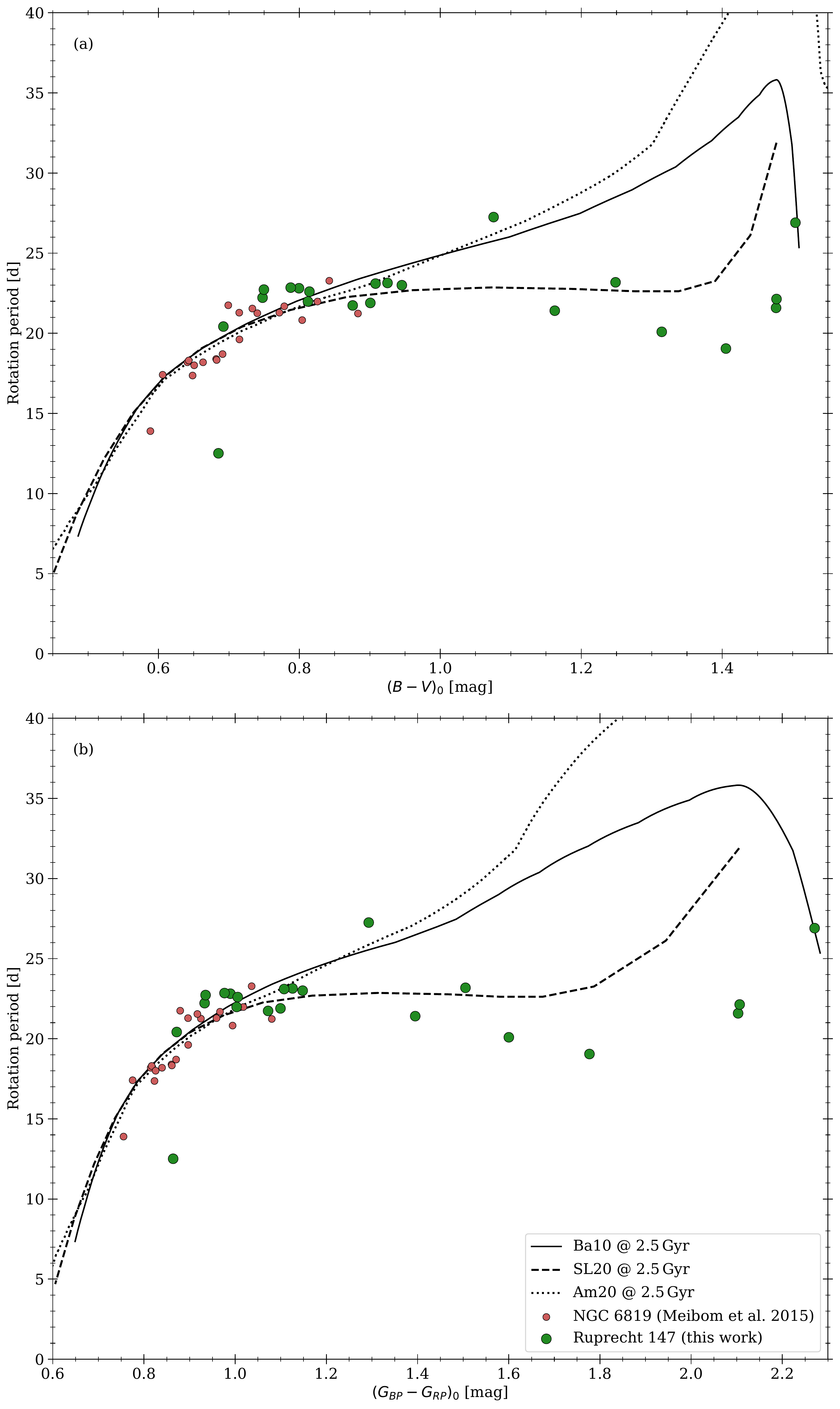}
    \caption{
        Comparison between the rotation period distributions in Ru\,147 (green) and NGC\,6819 (red) and rotational isochrones for 2.5\,Gyr from \citet[][dotted]{2019A&A...631A..77A}, \citet[][solid]{2010ApJ...722..222B}, and \citet[][dashed]{2020A&A...636A..76S}.
        The \citet[][]{2020A&A...636A..76S} models come closest to the measured Ru\,147 rotation periods in the mid-K to mid-M region, with a relatively flat morphology in the K\,star region and an upturn among the early M\,stars.
    }
    \label{fig_cpd_comparison} 
\end{figure}

 As can be seen in Fig.\,\ref{fig_cpd_comparison}, in the region of the CPD that represents Sun-like and warmer stars, their model is essentially indistinguishable from the Ba10 one, as expected because of the identical braking formalism. However, for cooler regions in the mid-K to mid-M spectral range, the SL20 isochrone is able to resist the spindown seen in the Ba10 and other comparable models and clearly comes the closest to describing the Ru147 rotation period data, including an increase in the rotation periods of mid-M stars. We presume that a small adjustment in the coupling parameter could push the models closer to the Ru\,147 data points, but we are not in a position to speculate whether such a change would also be compatible with the rotation period distributions of younger open clusters.

\subsubsection{The \cite{2019A&A...631A..77A} model}
 
 The \citet[][Am19 hereafter]{2019A&A...631A..77A} model has the largest number of degrees of freedom of the major models, in principle allowing for the most faithful reproduction of the data. It implements the spindown formulation of the \cite{2015ApJ...799L..23M} model, itself a modified version of the Ba10 model with more degrees of freedom, onto the stellar models of the Geneva-Montpelier group. However, the threshold for magnetic saturation has been modified from that in \cite{2015ApJ...799L..23M} and certain other choices have been made made both in the main sequence and pre-main sequence phases\footnote{See Fr20 for a summary.}. It should also be noted that this model provides a competitive description of the ZAMS rotation period data, as shown in Fr20, even if all features of those data are not reproduced.
 
 As can be seen in Fig.\,\ref{fig_cpd_comparison}, the Am19 isochrone for 2.5\,Gyr is located in approximately the same region as both the observations and the other models for Sun-like and warmer stars. However, it begins to diverge from the data at spectral type K0V and is significantly above the Ru147 rotation periods for later spectral types. The spindown formulation of this model is clearly over-aggressive in the K-M region, even more so than the Ba10 model. We note that the isochrones published by \cite{2019A&A...631A..77A} incorporate the slightly older color transformation for \emph{Gaia} colors from \cite{2018A&A...616A...4E}. This is not to blame here, because we instead recalculate these for $G_{Bp}-G_{Rp}$ from $B-V$ using our own transformation, as described above. Finally, it should be mentioned that we do not display separable models, such as those of \cite{2007ApJ...669.1167B}, \cite{2008ApJ...687.1264M}, or \cite{2020arXiv200509387A} for the detailed reasons given before. In particular, the fact that all of these lead to a mass dependence that does not change with age is a serious challenge in view of the observed time-varying morphology of open cluster CPDs.

\subsection{Implications for the modeling of rotating stars}

\begin{figure}
    \centering
    \includegraphics[width=\linewidth]{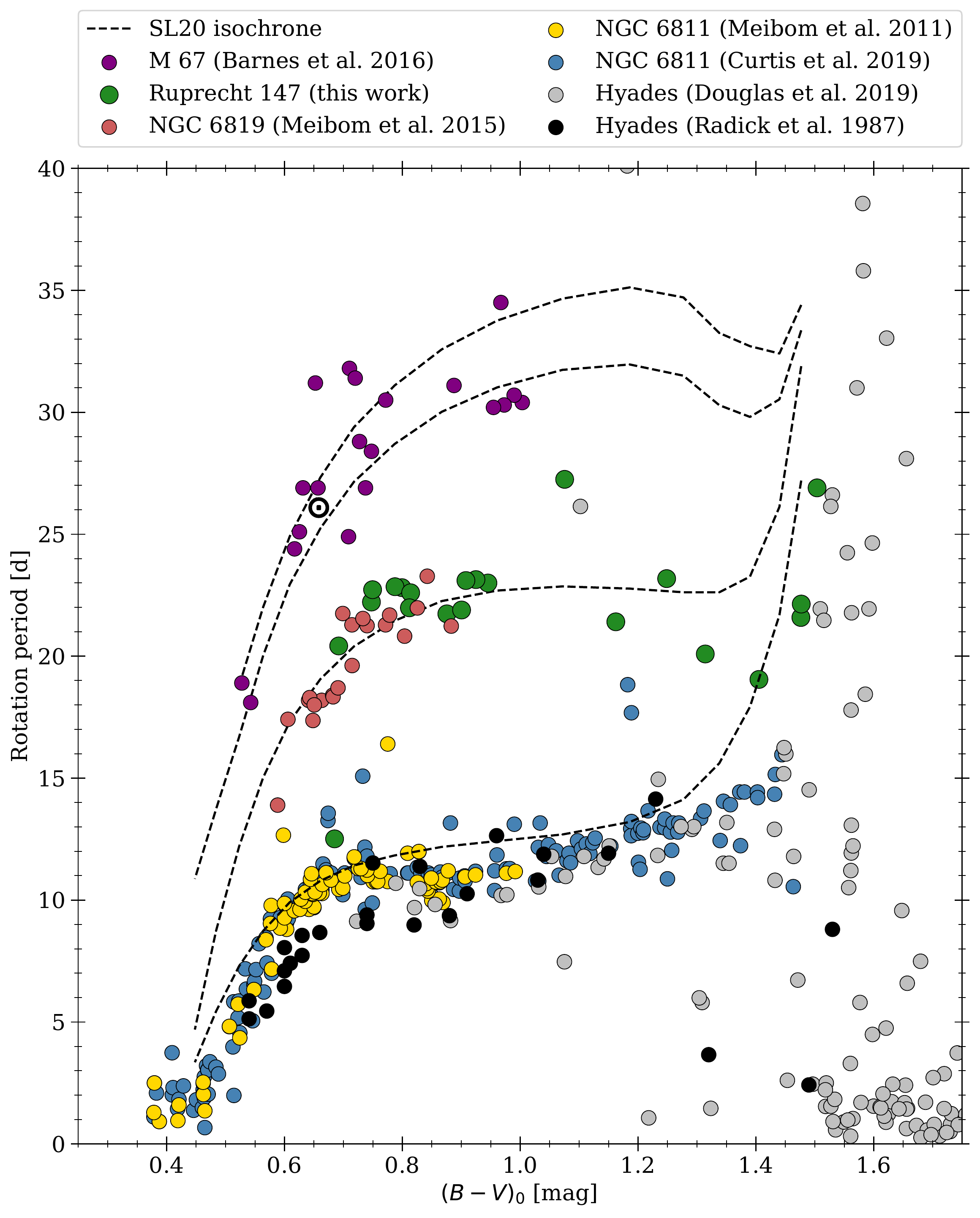}
    \caption{ 
        Color-Period Diagram displaying the same cluster rotation period data as Fig.\,\ref{fig_cpd_cluster}, compared with rotational isochrones (dashed lines) rotational isochrones from \citet[][SL20]{2020A&A...636A..76S} for younger- and older ages. The ages, from top to bottom, correspond to those of the Sun (4.57\,Gyr, shown with its usual symbol), M\,67 (4.0\,Gyr), NGC\,6819 \& Ru147 (both 2.5\,Gyr), NGC\,6811 (1.0\,Gyr), and the Hyades (600\,Myr). 
    }
    \label{fig_cpd_cluster_sl20}
\end{figure}

 In summary, we find that the new Ru\,147 rotation periods create significant challenges for theoretical rotational evolution models in the K-M spectral region that were not anticipated when only warmer middle-aged stars were measured in NGC\,6819 \citep{2015Natur.517..589M} and M\,67 \citep{ApJ...823..2016.16B}. Rotational isochrones for 2.5\,Gyr from the models of \cite{2010ApJ...722..222B} and \cite{2019A&A...631A..77A} which are in reasonable agreement with the data for stars bluer than early\,K-type stars appear to predict significantly longer rotation periods than actually measured in Ru147 among the mid-K to M\,stars. The isochrones of \cite{2020A&A...636A..76S}, which include a parameterized 2-zone model with associated angular momentum exchange, appear to perform considerably better (see Fig.\,\ref{fig_cpd_cluster_sl20}) with respect to the K-M stars in both Ru\,147 and prior measurements. We also show their isochrones for younger and older ages in Fig.\,\ref{fig_cpd_cluster_sl20}, so that the overall behavior of these models vis-a-vis other cluster measurements can be appreciated. It is possible that small adjustments to the parameters in their model might result in an even closer match to the observations. 
 
 More generally, our work here appears to confirm the existence of a single surface in rotation~period-mass-age space that is occupied by ``effectively single'' non-pathological rotating stars of roughly solar metallicity. The warmer (Sun-like) part of this surface appears to be asymptotically Skumanich-like in its behavior against age, although with a strong mass dependence, and can likely be modeled reasonably using just two or even one mass-dependent timescales, depending on the degree of fidelity desired, and whether or not fast rotators are included in the description. The cooler (K-M) part of this surface appears to exhibit more complex behavior and seems to require an additional (strongly) mass-dependent timescale to model it. Describing the spindown of cool stars on the main sequence therefore seems to require the invocation of three distinct physical processes.

\section{Conclusions }

 We have studied space-based photometric data from Campaign\,7 of the \emph{Kepler/K2} satellite for the 2.5\,Gyr-old open cluster Ruprecht\,147 in combination with prior membership work, to examine the calibration of gyrochronology for middle age stars, especially in the previously unexplored $K-M$\,star region. We have identified 102 cluster member stars that were observed by \emph{K2}. That target selection appears to be biased towards solar type stars, rather than being broadly representative of the entire Ru\,147 population. There are also certain technical issues with the nature of the \emph{Kepler/K2} light curves and the Campaign\,7 data from K2 that require additional efforts, as compared with those from the original Kepler field. Nevertheless, we have identified periodic behavior for 32 of these objects that can plausibly be associated with star spot modulation.

 Twenty one of these periods correspond to single stars and are unambiguous enough for a comparison with both previous open cluster studies and widely used rotational evolution models. We find that our results connect reasonably to prior measurements by \cite{2015Natur.517..589M} in NGC\,6819, a cluster of very similar age, verifying the behavior of 2.5\,--\,3\,Gyr rotating stars. Our data extend the measured rotation period sample to the previously unexplored K-\,and\,M-star region. We find that the Ru147 rotation periods are compatible with the idea that it, M\,67 (4\,Gyr), NGC\,6819 (2.5\,Gyr), NGC\,6811 (1\,Gyr), and the Hyades (625\,Myr) clusters all lie on a single surface in color-rotation~period-age space. This surface apparently extends to the ZAMS, where the ($\sim 130 - 150$\,Myr-old) Pleiades, NGC\,2516, M\,35, M\,50, and Blanco\,1 open clusters appear to have identical rotation period distributions \citep[][]{Fr2020}. However, the Ru\,147 and NGC\,6811 data for $K-M$\,type stars suggest that it has a a revised shape as compared with the original form proposed by \cite{2003ApJ...586..464B} and succeeding models.

 A comparison with the predictions of rotational evolution models shows that most models fail to predict the observed distribution of stars redder than spectral type K3. We find that the best current description of the spindown of stars beyond 1\,Gyr is provided by the model of \cite{2020A&A...636A..76S}, invoking a third mass-dependent timescale in addition to the two timescales in the model of \cite{2010ApJ...721..675B} and \cite{2010ApJ...722..222B}. Consequently, it appears that models describing the rotational evolution of solar metallicity cool main sequence stars need to include three distinct physical processes if they are to account for the fast, slow, and low mass rotators observed in open clusters to date.

\begin{acknowledgements} 
    We are grateful to the referee, Gibor Basri, for a timely response and helpful suggestions to improve readability.
    We thank Jason L. Curtis for the provision of period data from his team's independent study prior to final publication, enabling the relevant comparison.
    SAB gratefully acknowledges support from the Deutsche Forschungs Gemeinschaft (DFG) through project number STR645/7-1.
    Some of the data presented in this paper were obtained from the Mikulski Archive for Space Telescopes (MAST). STScI is operated by the Association of Universities for Research in Astronomy, Inc., under NASA contract NAS5-26555. Support for MAST for non-HST data is provided by the NASA Office of Space Science via grant NNX09AF08G and by other grants and contracts.
    This work has made use of data from the European Space Agency (ESA) mission {\it Gaia} (\url{https://www.cosmos.esa.int/gaia}), processed by the {\it Gaia} Data Processing and Analysis Consortium (DPAC, \url{https://www.cosmos.esa.int/web/gaia/dpac/consortium}). Funding for the DPAC has been provided by national institutions, in particular the institutions participating in the {\it Gaia} Multilateral Agreement. 
    This research has made use of the SIMBAD database, operated at CDS, Strasbourg, France, and the WEBDA database, operated at the Department of Theoretical Physics and Astrophysics of the Masaryk University, Czech Republic.
\end{acknowledgements}

\bibliographystyle{aa}
\bibliography{paper}

\appendix

\section{Relationship between \emph{Gaia} $G_{BP}-G_{RP}$ and Johnson $B-V$}  \label{sec_color_trafo}
  
 In this section of the Appendix, we present our empirical color transformation between $G_{BP}-G_{RP}$ and $B-V$. The necessity for an easy way to transform between those two colors systems was mentioned above. However, a reliable transformation has yet to be established. \cite{2018A&A...616A...4E} does not provide a direct transformation between the colors. It is possible to use the combination of two relations given there to construct a transformation. But this not only introduces additional uncertainties, a closer inspection reveals that it also fails for stars redder than $B-V=1.2$. Therefore we decide to derive an empirical relation from observed stars in both color systems
  
 We obtain photometric data for Hyades and Pleiades stars from the WEBDA cluster data base\footnote{\url{webda.physics.muni.cz}}. We use the photoelectric $B-V$ colors and $V$-band magnitudes, match the stars to the list of identifiers based on their internal reference number, query those in SIMBAD\footnote{\url{simbad.u-strasbg.fr/simbad}}, retrieve the \emph{Gaia} DR2 crossmatch from there, and lookup them up in the \emph{Gaia} DR2 catalog to obtain $G_{BP}-G_{RP}$, $G$, and their parallaxes. This procedure leaves us with 401 and 135 unique stars respectively representing 1044 and 324 magnitudes for the Hyades and Pleiades. We retain all instances of multiple occurrences of the same star. To those, we add the table of \cite{2013ApJS..208....9P} to cover redder colors. This provides an independent perspective, given that that relationship is constructed using field stars rather than those of clusters. We complement those with our Ru147 sample that have K2 counterparts.
  
 To obtain the intrinsic colors and absolute brightnesses, we deredden the Hyades data. Here we use 
 \[ E_{B-V,\text{ Hyades}} = 0.027 \quad\text{\&}\quad E_{B-V,\text{ Pleiades}} = 0.04 \]
 from \cite{2006AJ....132..111J} and \cite{1986ApJ...309..311B}, respectively. Extinction is calculated as
 \[ A_V=R_V\cdot E_{B-V}  \]
 with $R_V=3.1$ for the Johnson colors and magnitudes and
 \begin{equation}
     E_{G_{BP}-G_{RP}} = k \cdot E_{B-V} \label{eq_reddening_relation} 
 \end{equation}
 and
 \[ A_G=R_G\cdot E_{G_{BP}-G_{RP}} \]
 with $k=1.339$ \citep[using the mean extinction coefficients from ][; their Table\,2]{2018MNRAS.479L.102C} and $R_G=2.0$ (as found in Sect.\ref{sec_color_transformation} and Fig.\,\ref{reddening_effects}) for the \emph{Gaia} colors and magnitudes.  
  
\begin{figure*}[ht!]
    \centering
    \includegraphics[width=\linewidth]{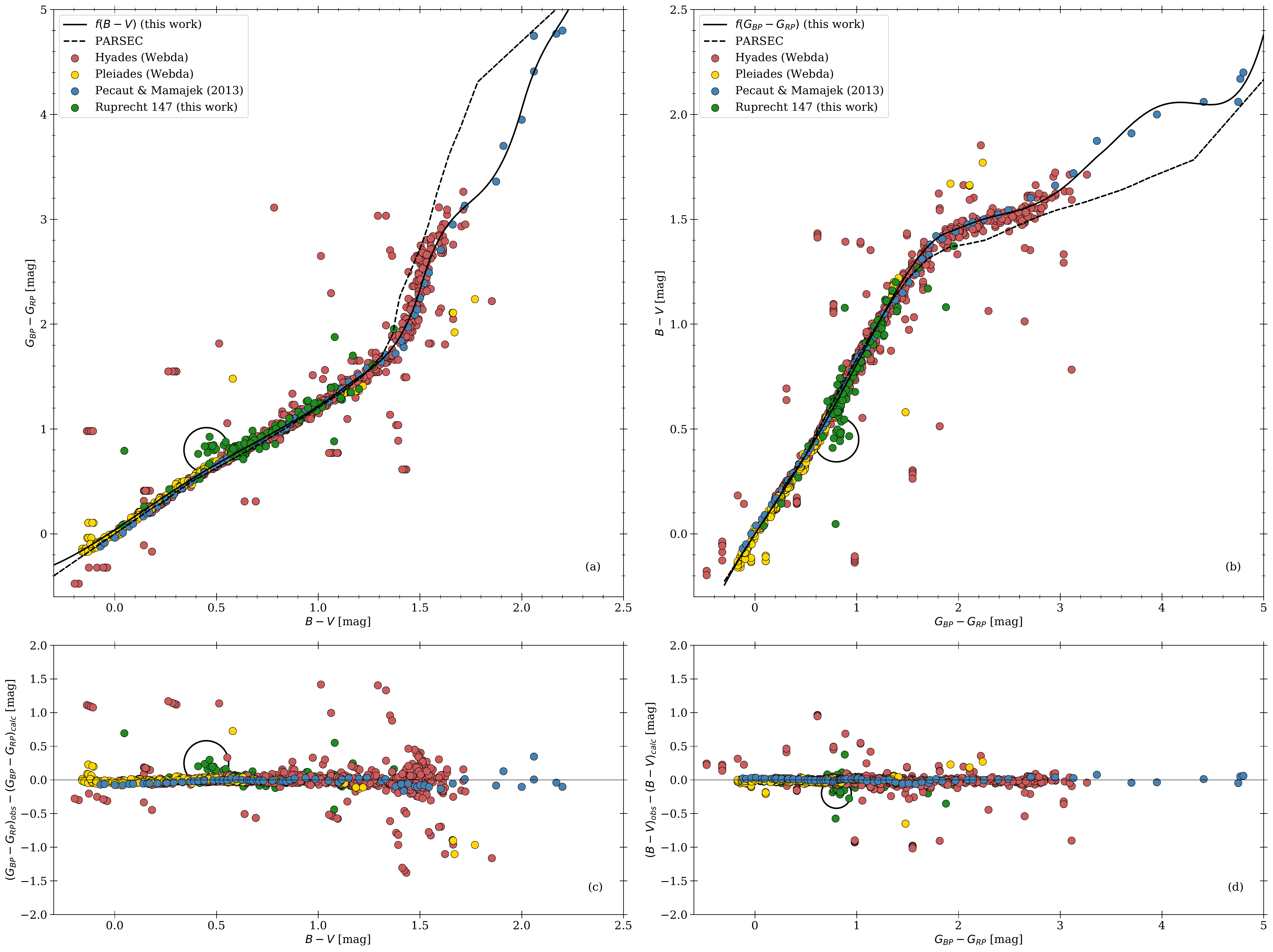}
    \caption{
        Calibration of the color transformations from $B-V$ to $G_{BP}-G_{RP}$ in Eq.\,\eqref{eq_bv2bprp} and vice versa in Eq.\,\eqref{eq_bprp2bv}. 
        Panel (a) shows a Color-Color diagram for the Hyades (red), the Pleiades (yellow), and the calibration by PM13 (blue). 
        The same Color-Color diagram, but with switched axes, is shown in panel (b). 
        Overplotted are the relationsships between the two colors as derived. 
        Panels (c) and (d) show the residual between the calculated and measured color for both transformations each. 
        The subgiants of Ru147 (encircled) do not follow the same relation, as expected. 
        We note that the small horizontal structures in in panel (a), and their equivalents in the other panels, originate from stars with multiple entries in WEBDA.
    }
    \label{color_trafo}
\end{figure*}

 The distribution of stars in a Color-Color diagram (cf. panel (a) and (b) in Fig.\,\ref{color_trafo}) shows that the relation between both color is complex. Despite this complexity, we find very good agreement between the Hyades, Pleiades, the averages of local dwarfs in PM13, and our sample of Ru147 members, together providing confidence in the applicability of our analytical description. However, given the complexity of the relation, it cannot easily be represented with a simple polynomial approximation.

\subsection{Forward transformation}
  
 Hence, we decide to use a function with multiple components, each itself a polynomial with a different range of validity. The goal is a one-to-one function
 \begin{equation}
      f(B-V)=G_{BP}-G_{RP}
 \end{equation}
 that can be used piece-wise, depending on how red one needs to go, and is both continuous and continuously differentiable, even at join points. For brevity, we substitute $x=B-V$ and $y=G_{BP}-G_{RP}$ in the following. It turns out that we are not able to describe the observed distribution with only one or two polynomials sufficiently\footnote{Testing up to 12th order.}. However, we can describe the distribution sufficiently with four polynomials $f_i$ such that, symbolically,
 \[      y=f(x)=f_1(x)+f_2(x)+f_3(x)+f_4(x), \]
 with successive terms added as required when the desired color is redder. We tested various combinations of functions to reproduce the observed distribution and the aforementioned combination provided the best and most simple result we could find. We tested polynomials of different orders, logarithmic-,  exponential-, and trigonometric functions. We find that the combination of 4th-order polynomials results in the best description of the observed distribution without invoking too many parameters.
 \begin{equation}
    y = f(x) = f_1 + 
    \begin{cases}
        0 & x \leq 1.11 \\
        f_2 & 1.11 < x \leq 1.52 \\
        f_2 + f_3  & 1.52 < x \leq 2.04 \\
        f_2 + f_3 + f_4  & x > 2.04
    \end{cases}
    \label{eq_bv2bprp}
 \end{equation}
 where
 \begin{align*}
    f_1(x') &= m_{4,1}\cdot x'^4 + m_{3,1}\cdot x'^3 + m_{2,1}\cdot x^2 + m_{1,1}\cdot x', \\
    f_2(x') &= m_{4,2}\cdot x'^4 + m_{2,2}\cdot x'^2,\\
    f_3(x') &= m_{4,3}\cdot x'^4 + m_{2,3}\cdot x'^2, \\
    f_4(X) &= m_{4,4}\cdot x'^4 + m_{2,4}\cdot x'^2,
 \end{align*}
 with
 \[x' = s_i\cdot(x-x_{0,i})\]
 and the coefficients given in Table\,\ref{tab_color_trafo}. The resulting curve is displayed in panel (a) in Fig\,\ref{color_trafo}. Panel (c) shows the residual of the fit.
  
 To ensure continuity of the function and its slope, we restrict the fit to the parts of the polynomial with even exponents (with the exception of $f_1$) and use only a shift $x_0$ and re-scaling $s$ in $x$ and no shift in $y$. This creates a function that, at each breaking point in Eq.\,\eqref{eq_bv2bprp}, is a combination of the ones before plus a function whose value and slope are 0.

\subsection{Inverse transformation}
  
 We do not only want to have a transformation from $B-V$ to $G_{BP}-G_{RP}$ but also the back transformation. Given the complex shape of $f(x)$ it is not practical to calculate its inverse. It is also not very convenient to solve $f(x)$ numerically for this purpose every time. We decide to derive a completely independent transformation $g(G_{BP}-G_{RP})=B-V$ in the same way as before. We find the following representation: 
 \begin{equation}
    x = g(y) = g_1 + 
    \begin{cases}
        0 & x\leq 1.88\\
        g_2 & 1.88 < x \leq 2.67 \\
        g_2 + g_3  & 1.88 < x \leq 3.04 \\
        g_2 + g_3 + g_4  & x > 3.40 \label{eq_bprp2bv}
    \end{cases}
 \end{equation}
 where
 \begin{align*}
    g_1(y) &= n_{4,1}\cdot y'^4 + n_{3,1}\cdot y'^3 + n_{2,1}\cdot y'^2 + n_{1,1}\cdot y', \\
    g_2(y) &= n_{4,2}\cdot y'^4 + n_{2,2}\cdot y'^2,\\
    g_3(y) &= n_{4,3}\cdot y'^4 + n_{2,3}\cdot y'^2, \\
    g_4(y) &= n_{4,4}\cdot y'^4 + n_{2,4}\cdot y'^2,
    \intertext{with}
    y' &= s\cdot(y-y_{0})
 \end{align*}
 and the coefficients given in Table\,\ref{tab_color_trafo}. The resulting curve is displayed in panel (b) in Fig\,\ref{color_trafo}. Panel (d) shows the residual of the fit.
  
 Because the two transformations are not inverses in a mathematical sense, a forward-and-reverse transformation will not generally yield the initial color perfectly again. Given the limited range of colors covered by the stars we used for a fit here, parts of the relation are speculative (and prone to overfitting) and have to be adopted carefully. The stars merit confidence in the relationship described here for 
 \begin{equation}
    0.0 \leq B-V \leq 1.6 \quad\text{and}\quad 0.0 \leq G_{BP}-G_{RP} \leq 3.0.
 \end{equation}
 We note further that the here found relation is only calibrated on dwarf stars, it may look different for (sub-)giants or white dwarfs. This can be seen for the few (sub-)giants in the Ru147 sample in Fig.\,\ref{color_trafo}. However, blue stragglers follow the here found relation.

\begin{table}[ht!]
    \caption{Coefficients used to construct the color transformation from $B-V$ to $G_{BP}-G_{RP}$ in Eq.\,\eqref{eq_bv2bprp} and vice verse in Eq.\,\eqref{eq_bprp2bv}.}
    \centering
    \begin{tabular}{c|cccccc}
        \hline
        \hline
            &\\[-0.7em]
            & $m_{4}$   & $m_{3}$ & $m_{2}$    & $m_{1}$& $x_0$   & $s$ \\
        \hline
        $f_1$   & $0.55$    & $-0.97$ & $0.33$     & $1.27$ & $0.0$   & $1.0$     \\
        $f_2$   & $20.94$    & --      & $-2.70$    & --     & $1.11$  & $1.00$ \\
        $f_3$   & $-8.62$  & --      & $-23.44$   & --     & $1.51$  & $1.36$ \\
        $f_4$   & $5.19$    & --      & $-11.90$   & --     & $2.00$  & $1.79$ \\
        \hline
        \hline
            &\\
            & $n_{4}$  &  $n_{3}$ & $n_{2}$  & $n_{1}$  &   $y_0$ & $s$ \\
        \hline
        $g_1$   & $-0.12$ & $0.29$  &  $-0.11$ & $0.74$  &   $0.0$ &  $1.0$ \\
        $g_2$   & $8.06$       & ---    & $4.68$    & ---   &  $1.88$ & $0.52$ \\
        $g_3$   & $-28.70$      & ---  & $-2.05$    & ---    &  $2.67$ & $0.45$ \\
        $g_4$   & $16.26$       & ---  & $3.5$    & ---    &  $3.40$ & $0.53$ \\
        \hline
    \end{tabular}
    \label{tab_color_trafo}
\end{table}

\begin{figure*}[ht!]
    \centering
    \includegraphics[width=\linewidth]{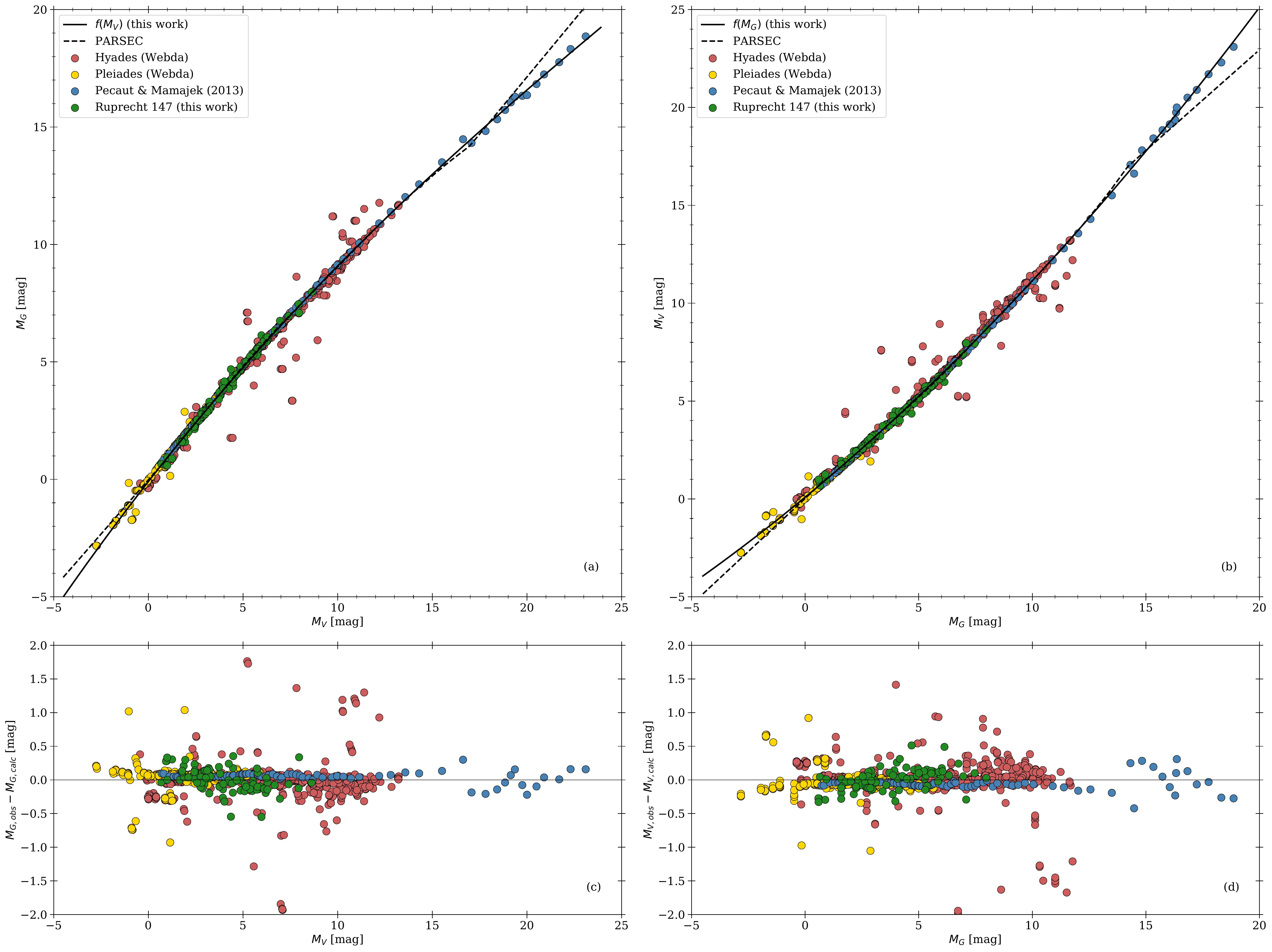}
    \caption{
        Calibration of the brightness transformations from $M_V$ to $M_G$ in Eq.\,\eqref{eq_v2g} and vice verse in Eq.\,\eqref{eq_g2v}. 
        Panel (a) shows a magnitude-magnitude diagram for the Hyades (red), the Pleiades (yellow), and the calibration by PM13 (blue). The same magnitude-magnitude diagram, but with switched axes, is shown in panel (b). Overplotted are the found relation between the two magnitudes each. 
        Panels (c) and (d) show the residual between the calculated and measured magnitudes for both transformations each.
    }
    \label{magnitude_trafo}
\end{figure*}

 Additionally to a color transformation, we also need one for the brightnesses to create a CMD. We adopt the same sample as above and use the Gaia parallaxes to calculate absolute magnitudes. As can be seen in Fig.\,\ref{magnitude_trafo} (panels a and b), the relation between $M_V$ and $M_G$ is much simpler and can be described with simple polynomials:
 \begin{equation}
    M_G = f(M_V) =   m_{3}\cdot M_V^3 + m_{2}\cdot M_V^2 + m_{1}\cdot M_V + b \label{eq_v2g}
 \end{equation}
 and
 \begin{equation}
    M_V = g(M_G) =   m_{3}\cdot M_G^3 + m_{2}\cdot M_G^2 + m_{1}\cdot M_G + b \label{eq_g2v}
 \end{equation}
 and the coefficients $m_i$ and $n_i$ given in Table\,\ref{tab_mag_trafo}.

\begin{table}[ht!]
    \caption{
        Coefficients used to construct the transformation from $M_V$ to $M_G$ in Eq.\,\eqref{eq_v2g} and vice verse in Eq.\,\eqref{eq_g2v}.
    }
    \label{tab_mag_trafo}
    \centering
    \begin{tabular}{c|ccccc}
        \hline
        \hline
              & $f(M_V)$   & $M_G$ \\
        \hline
        $m_3$ & $0.00017$  & $-0.0001$ \\
        $m_2$ & $-0.01339$ & $0.01684$ \\
        $m_1$ & $1.03269$  & $0.9465$  \\
        $b$   & $-0.07834$ & $0.09871$ \\
        \hline
    \end{tabular}
\end{table}

 We note that there are two approaches to the comparison: (1) on absolute brightnesses and intrinsic colors, or (2) on apparent colors and brightnesses. The latter is intrinsically more correct for the cluster it is calibrated on, since the spectral energy distribution (SED) of a star is reddened prior to the filter. However this limits any relation found to stars with the same reddening and is therefore impratical for an easy comparison. We decided in favor of the former approach, one which may introduce an additional error, but can be applied to every cluster independently of its particular reddening. The introduced error becomes larger for greater reddening. However, this problem can be overcome when the stellar SED is taken into account for the dereddening. \cite{2018MNRAS.479L.102C} derived extinction parameters that depend on the stellar parameters ($T_\text{eff}$ and [Fe/H]) and those can be used to describe a color-dependend reddening. The reddening of the Hyades is small and a uniform extinction is a good approxmiation.

\section{Details of the Principal Component Analysis }\label{sec_pca} 
  
 The basic idea of Principal Component Analysis (PCA) is a reduction in data dimensionality by identifying common patterns in the data and creating a new set of $k$ $m$-dimensional basis vectors $\vec v_j$ (PCA components) from the data $a$ with $k$ $m$-dimensional datapoints $a_i$. Each datapoint $a_i$ can then be described by a new $k$-dimensional coordinate $\vec b_i$ with $a_i=\sum_j^k b_{i,j}\cdot \vec v_j$ in this new basis. The basis is created by $k$-times successively finding the vector that explains the largest variation in the data (in principle the minimization of the average distance) and removing its contribution to the data. At this point, the dimensionality has changed from $m$ to $k$, generally not a reduction. The reduction in dimensionality is achieved by simply truncating the dimensionality (from $k$) in the calculation of a datapoint in the new basis, based on the assumption that the first few components provide a reasonably good approximation of the data.
  
 In our case, each light curve is an $m$-dimensional datapoint, with $m=4043$ being the number of points in each light curve\footnote{The sampling rate of the light curves is irrelevant as long as all are sampled in the same way}. Trend correction with PCA is based on the assumption that trends in the data are visible in a large number or all lightcurves and are therefore represented in the first few components. This means that when calculating a light curve from the new basis using the first components (typically 2-5), only the data systematics, but not the individual variability that changes from lightcurve to lightcurve, is reproduced. The observed light curve is then corrected by its (purposefully incomplete) reconstruction and the residual is the detrended lightcurve.
  
 Unfortunately, reality is almost always more complex. The $\vec v_j$ are not individual components of the systematics, with each representing one kind of trending, but averages of the data as a whole. This also means that other effects such as noise and pulsations can intrude into the first few components. Additional difficulties arise when the data systematics themselves show variations. Both of these problems are present in the Ru\,147 data. The challenge is to find the correct number of PCA components for the reconstruction to account for (enough of) the systematics present, while also not including the intrinsic stellar variability that is our signal, and which we obviously would like to retain. There is no metric to choose the number of components necessary for a given light curve in our case because those instrumental characteristics vary across the field, CCD, channel, etc. Consequently, the procedure becomes intrinsically somewhat subjective.
    
 We believe that we have been able to bound these problems for a significant number of stars of interest. Numerous lightcurves show variations that clearly originate in stellar pulsations, with remnants of this behavior finding their way into the prominent components. Similar effects can be observed for starspot induced variations. This can lead to over-fitting and the introduction of high frequency variations. We cross-checked by eliminating the pulsating stars from the sample and redoing the PCA; the problematic components disappeared as expected. This issue could not be resolved by omitting those components from the reconstruction because those still carry parts of the global trends necessary for the elimination of systematics. 
  
 Therefore, we use a different procedure. We select a subset of our \emph{full sample} by omitting all lightcurves that clearly show pulsations, spot induced variation, artifacts, such as sudden jumps, or eclipses. Furthermore, we eliminate very noisy lightcurves from our sample. Moderately noisy data are retained, but we smooth all lightcurves with 1\,d binning. This does not affect our analysis because variations on this timescale are irrelevant for our work. The smoothing is only applied for the purposes of PCA; the final lightcurves are provided on the original sampling corrected by the smoothed PCA solution. At this point, we are left with $k\approx 3000$ lightcurves as the basis for the PCA.

\begin{figure}
    \centering
    \includegraphics[width=\linewidth]{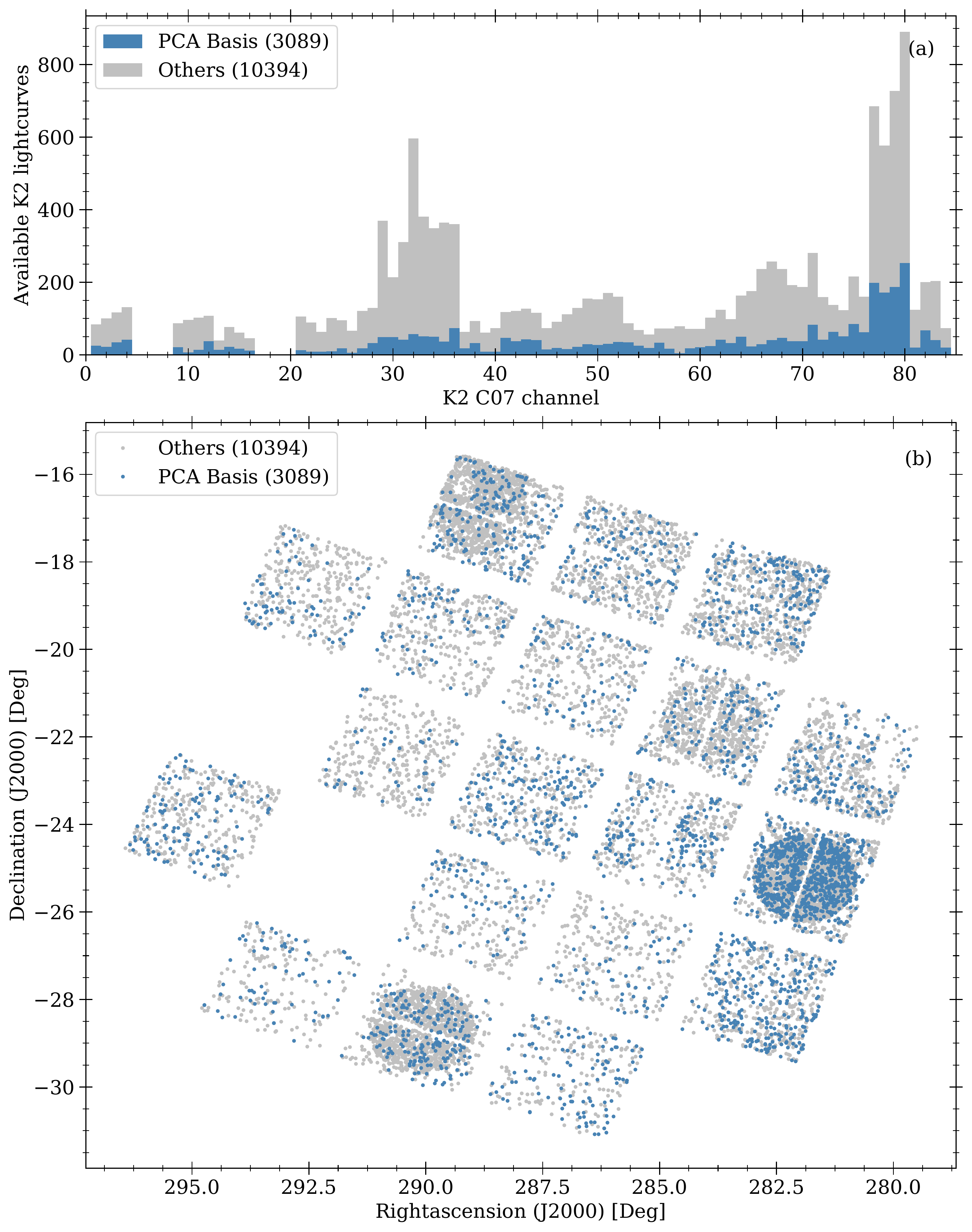}
    \caption{
        Selection of the ($\approx 3000$) lightcurves for the PCA basis. 
        Panel (a) shows histograms of lightcurves available (gray), and those taken for the PCA basis (blue), as distributed over the K2 channels. 
        Panel (b) shows the spatial distribution across the entire K2 C07 field. 
    }
    \label{fig_pca_statistics}
\end{figure}

 As can be seen in Fig.\,\ref{fig_pca_statistics}, our selection of lightcurves for the PCA basis is more or less evenly distributed across the C07 field of view and across the CCD channels, at least to the extent that the K2 target selection allows. Furthermore, it can be seen that the basis is not biased with respect to the position of the Target Pixel Files (TPFs) inside a Kepler module and channel. Finally, the boundaries of the lightcurve must be addressed. The PCA and the smoothing can introduce artifacts at the beginning and end of the data stream.  To eliminate those effects, we censor the first and last 80 data points from each lightcurve for the purposes of period analysis. 
  
 PCA requires a set of identically sampled light curves. We also work with light curves that are normalized to their respective medians. Each of the relevant K2 light curves contains 4043 data points, but not all of these have meaningful values stored. For instance there are both outliers and NaNs. \cite{2016AJ....152..100L} provide a mask that lists outliers for each data set. These are virtually identical for all datasets, but the NaNs are not. To perform the PCA we replace both the NaNs and masked values by a linear interpolation using neighboring data points. If these are unavailable, that is, at the beginning or the end of the data, we set the flux to unity. The interpolations occur on time scales ($\leq 0.5$\,d) that are irrelevant to our expected periodicity timescales of many days. This procedure provides a set of identically sampled lightcurves for the PCA.

\begin{figure}
    \centering
    \includegraphics[width=\linewidth]{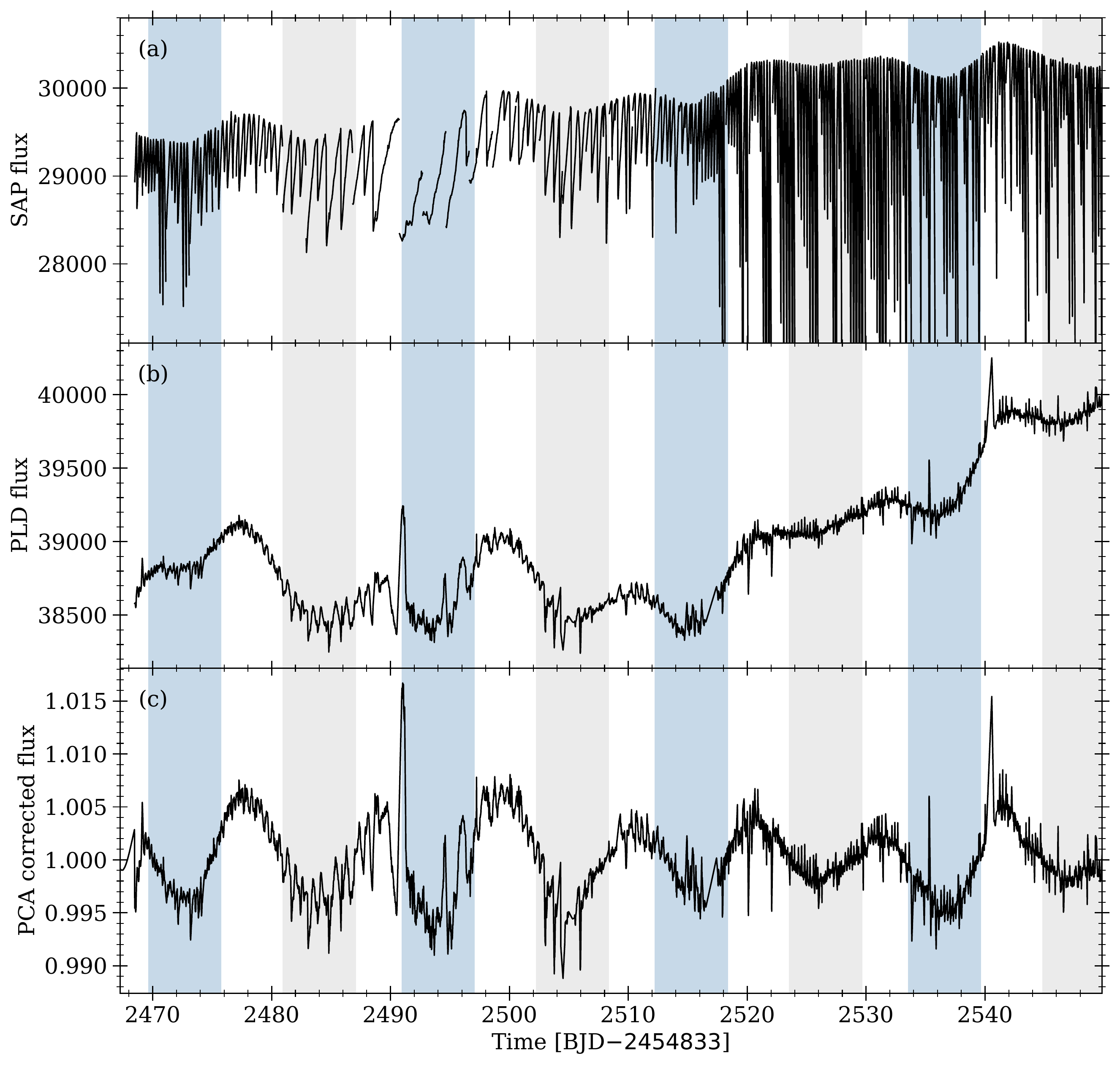}
    \caption{
        Example of the lightcurve correction process using EPIC~219722781. Panel (a) shows the original K2 lightcurve from simple aperture photomerty (SAP). Panel (b) shows the EVEREST lightcurve after the pixel level de-correlation (PLD). Panel (c) displays the lightcurve after both normalization and principal component analysis (PCA). In all panels, the colored regions mark identified flux variations assigned to two different spots with a phase shift of 0.47 and a period of 21.3\,d. 
    }
    \label{lc_correction}
\end{figure}

 We perform the PCA for each lightcurve individually by comparing the relevant lightcurve with the basis, as constructed above. The prominent components determined by PCA on just the basis, as compared with the basis plus one lightcurve, are virtually identical. The calculation is performed using the \verb+python+ implemetation in the \verb+sklearn+ package, which itself is based on the method outlined in \cite{2009arXiv0909.4061H}.  Figure\,\ref{lc_correction} displays a comparison of three lightcurves for the same object, EPIC~219722781, (a), from the Kepler archive based on simple aperture photometry, (b), the \verb+EVEREST+ lightcurves, and (c), our detrended lightcurve. Flux dips identified as belonging to two different starspots are marked with their periodic re-occurrence. As can be seen, these features are visible to the experienced eye in all stages of the processed lightcurve. 
  
 The degree of reproduction from the PCA is crucial for our final lightcurves and the derived results. If only a small number of components is used, the reproduction is clearly insufficient in suppressing observable trends. If too many components are used, we risk overfitting, and destroy clear signals from stellar variation. As can be seen, the lightcurve trends occur on the same timescale (10\,d to 40\,d) as the spot-induced variability. This complicates the identification of stellar flux variations. The slowest rotators tend to be impacted more by this, given their low amplitude of brightness variation.
  
\begin{figure}
    \centering
    \includegraphics[width=\linewidth]{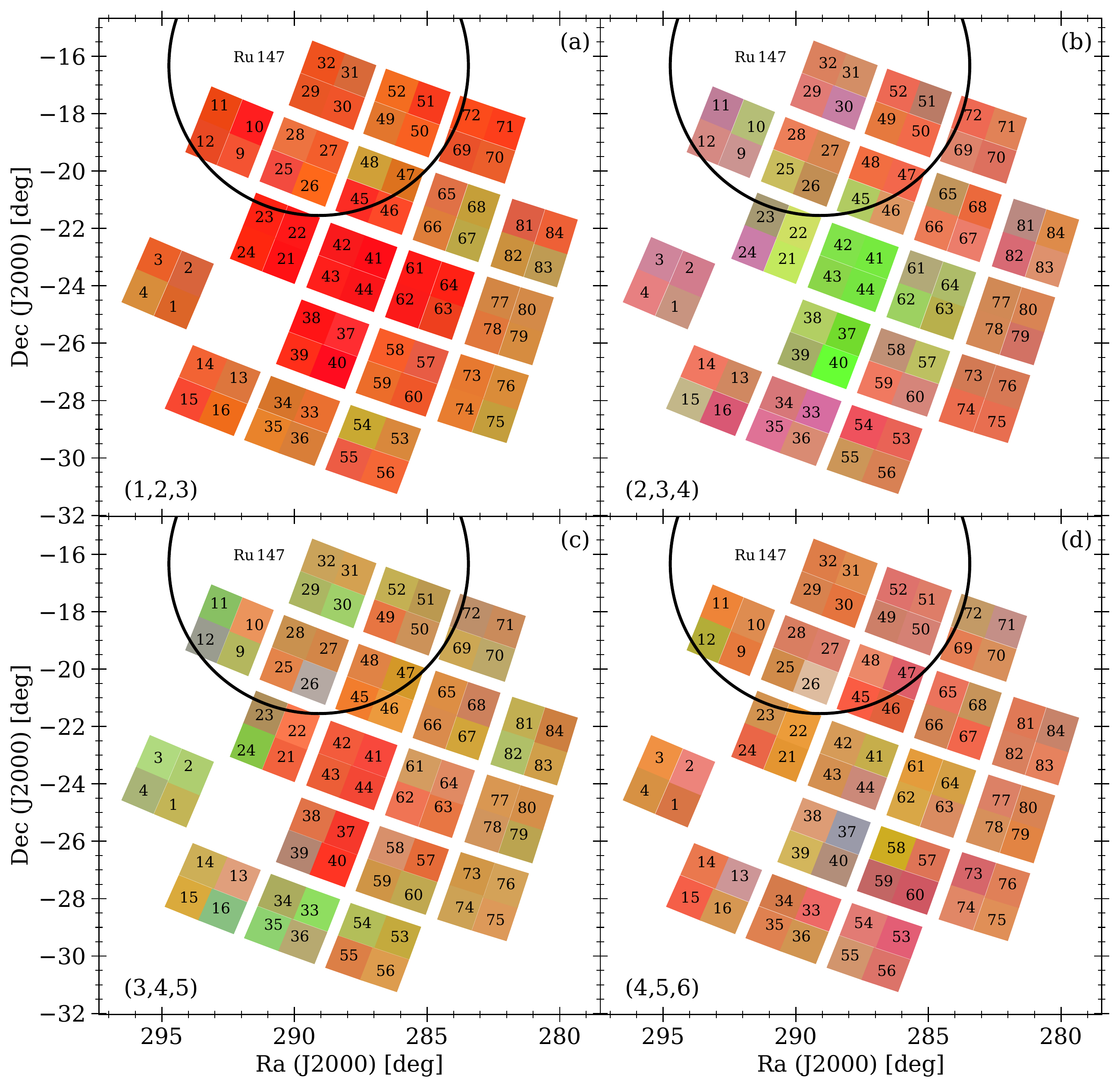}
    \caption{
        Spatial distribution of the observed trending in the EVEREST lightcurves of K2 C07. Each colored patch represents a channel with the corresponding number labeled. The color of each patch is based on the contribution of individual components. Here, the calculation works as follows: For each color, three components are adopted with their scaling factors from the PCA as RGB values. The numbers for each color are normalized to their maximum to guarantee a color range of 0\,--\,1. This is calculated for each lightcurve that is part of our basic sample (see text, Sect.\,\ref{sec_pca}). We calculate a mean color and mean position of all lightcurves from each channel, which is then plotted. The channel number is plotted at the mean position. The difference between the individual panels are the components used for the RGB color. In panel (a), red is given by component one, green by component two and blue by component three. For (b), (c), and (d), the components used are $(2,3,4)$, $(3,4,5)$, and $(4,5,6)$, respectively. The position of Ru147 is indicated.  
    }
    \label{trend_distribution}
\end{figure}

 We have compared the contribution of the components for the individual lightcurves to their origin on the Kepler/K2 CCD and have not found any correlation with the channel, module, or location in each channel. However, there appears to be a correlation with the location in the K2 C07 field itself; see Fig.\,\ref{trend_distribution}. The further away from the center of the K2 field a lightcurve was extracted, the more that extraction shifts the contribution from the first to the second and third components. When one considers higher components, this trend is reversed.

 As a consequence of the foregoing considerations, we adjust the number of components used for each star individually. We require that a variation that is adopted as a spot-induced feature has to be visible in the original light curve as well as for a PCA corrected with a high number of components. Its specific form may vary because of the presence of data systematics or overfitting, but it will still be visible.   It turns out that 5 to 8 components are usually used for the reconstruction. We do not skip individual components up the selected one, that is, all components of lower order than the final number are used. The resulting (de-trended) light curves are subjected to periodicity analysis, as described in the main text in Sect.\,\ref{sec_analysis}.

\section{Sample table } \label{sec_apx_table}

\begin{sidewaystable*}\small
    \centering
    \caption{Identifiers and basic information for the 32 periodic Ru\,147 stars. }
    \label{tab_sample_overview}
    \begin{tabular}{ccc|cccc|ccccc|cccc}
        \hline
        \hline
            &&&&&&&&&&&&&\\[-0.7em]
        Gaia & EPIC & 2MASS & $G$ & $(G_{BP}-G_{RP})_0$ & Plx & $M_G$ &$B$&$V$&$J$&$H$&$K$ & Cu13\tablefootmark{b} & GC18\tablefootmark{c} & CC18\tablefootmark{d} & Ol19\tablefootmark{d} \\
            &      &       & [mag]  &[mag]&[mas]&[mag]  &[mag]                  &[mag]&[mas]&[mag]&[mag]                  &                       &\\
            &&&&&&&&&&&&&\\[-0.7em]
        \hline &&&&&&&&&&&&&\\[-0.7em]
        4084645500601105536 & 218933140 & 19180938-1752498 & 12.36 & 0.87 & 3.31 & 4.91 & 13.28 & 12.52 & 11.11 & 10.82 & 10.74 & - & - & 1.0 & 0.35 \\
        4087622153460553216 & 219037489 & 19144383-1739427 & 13.03 & 0.99 & 3.32 & 5.59 & 14.13 & 13.22 & 11.63 & 11.28 & 11.15 & - & Y & 1.0 & 0.99 \\
        4087725262729508736 & 219141523 & 19153691-1726070 & 16.56 & 2.27 & 3.46 & 9.21 & - & - & 13.91 & 13.31 & 13.1 & - & Y & 0.7 & 0.31 \\
        4087736159069458304 & 219238231 & 19163672-1713101 & 12.09 & 0.78 & 3.25 & 4.6 & 13.37 & 12.3 & 10.95 & 10.68 & 10.62 & Y & Y & 1.0 & 0.81 \\
        \hline &&&&&&&&&&&&&\\[-0.7em]
        4087748833503644800 & 219275512 & 19155912-1708032 & 11.86 & 0.88 & 3.3 & 4.4 & 12.54 & 11.87 & 10.61 & 10.28 & 10.19 & - & Y & 1.0 & 0.0 \\
        4087714409360276864 & 219280168 & 19133741-1707261 & 13.33 & 1.15 & 3.35 & 5.9 & 14.58 & 13.57 & 11.77 & 11.27 & 11.17 & - & Y & 0.7 & 0.92 \\
        4087799655867454720 & 219297228 & 19152010-1705038 & 13.61 & 1.13 & 3.2 & 6.08 & 14.47 & 13.45 & 12.11 & 11.62 & 11.56 & - & Y & 1.0 & 0.92 \\
        4087762371240557696 & 219306354 & 19172705-1703472 & 13.03 & 0.98 & 3.2 & 5.5 & - & 13.2 & 11.66 & 11.29 & 11.21 & - & Y & 1.0 & 0.99 \\
        \hline &&&&&&&&&&&&&\\[-0.7em]
        4087769075699852416 & 219333882 & 19173541-1659580 & 13.56 & 1.16 & 3.03 & 5.92 & 14.88 & 13.89 & 11.96 & 11.47 & 11.37 & - & - & 0.7 & 0.95 \\
        4180839161574392704 & 219341906 & 19193373-1658514 & 9.76 & 0.75 & 3.29 & 2.3 & 10.48 & 9.85 & 8.68 & 8.4 & 8.33 & Y & - & - & - \\
        4087770239621640704 & 219353203 & 19173100-1657159 & 16.24 & 2.1 & 3.3 & 8.78 & - & - & 13.75 & 13.09 & 12.84 & - & - & 1.0 & 0.99 \\
        4087782677845919744 & 219388192 & 19173402-1652177 & 12.36 & 0.86 & 3.25 & 4.87 & 13.28 & 12.53 & 11.07 & 10.73 & 10.67 & P & Y & 1.0 & 0.93 \\
        \hline &&&&&&&&&&&&&\\[-0.7em]
        4087786874044570880 & 219404735 & 19170954-1649540 & 11.48 & 0.79 & 3.45 & 4.12 & 12.3 & 11.61 & 10.34 & 10.05 & 9.97 & - & Y & 0.8 & 0.0 \\
        4088004611707768320 & 219409830 & 19134334-1649109 & 12.28 & 0.83 & 3.19 & 4.75 & 13.07 & 12.35 & 11.13 & 10.77 & 10.72 & P & Y & 1.0 & 1.0 \\
        4183850105448920576 & 219422386 & 19182218-1647232 & 13.17 & 1.01 & 3.16 & 5.62 & 14.26 & 13.4 & 11.79 & 11.4 & 11.31 & - & Y & 1.0 & 0.98 \\
        4087838104415873920 & 219479319 & 19164975-1638577 & 14.94 & 1.6 & 3.19 & 7.41 & 16.59 & 15.42 & 12.9 & 12.32 & 12.13 & - & Y & 1.0 & 0.98 \\
        \hline &&&&&&&&&&&&&\\[-0.7em]
        4183867079159884672 & 219489683 & 19175045-1637260 & 15.38 & 1.78 & 3.31 & 7.92 & 17.11 & 16.03 & 13.18 & 12.51 & 12.32 & - & Y & 1.0 & 0.97 \\
        4183915388953023616 & 219515762 & 19172865-1633313 & 9.99 & 0.57 & 3.17 & 2.44 & 10.62 & 10.16 & 9.11 & 8.97 & 8.9 & Y & - & - & - \\
        4088042888457322624 & 219545563 & 19135496-1628553 & 12.79 & 0.93 & 3.22 & 5.28 & 13.79 & 13.01 & 11.52 & 11.12 & 11.05 & - & Y & 1.0 & 0.97 \\
        4088034161083427968 & 219551103 & 19123785-1628037 & 13.03 & 1.0 & 3.2 & 5.51 & 14.1 & 13.24 & 11.63 & 11.21 & 11.16 & - & Y & 1.0 & 1.0 \\
        \hline &&&&&&&&&&&&&\\[-0.7em]
        4087858819031214336 & 219566703 & 19153354-1625368 & 14.46 & 1.51 & 3.17 & 6.91 & 16.17 & 14.9 & 12.52 & 11.89 & 11.74 & Y & Y & 1.0 & 0.93 \\
        4088060686802235392 & 219610232 & 19133109-1618401 & 13.32 & 1.17 & 3.2 & 5.8 & 14.65 & 13.7 & 11.63 & 11.13 & 11.01 & - & - & 1.0 & 0.98 \\
        4088051783318575488 & 219610822 & 19144049-1618344 & 13.52 & 1.11 & 3.16 & 5.97 & 14.72 & 13.7 & 12.0 & 11.57 & 11.42 & - & Y & 1.0 & 0.99 \\
        4088060892960421248 & 219619241 & 19133215-1617120 & 16.22 & 2.11 & 3.36 & 8.8 & - & - & 13.77 & 13.08 & 12.84 & - & Y & 1.0 & 0.99 \\
        \hline &&&&&&&&&&&&&\\[-0.7em]
        4183930777809206656 & 219634222 & 19181352-1614496 & 13.64 & 1.29 & 3.46 & 6.28 & 15.04 & 13.98 & 11.87 & 11.32 & 11.2 & N & Y & 1.0 & 0.96 \\
        4183942670575016320 & 219646472 & 19172382-1612488 & 9.99 & 0.61 & 3.23 & 2.48 & 10.66 & 10.03 & 9.26 & 8.9 & 8.82 & P & Y & 0.7 & 0.78 \\
        4183968401721906048 & 219683737 & 19175075-1606406 & 13.31 & 1.07 & 3.27 & 5.84 & 14.53 & 13.54 & 11.84 & 11.41 & 11.31 & - & Y & 1.0 & 0.99 \\
        4184140406570650496 & 219721519 & 19161757-1600177 & 13.08 & 1.1 & 3.93 & 6.0 & 14.27 & 13.31 & 11.52 & 11.07 & 10.94 & P & - & - & - \\
        \hline &&&&&&&&&&&&&\\[-0.7em]
        4184146900561610880 & 219722212 & 19152141-1600107 & 12.77 & 0.94 & 3.11 & 5.18 & 13.72 & 12.9 & 11.43 & 11.09 & 11.05 & Y & Y & 1.0 & 0.98 \\
        4184140994996369152 & 219722781 & 19163732-1600050 & 13.89 & 1.39 & 4.01 & 6.86 & - & - & 12.02 & 11.45 & 11.35 & Y & - & - & 0.64 \\
        4184200501759138688 & 219755108 & 19141294-1554291 & 11.95 & 0.94 & 3.2 & 4.42 & 12.96 & 12.14 & 10.61 & 10.25 & 10.17 & Y & - & - & - \\
        4184182737768311296 & 219800881 & 19162203-1546159 & 12.56 & 0.9 & 3.09 & 4.96 & 13.5 & 12.71 & 11.29 & 11.0 & 10.86 & Y & Y & 1.0 & 0.45 \\
        \hline
    \end{tabular}
    \tablefoot{
        \tablefoottext{a}{Reddening and extinction adopted from closeby source}\\ 
        \tablefoottext{b}{Y = member, P = likely member, N = no member, Cu13 = \cite{2013AJ....145..134C}} \\ 
        \tablefoottext{c}{Y = member, GC18 =\cite{2018A&A...616A..10G}}  \\  
        \tablefoottext{d}{Membership probability, DANCe = \cite{2019A&A...625A.115O}, CC18 = \cite{2018A&A...618A..93C}}
    }
\end{sidewaystable*}

\section{Lightcurves and phase diagrams}\label{apx_lightcurves}

\begin{figure*} 
    \centering
    \includegraphics[width=\linewidth]{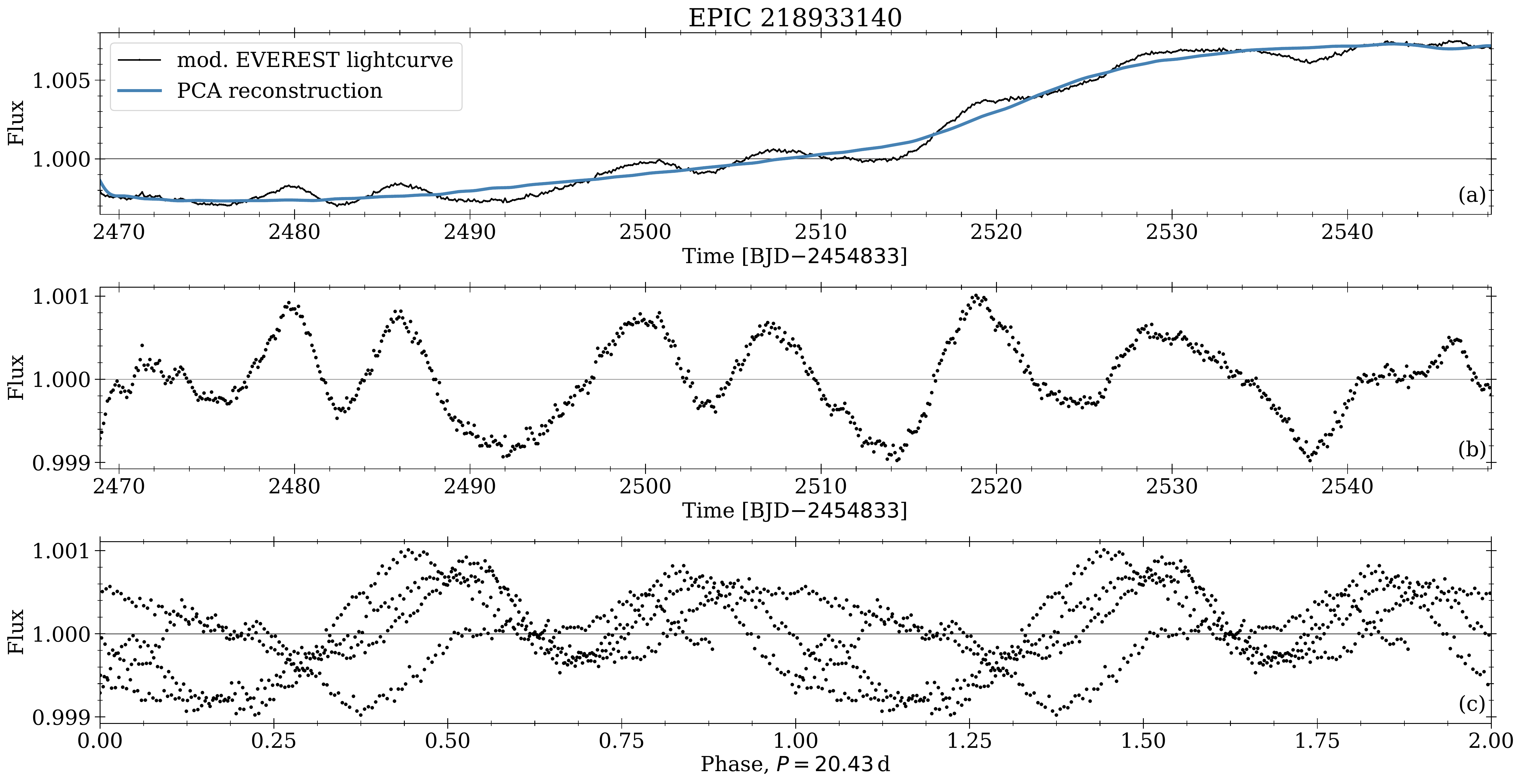} 
    \caption{Light curve, PCA correction and phase plot for EPIC\,218933140. Panel (a) shows the modified \texttt{EVEREST} light curve (black) and the reconstruction from the PCA (blue). The modifications for the PCA are applied as outlined Sect.\,\ref{sec_pca_1}. Panel (b) shows the corrected lightcurve and panel (c) the phase plot. For visibility reasons all data are displayed with a 0.1\,d binning.  \label{fig_lc_1}} 
\end{figure*}  

\begin{figure*} 
    \centering
    \includegraphics[width=\linewidth]{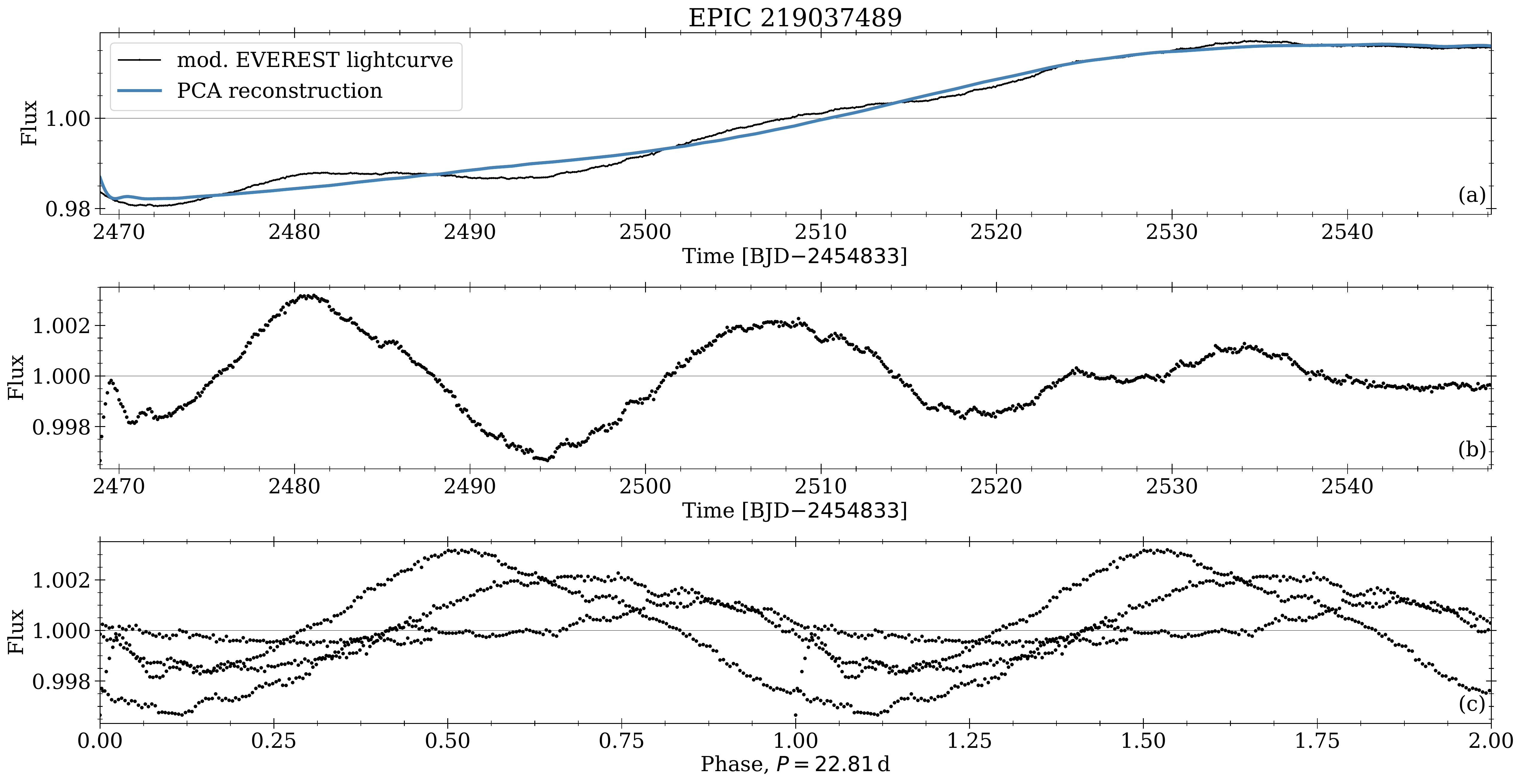} 
    \caption{Same as Fig.\,\ref{fig_lc_1}, but for EPIC 219037489 \label{fig_lc_2}} 
\end{figure*}

\begin{figure*} 
    \centering
    \includegraphics[width=\linewidth]{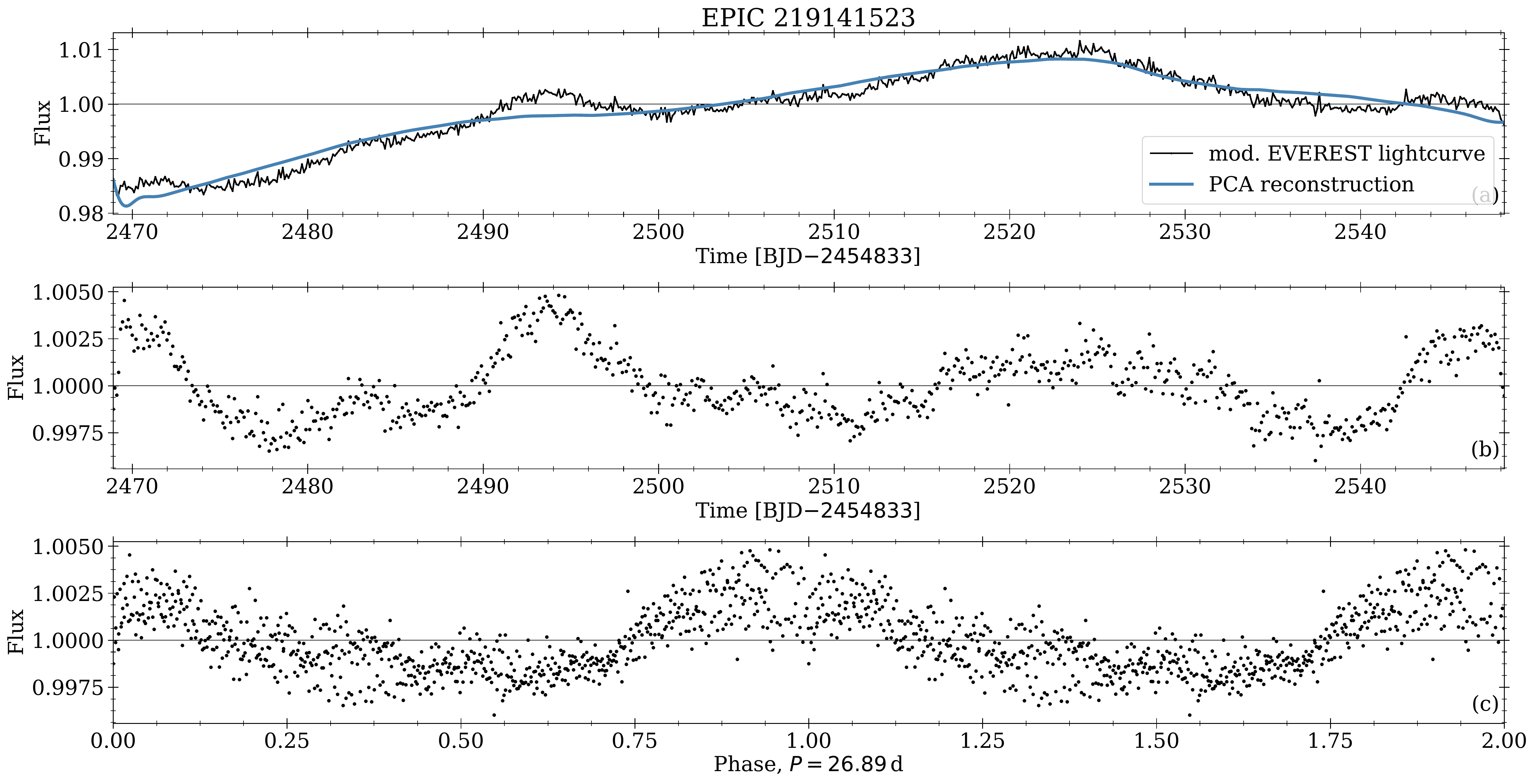} 
    \caption{Same as Fig.\,\ref{fig_lc_1}, but for EPIC 219141523 \label{fig_lc_3}} 
\end{figure*}  

\begin{figure*} 
    \centering
    \includegraphics[width=\linewidth]{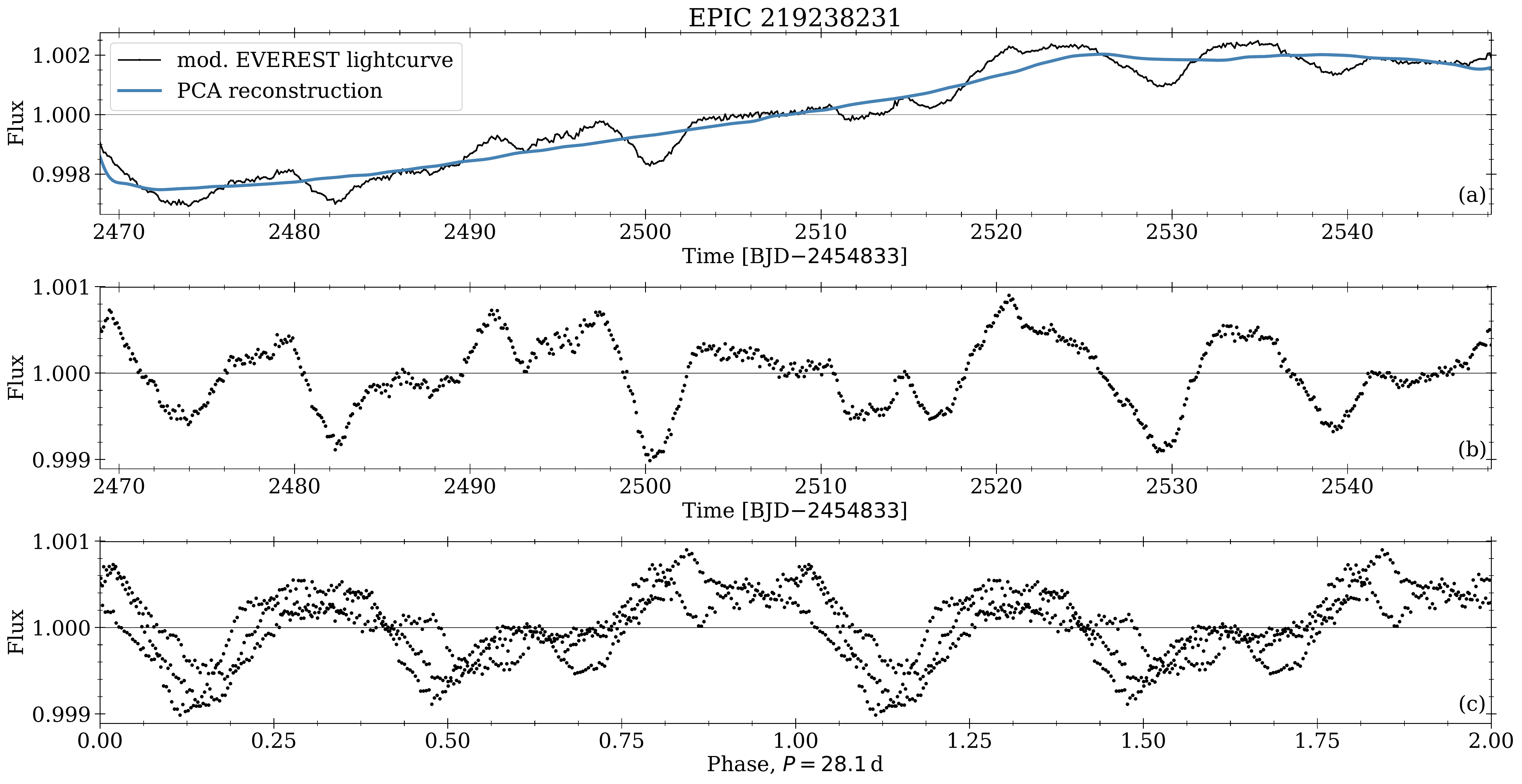} 
    \caption{Same as Fig.\,\ref{fig_lc_1}, but for EPIC 219238231 \label{fig_lc_4}} 
\end{figure*}

\begin{figure*} 
    \centering 
    \includegraphics[width=\linewidth]{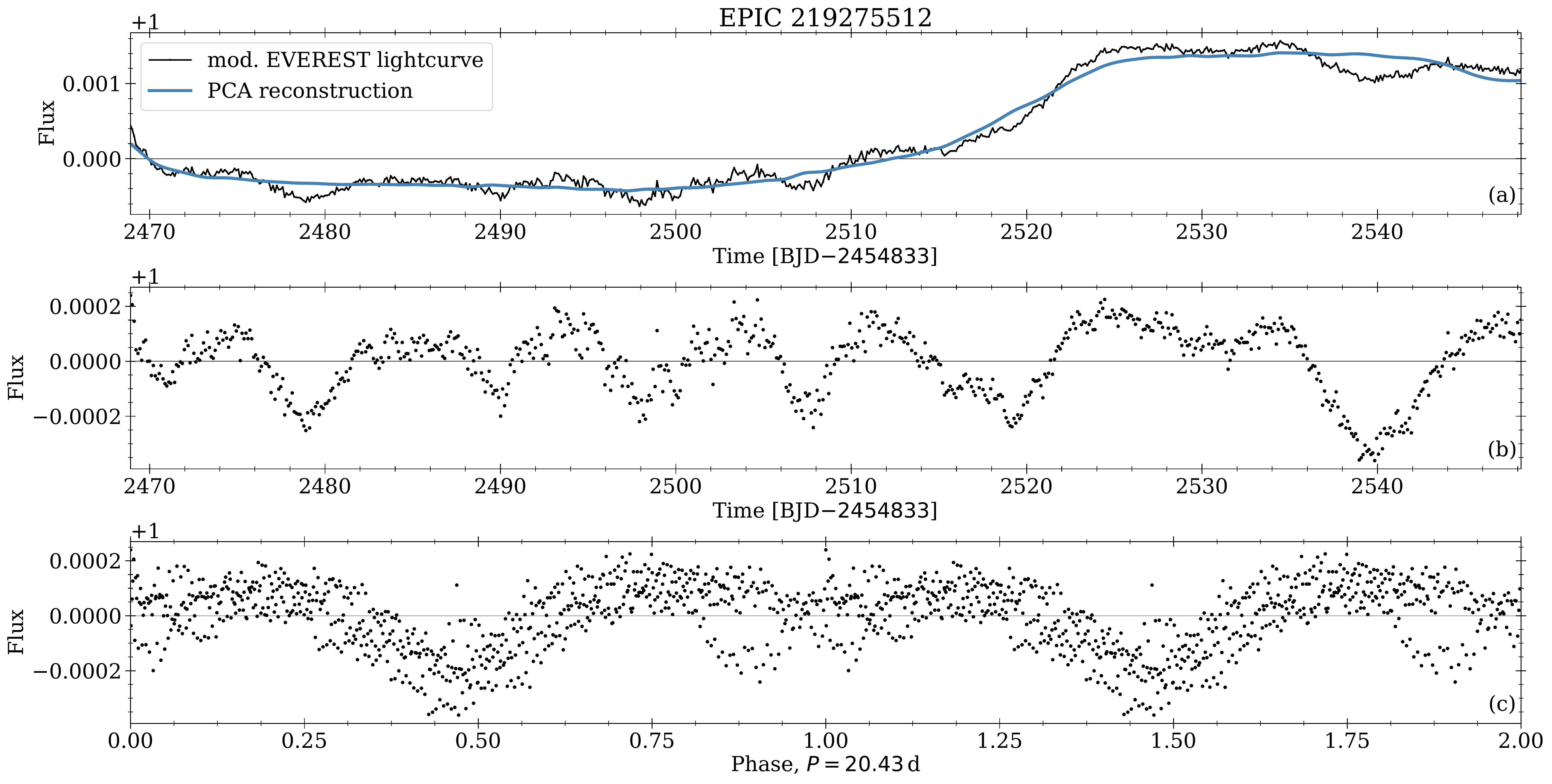} 
    \caption{Same as Fig.\,\ref{fig_lc_1}, but for EPIC 219275512 \label{fig_lc_5}} 
\end{figure*}  

\begin{figure*} 
    \centering 
    \includegraphics[width=\linewidth]{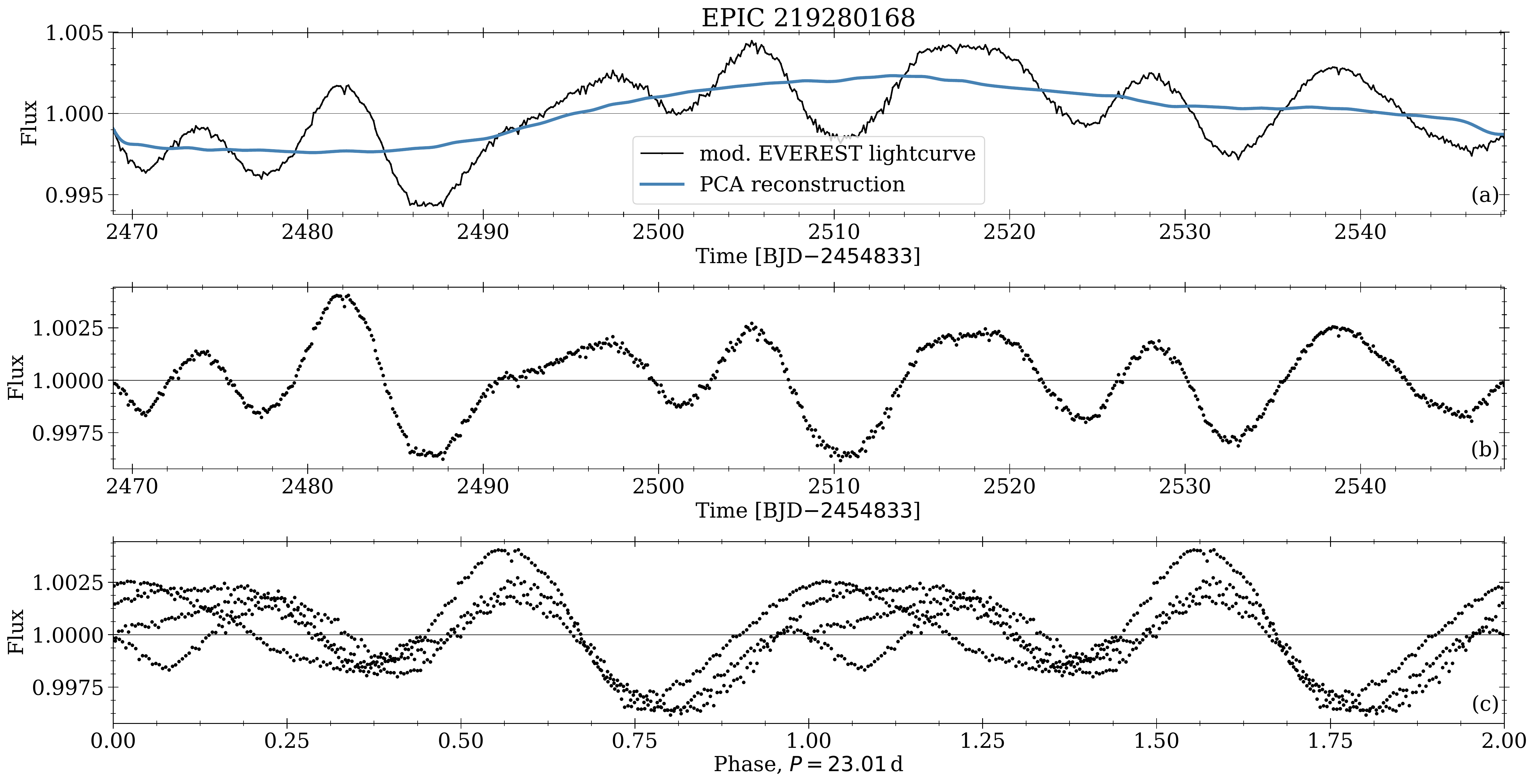} 
    \caption{Same as Fig.\,\ref{fig_lc_1}, but for EPIC 219280168 \label{fig_lc_6}} 
\end{figure*}

\begin{figure*} 
    \centering 
    \includegraphics[width=\linewidth]{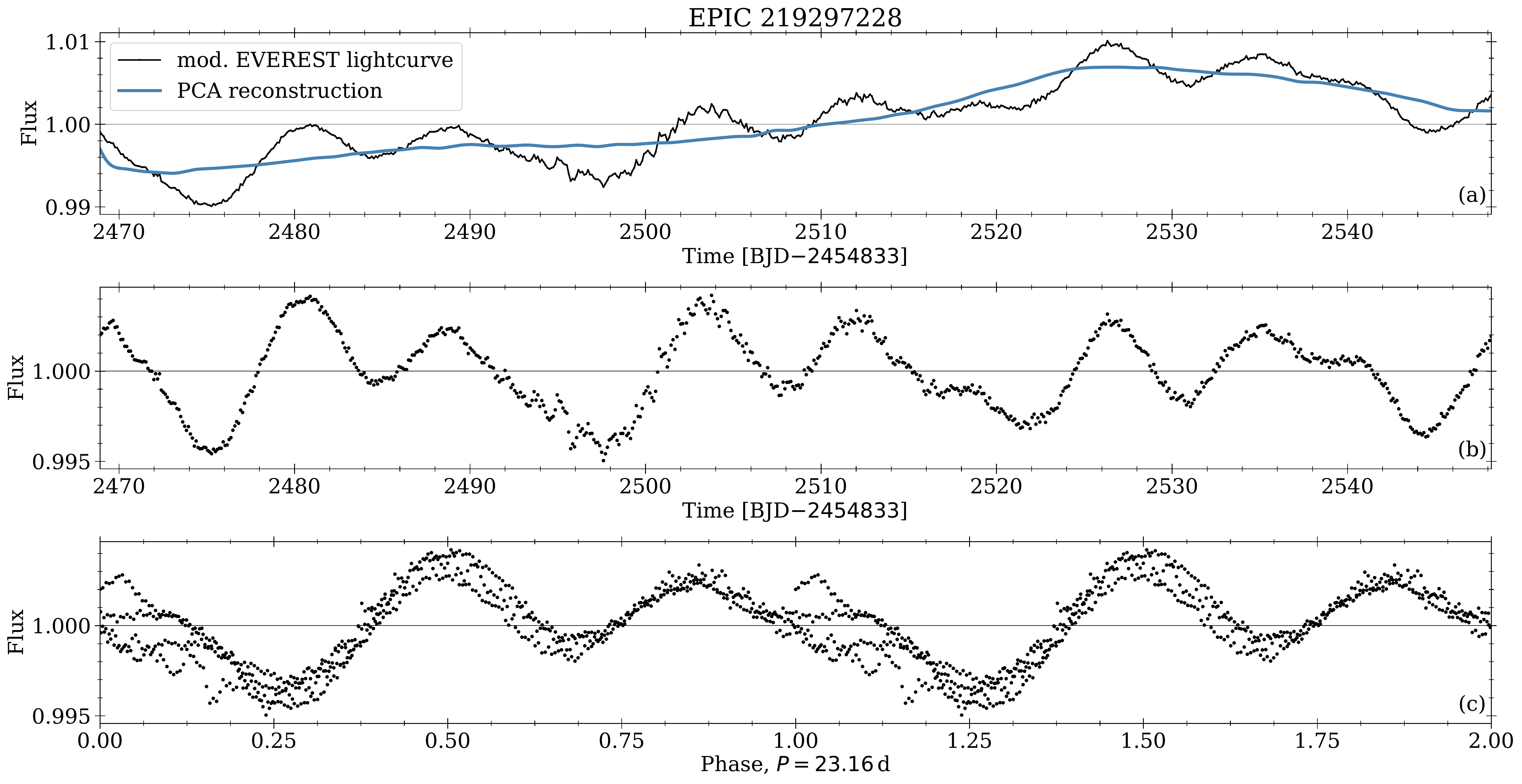} 
    \caption{Same as Fig.\,\ref{fig_lc_1}, but for EPIC 219297228 \label{fig_lc_7}} 
\end{figure*}  

\begin{figure*} 
    \centering 
    \includegraphics[width=\linewidth]{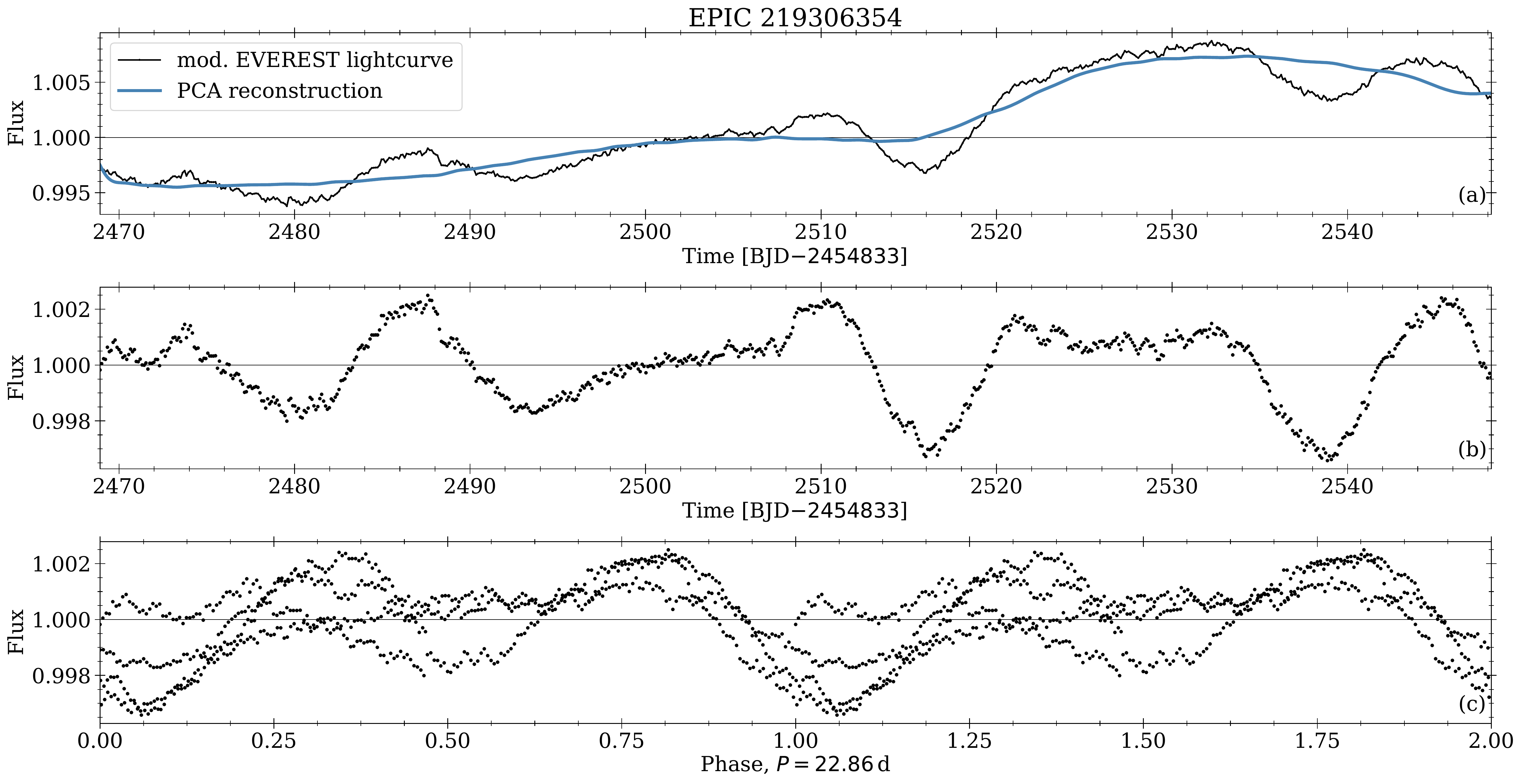} 
    \caption{Same as Fig.\,\ref{fig_lc_1}, but for EPIC 219306354 \label{fig_lc_8}} 
\end{figure*}

\begin{figure*} 
    \centering 
    \includegraphics[width=\linewidth]{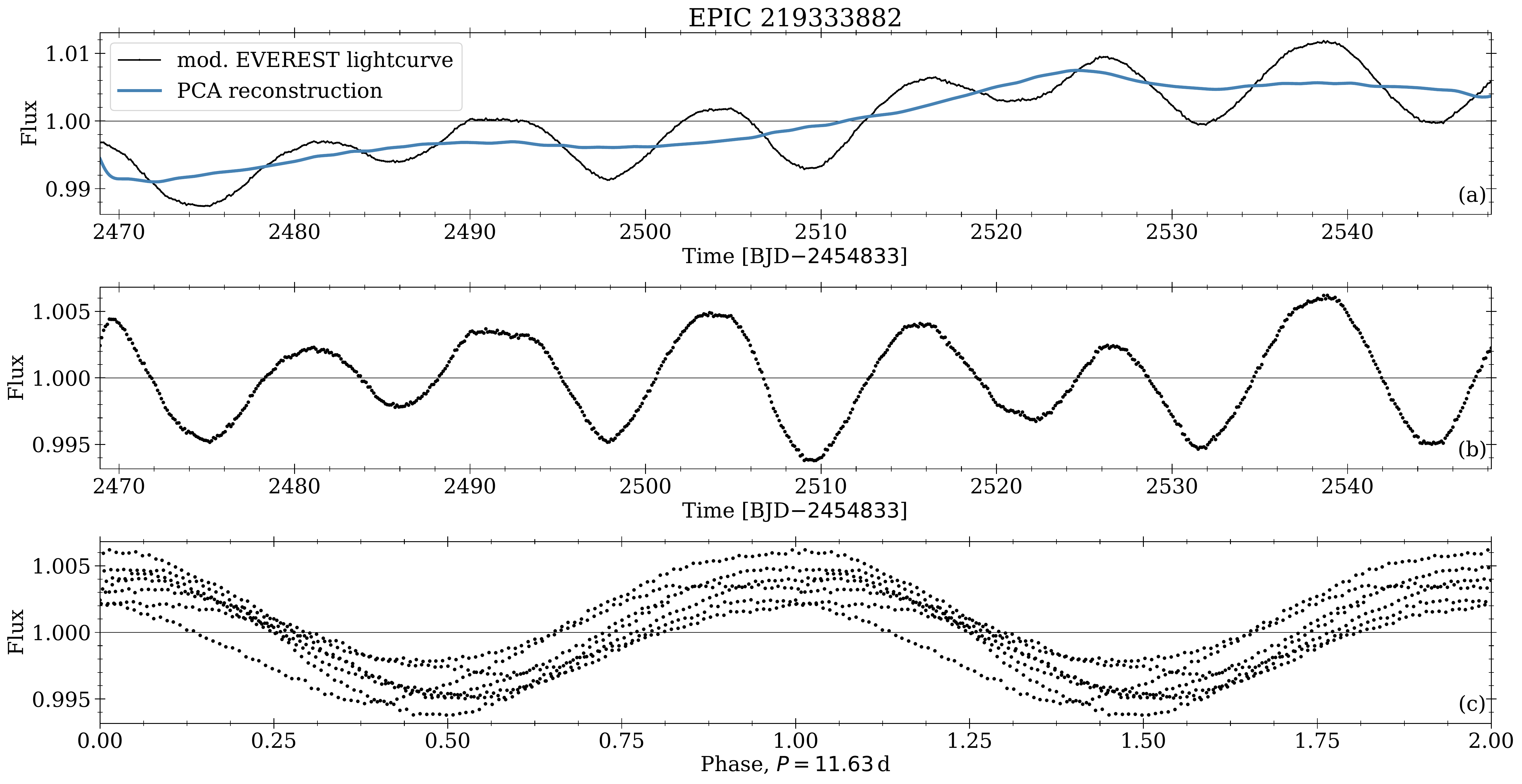} 
    \caption{Same as Fig.\,\ref{fig_lc_1}, but for EPIC 219333882 \label{fig_lc_9}} 
\end{figure*}  

\begin{figure*} 
    \centering 
    \includegraphics[width=\linewidth]{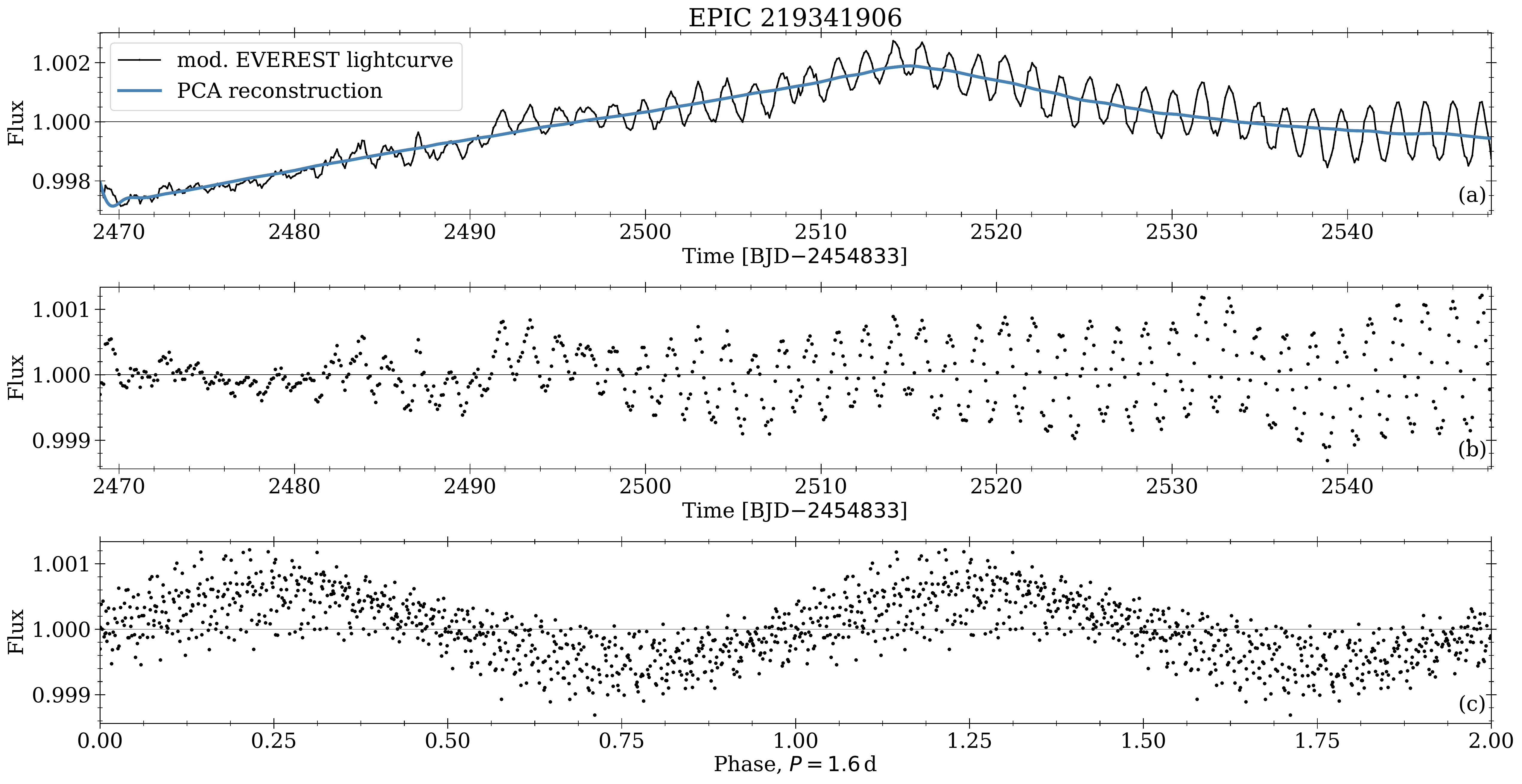} 
    \caption{Same as Fig.\,\ref{fig_lc_1}, but for EPIC 219341906 \label{fig_lc_10}} 
\end{figure*}

\begin{figure*} 
    \centering 
    \includegraphics[width=\linewidth]{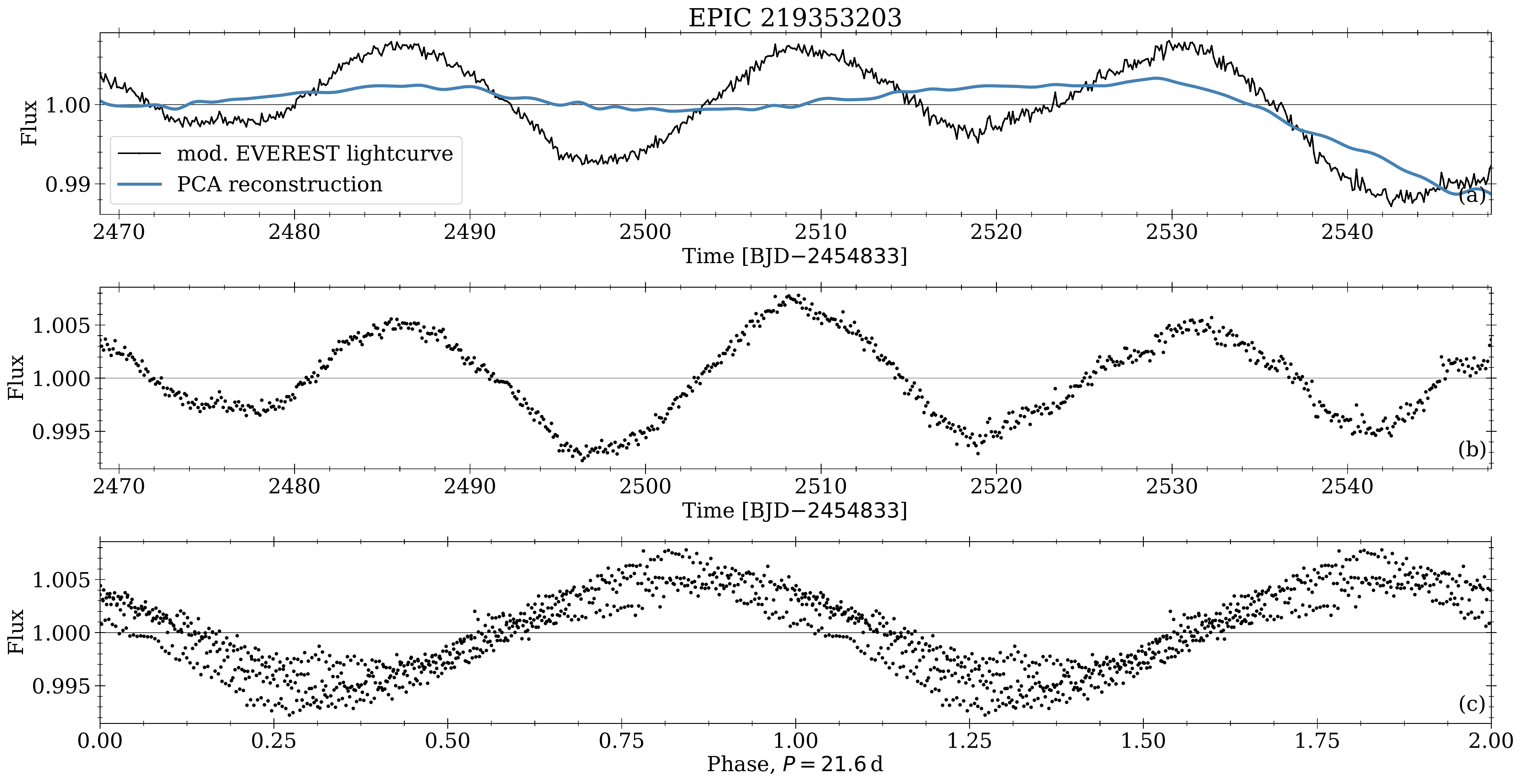} 
    \caption{Same as Fig.\,\ref{fig_lc_1}, but for EPIC 219353203 \label{fig_lc_11}} 
\end{figure*}  

\begin{figure*} 
    \centering 
    \includegraphics[width=\linewidth]{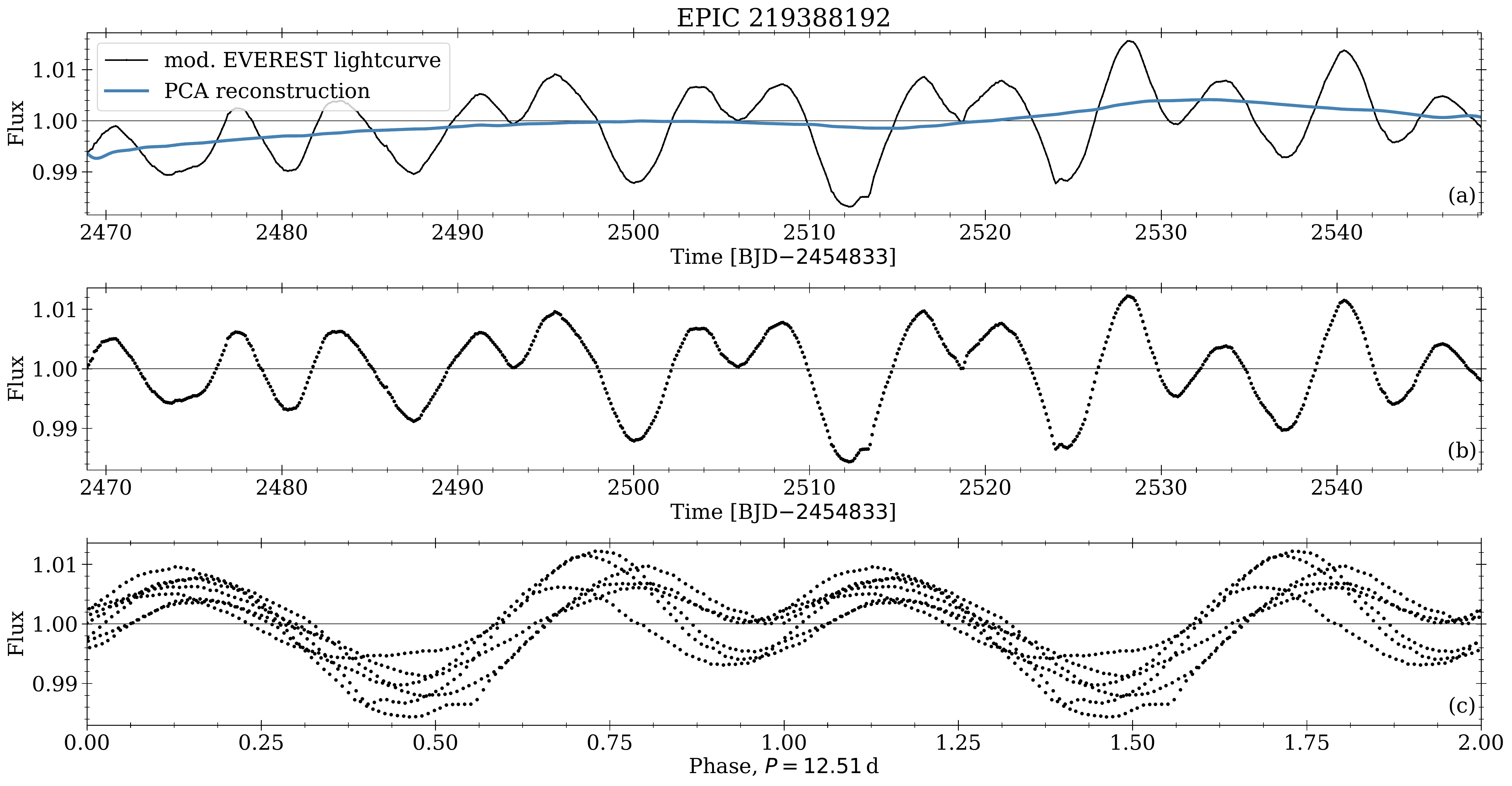} 
    \caption{Same as Fig.\,\ref{fig_lc_1}, but for EPIC 219388192 \label{fig_lc_12}} 
\end{figure*}

\begin{figure*} 
    \centering 
    \includegraphics[width=\linewidth]{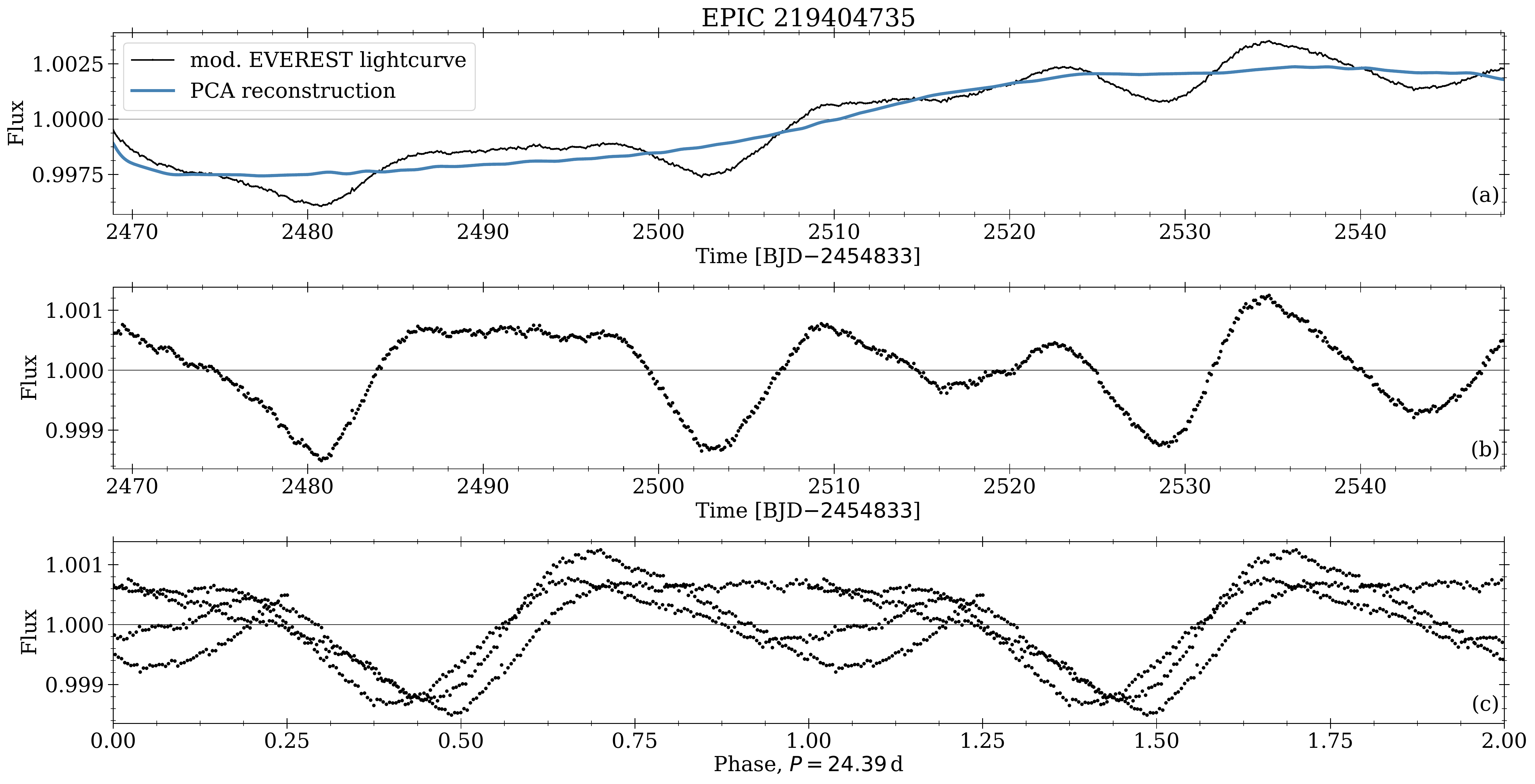} 
    \caption{Same as Fig.\,\ref{fig_lc_1}, but for EPIC 219404735 \label{fig_lc_13}} 
\end{figure*}  

\begin{figure*} 
    \centering 
    \includegraphics[width=\linewidth]{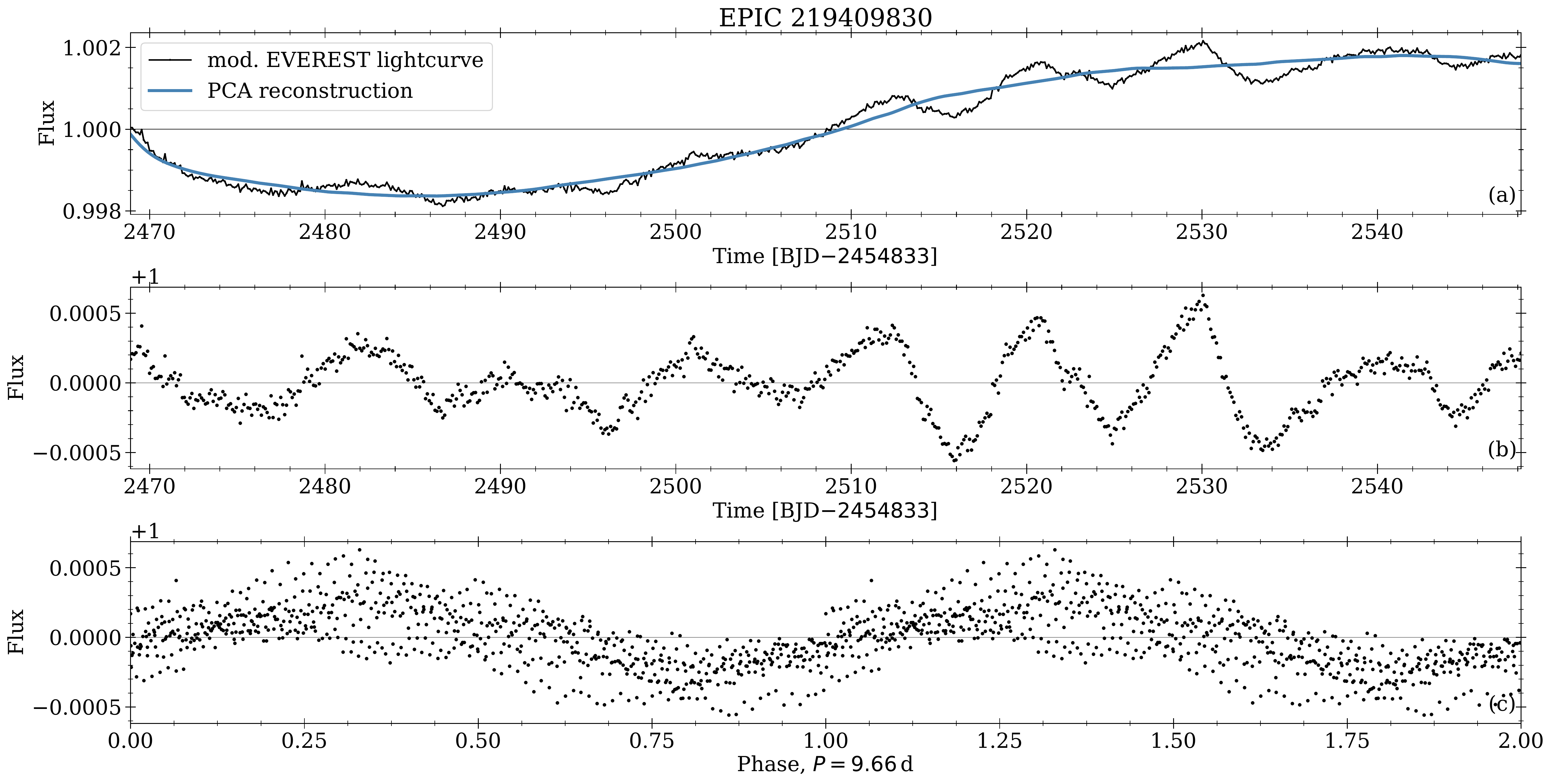} 
    \caption{Same as Fig.\,\ref{fig_lc_1}, but for EPIC 219409830 \label{fig_lc_14}} 
\end{figure*}

\begin{figure*} 
    \centering 
    \includegraphics[width=\linewidth]{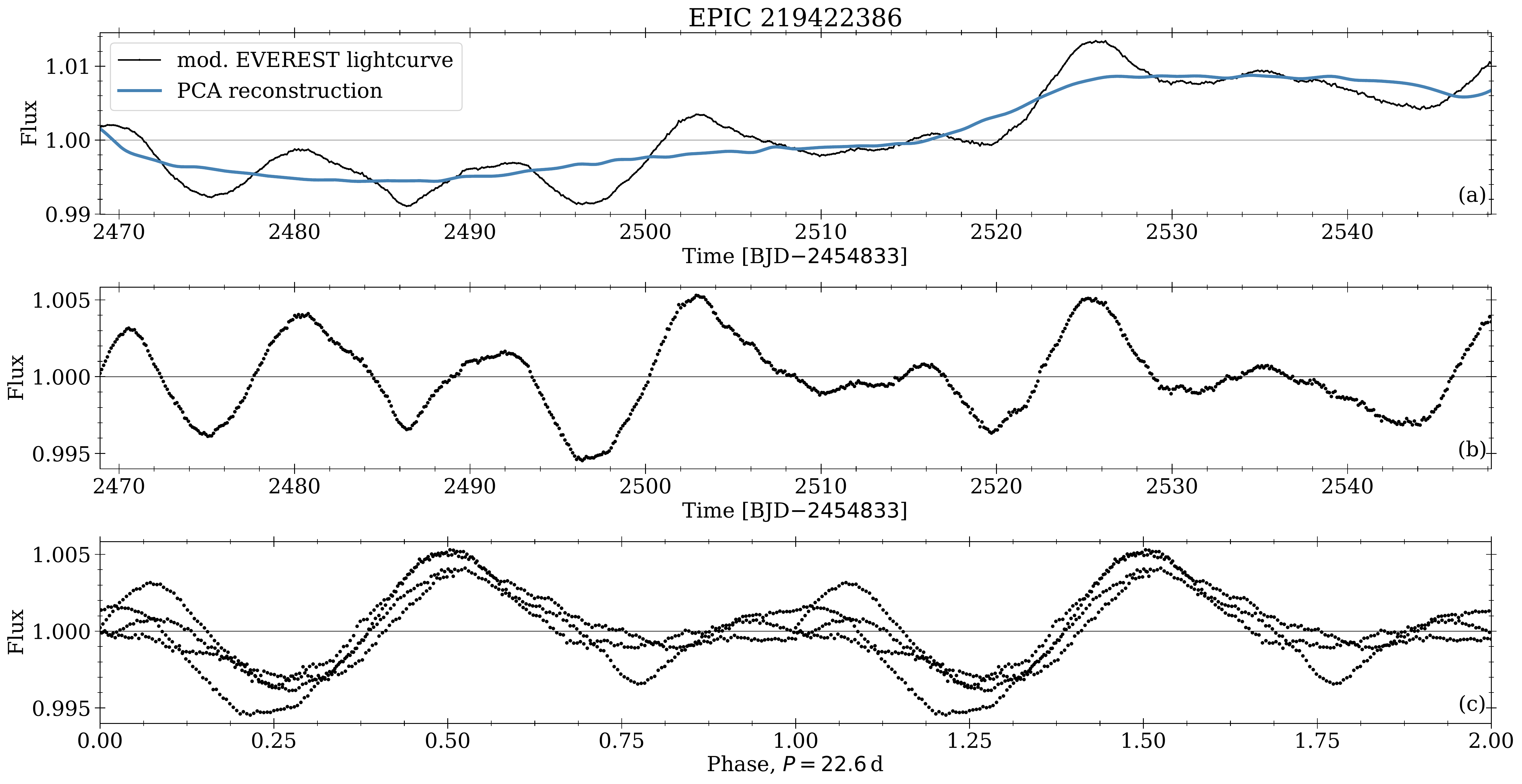} 
    \caption{Same as Fig.\,\ref{fig_lc_1}, but for EPIC 219422386 \label{fig_lc_15}} 
\end{figure*}  

\begin{figure*} 
    \centering 
    \includegraphics[width=\linewidth]{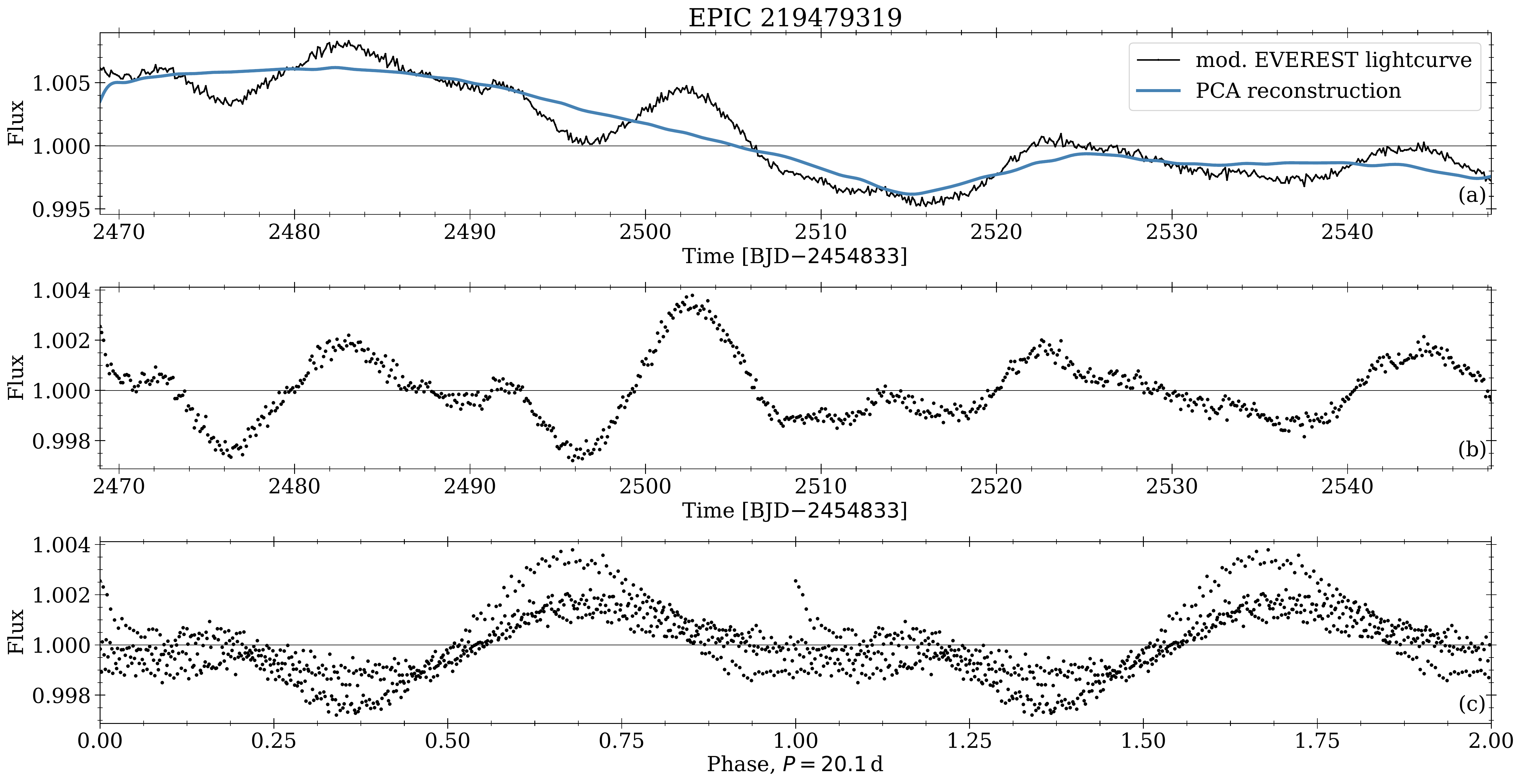} 
    \caption{Same as Fig.\,\ref{fig_lc_1}, but for EPIC 219479319 \label{fig_lc_16}} 
\end{figure*}

\begin{figure*} 
    \centering 
    \includegraphics[width=\linewidth]{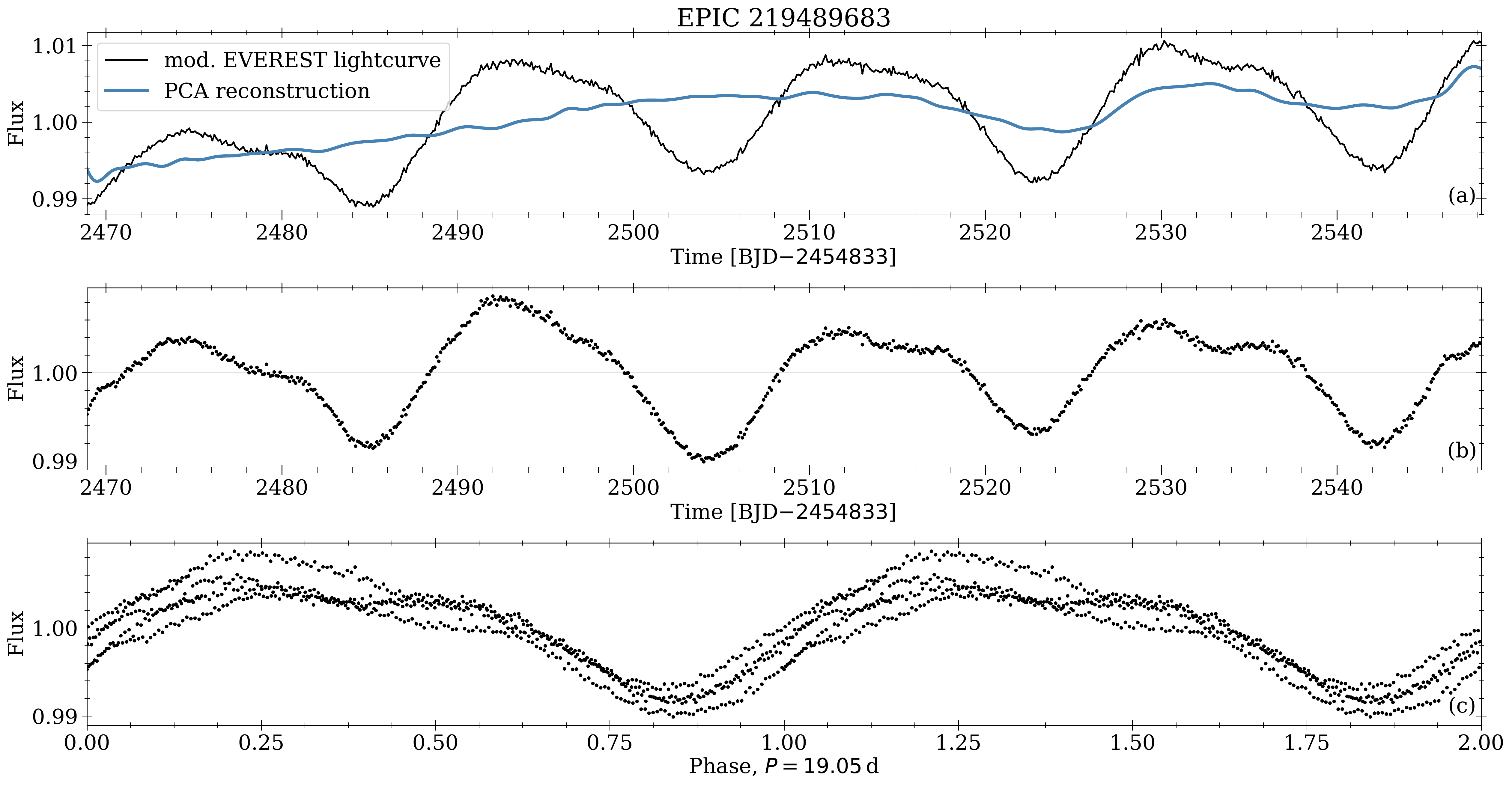} 
    \caption{Same as Fig.\,\ref{fig_lc_1}, but for EPIC 219489683 \label{fig_lc_17}} 
\end{figure*}  

\begin{figure*} 
    \centering 
    \includegraphics[width=\linewidth]{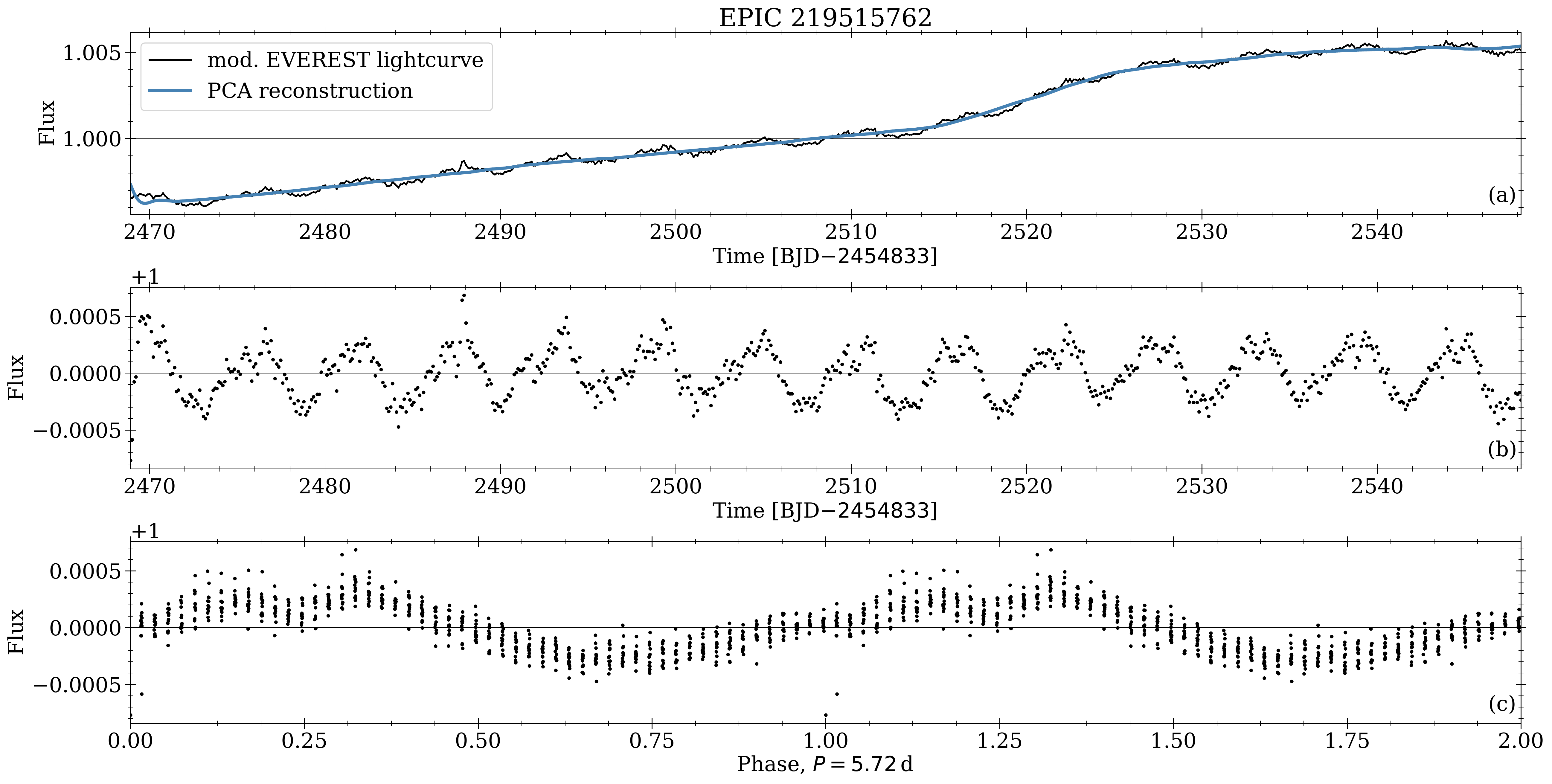} 
    \caption{Same as Fig.\,\ref{fig_lc_1}, but for EPIC 219515762 \label{fig_lc_18}} 
\end{figure*}

\begin{figure*} 
    \centering 
    \includegraphics[width=\linewidth]{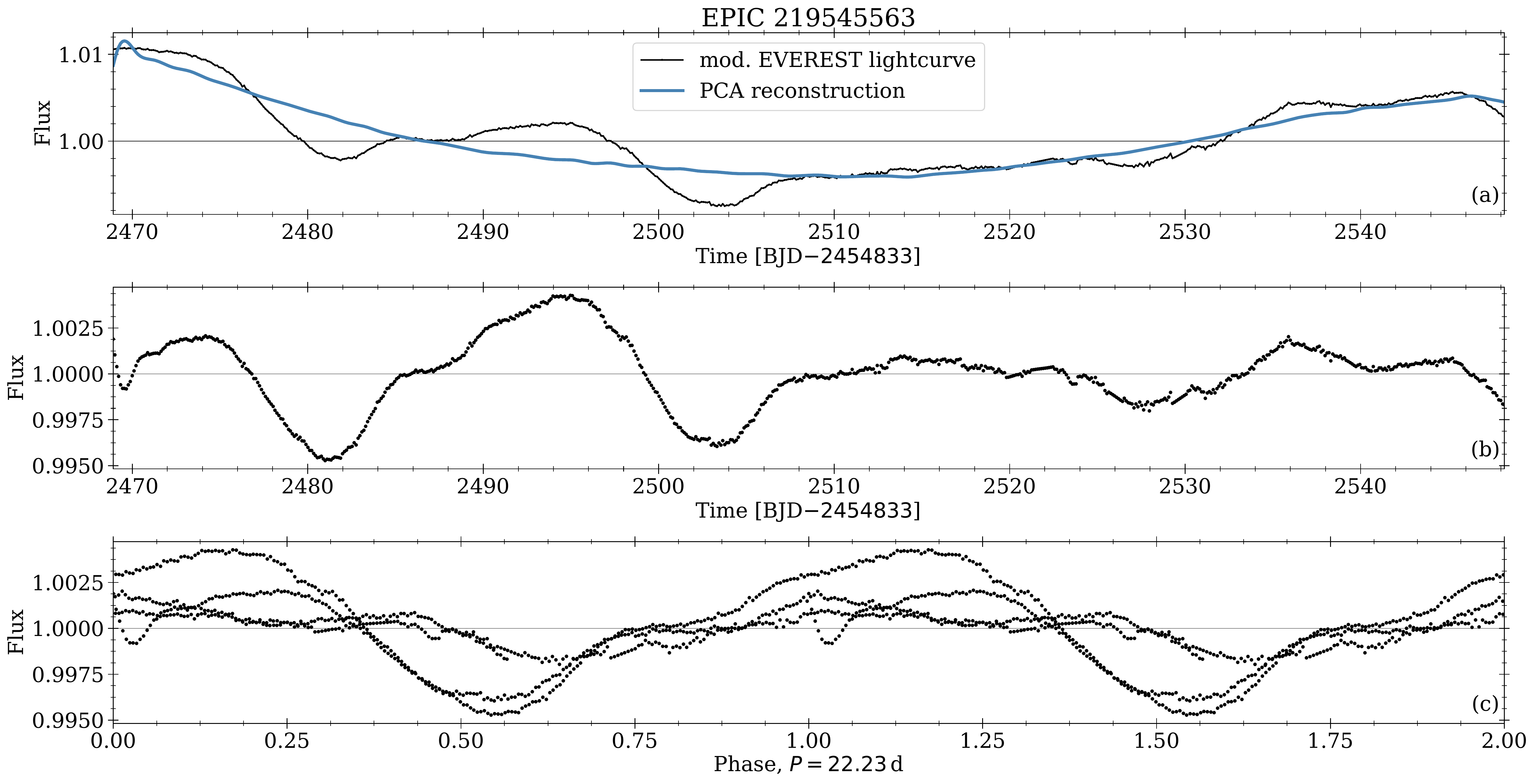} 
    \caption{Same as Fig.\,\ref{fig_lc_1}, but for EPIC 219545563 \label{fig_lc_19}} 
\end{figure*}  

\begin{figure*} 
    \centering 
    \includegraphics[width=\linewidth]{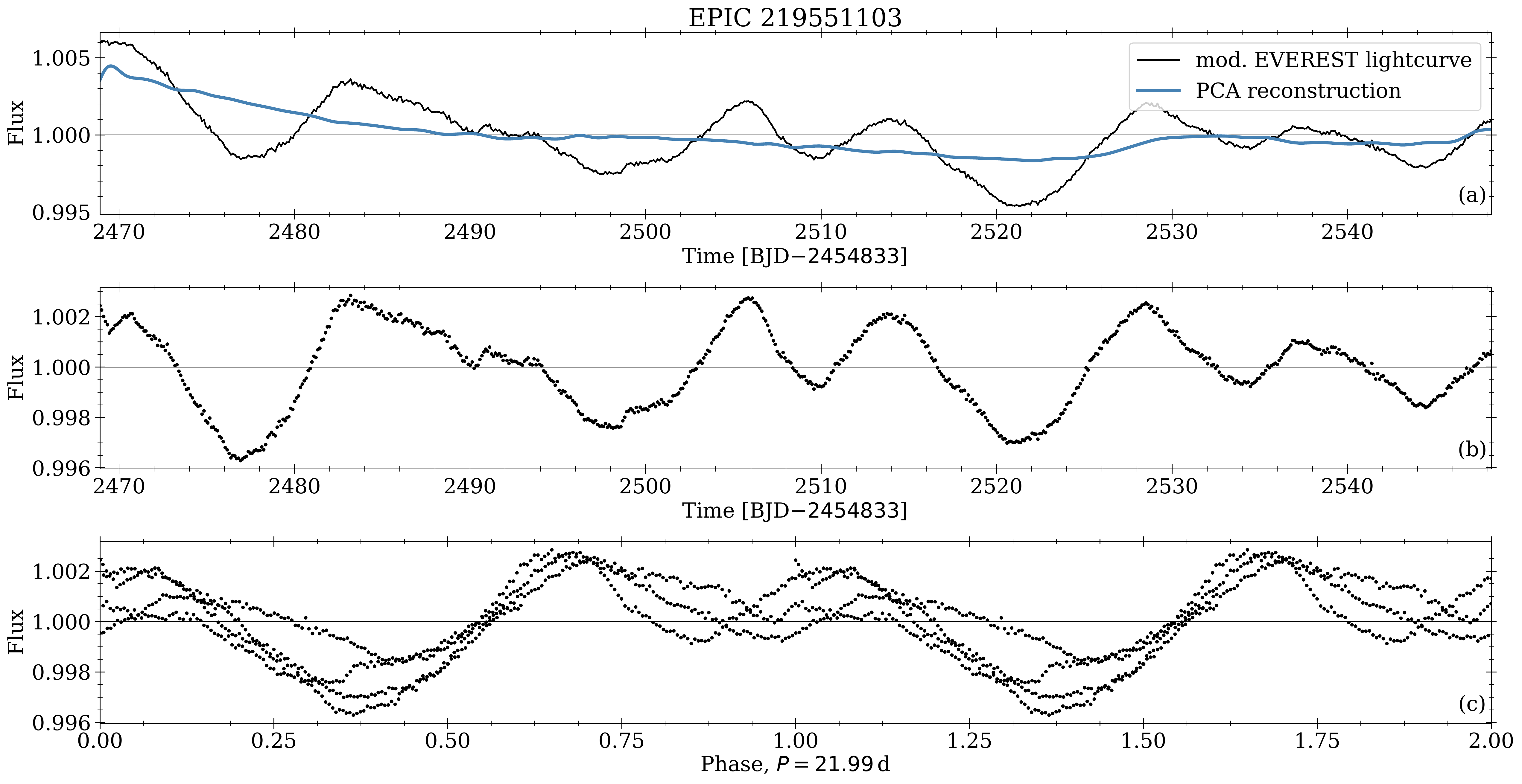} 
    \caption{Same as Fig.\,\ref{fig_lc_1}, but for EPIC 219551103 \label{fig_lc_20}} 
\end{figure*}

\begin{figure*} 
    \centering 
    \includegraphics[width=\linewidth]{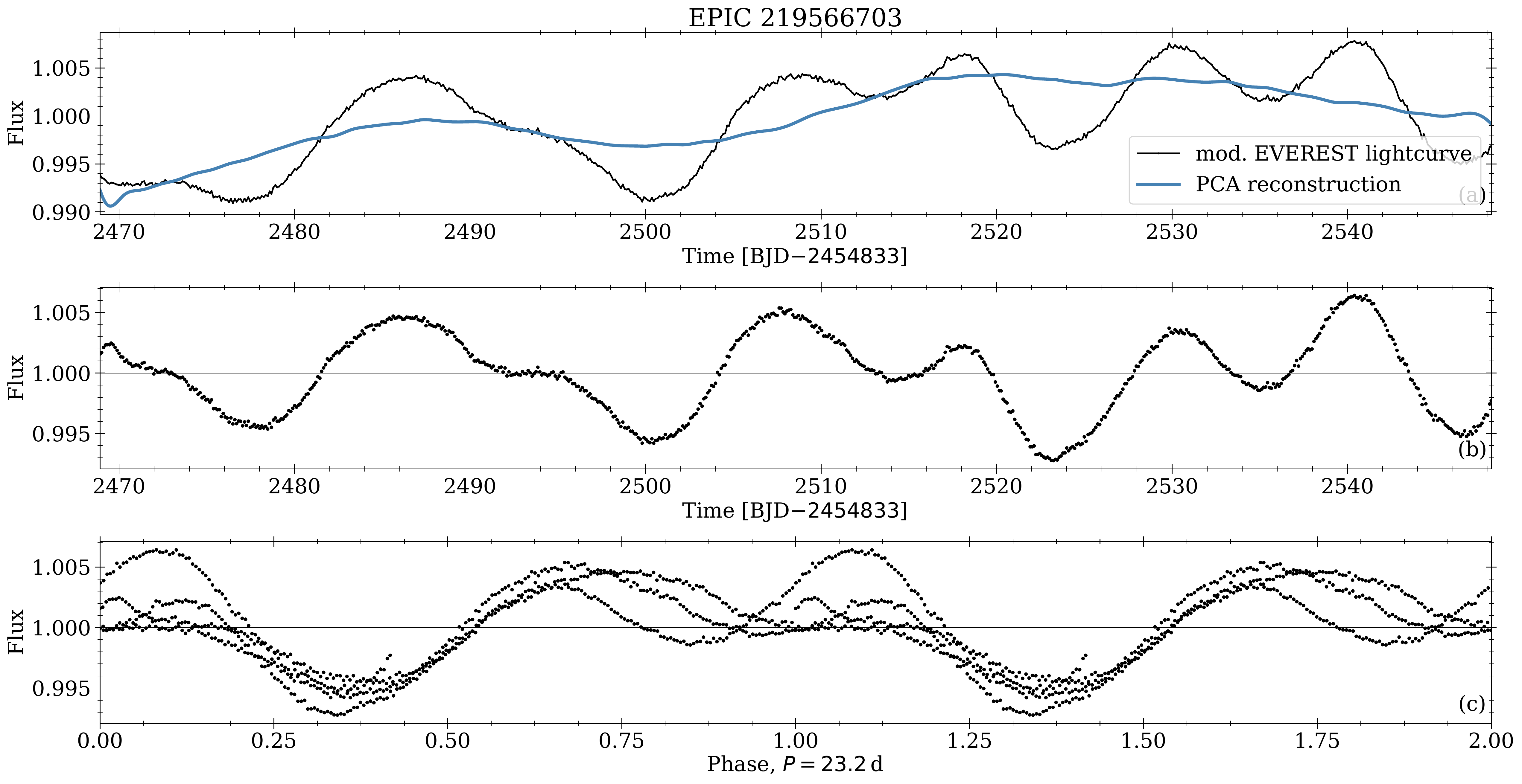} 
    \caption{Same as Fig.\,\ref{fig_lc_1}, but for EPIC 219566703 \label{fig_lc_21}} 
\end{figure*}  

\begin{figure*} 
    \centering 
    \includegraphics[width=\linewidth]{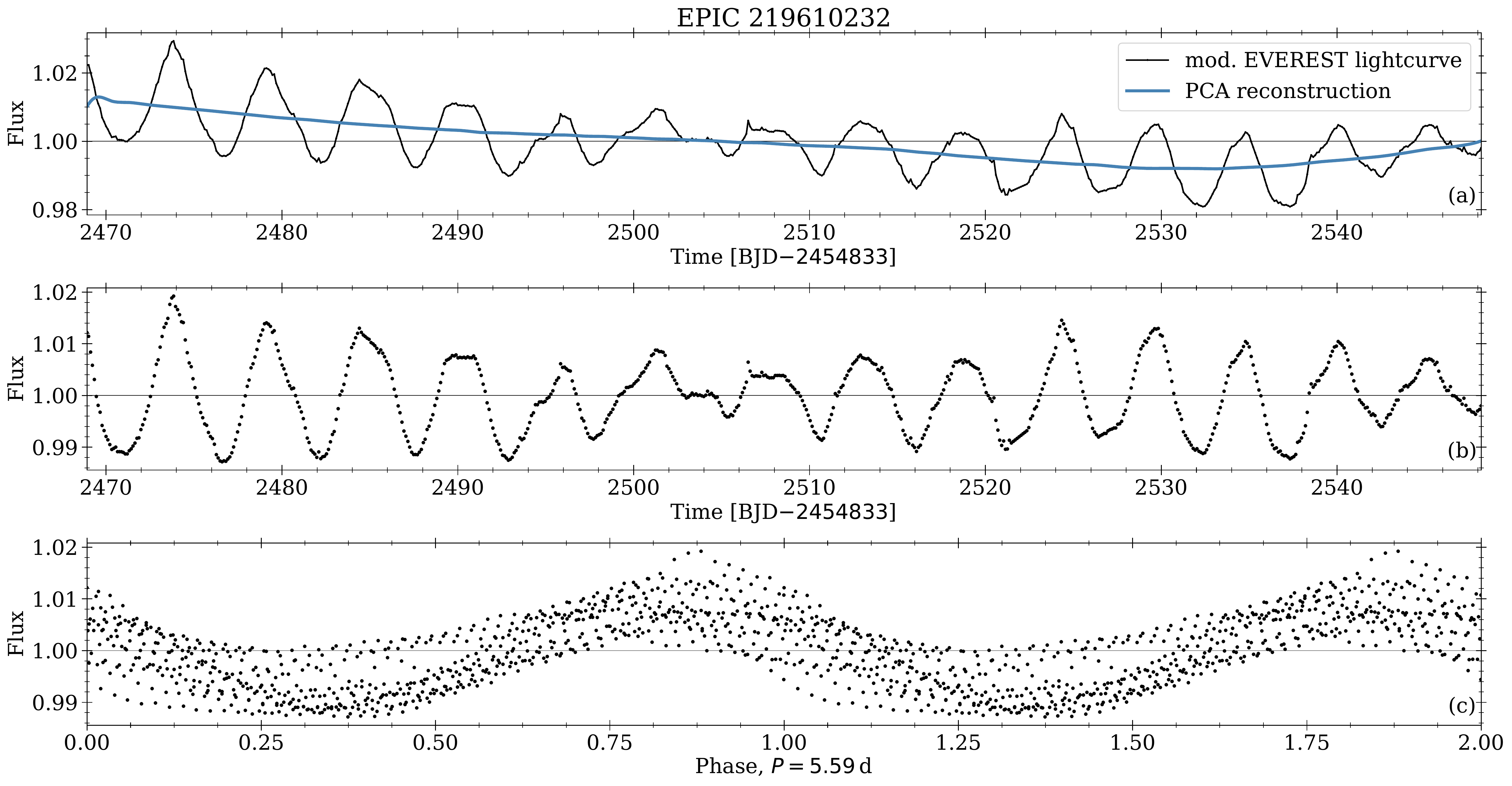} 
    \caption{Same as Fig.\,\ref{fig_lc_1}, but for EPIC 219610232 \label{fig_lc_22}} 
\end{figure*}

\begin{figure*} 
    \centering 
    \includegraphics[width=\linewidth]{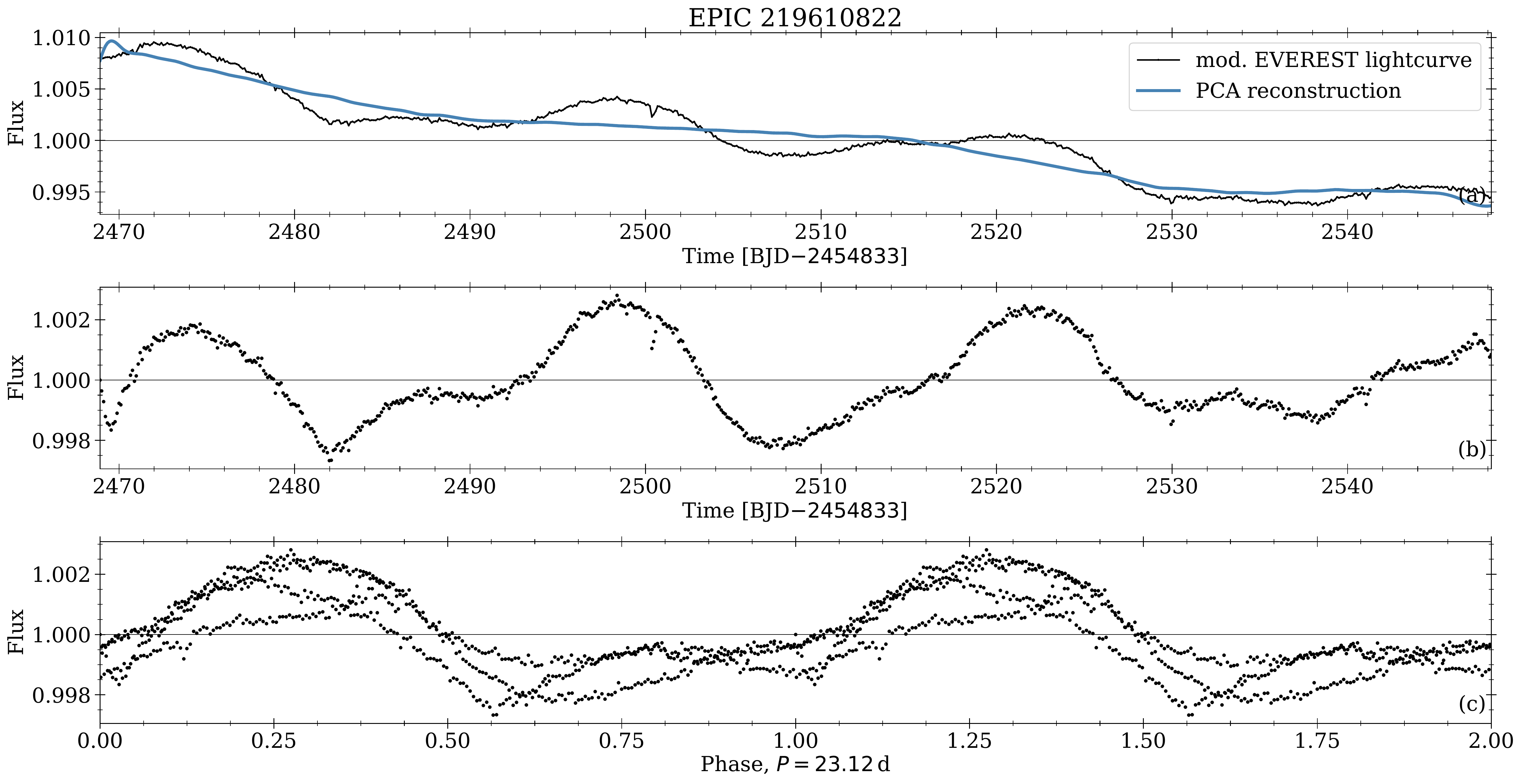} 
    \caption{Same as Fig.\,\ref{fig_lc_1}, but for EPIC 219610822 \label{fig_lc_23}} 
\end{figure*}  

\begin{figure*} 
    \centering 
    \includegraphics[width=\linewidth]{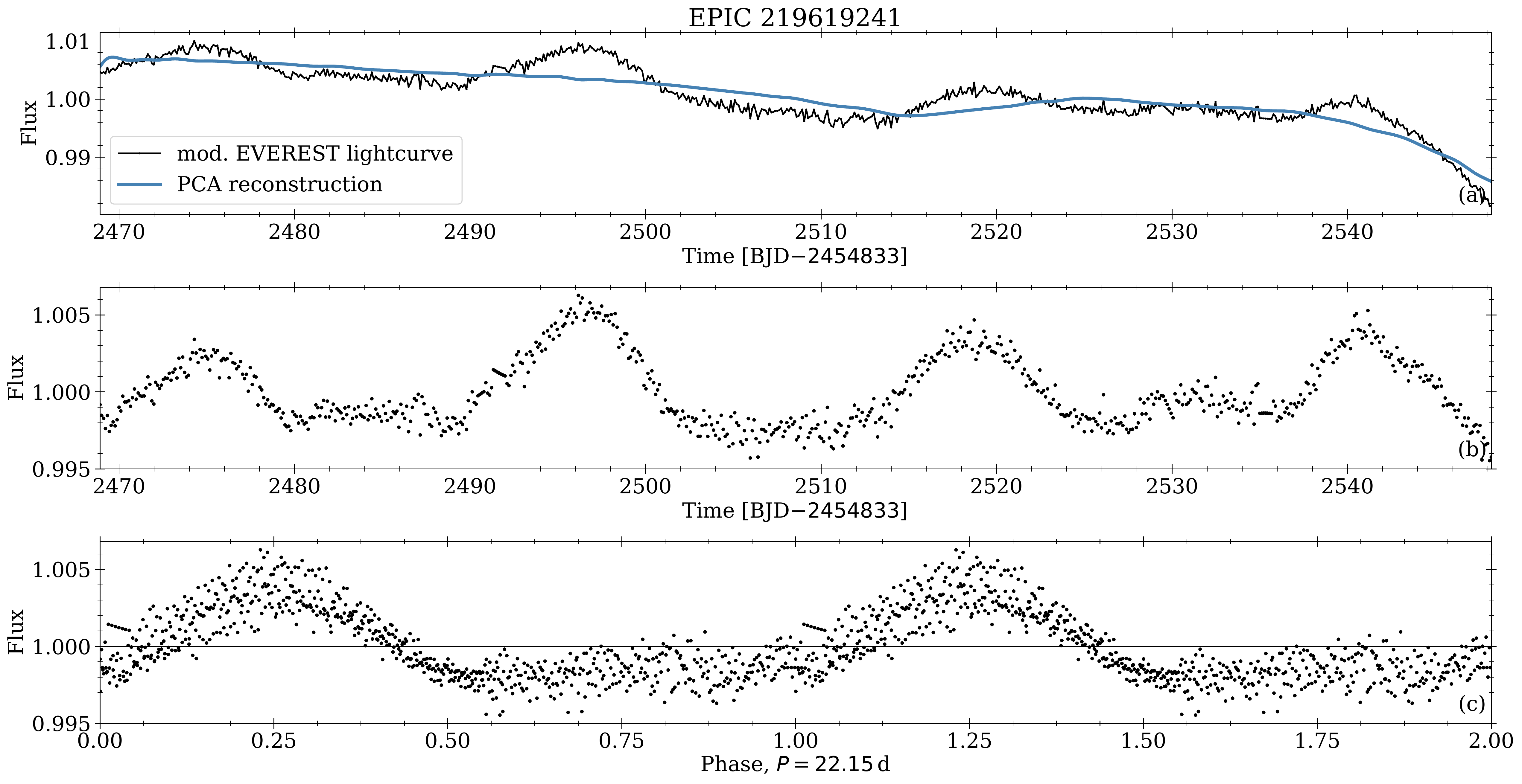} 
    \caption{Same as Fig.\,\ref{fig_lc_1}, but for EPIC 219619241 \label{fig_lc_24}} 
\end{figure*}

\begin{figure*} 
    \centering 
    \includegraphics[width=\linewidth]{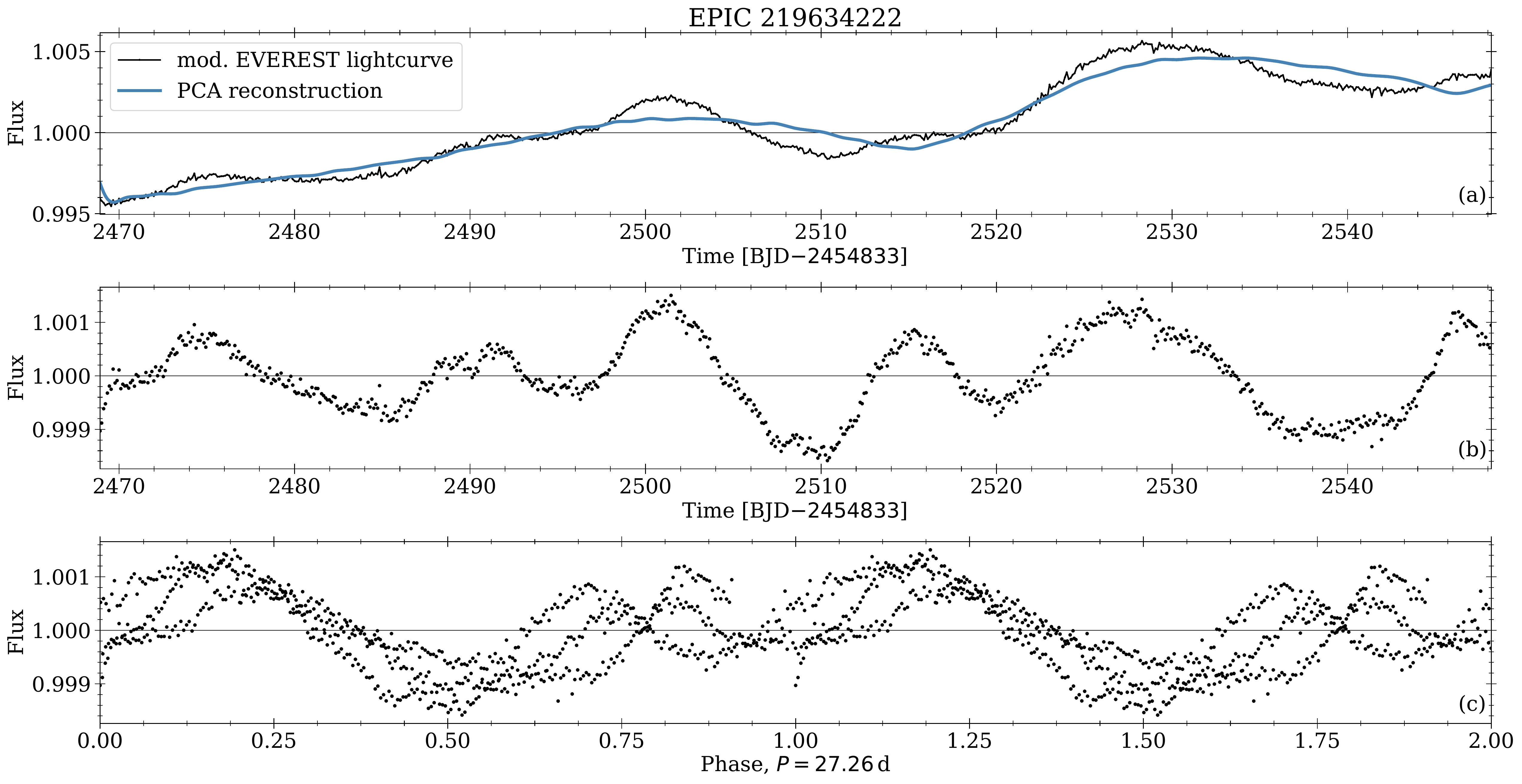} 
    \caption{Same as Fig.\,\ref{fig_lc_1}, but for EPIC 219634222 \label{fig_lc_25}} 
\end{figure*}  

\begin{figure*} 
    \centering 
    \includegraphics[width=\linewidth]{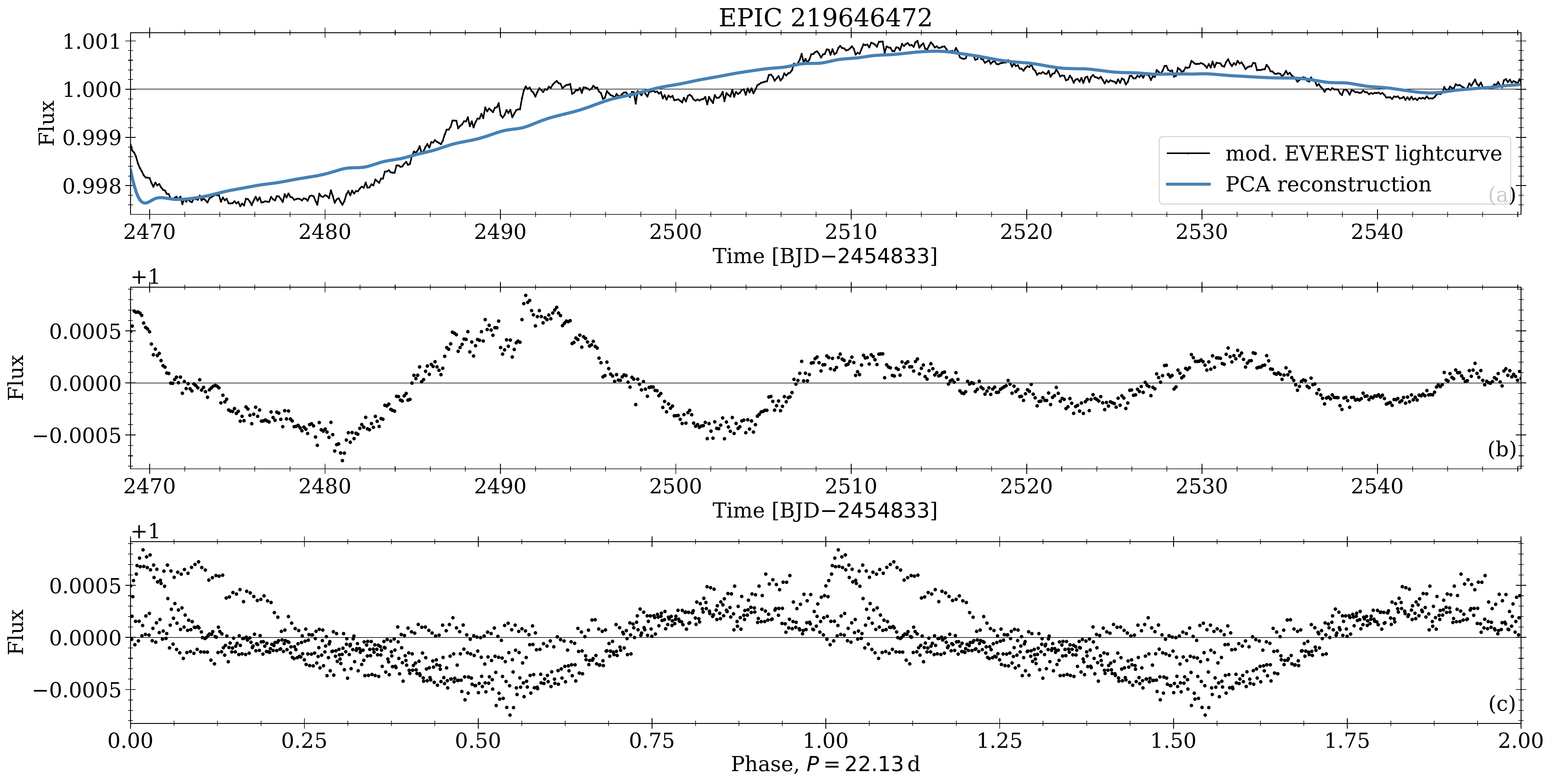} 
    \caption{Same as Fig.\,\ref{fig_lc_1}, but for EPIC 219646472 \label{fig_lc_26}} 
\end{figure*}

\begin{figure*} 
    \centering 
    \includegraphics[width=\linewidth]{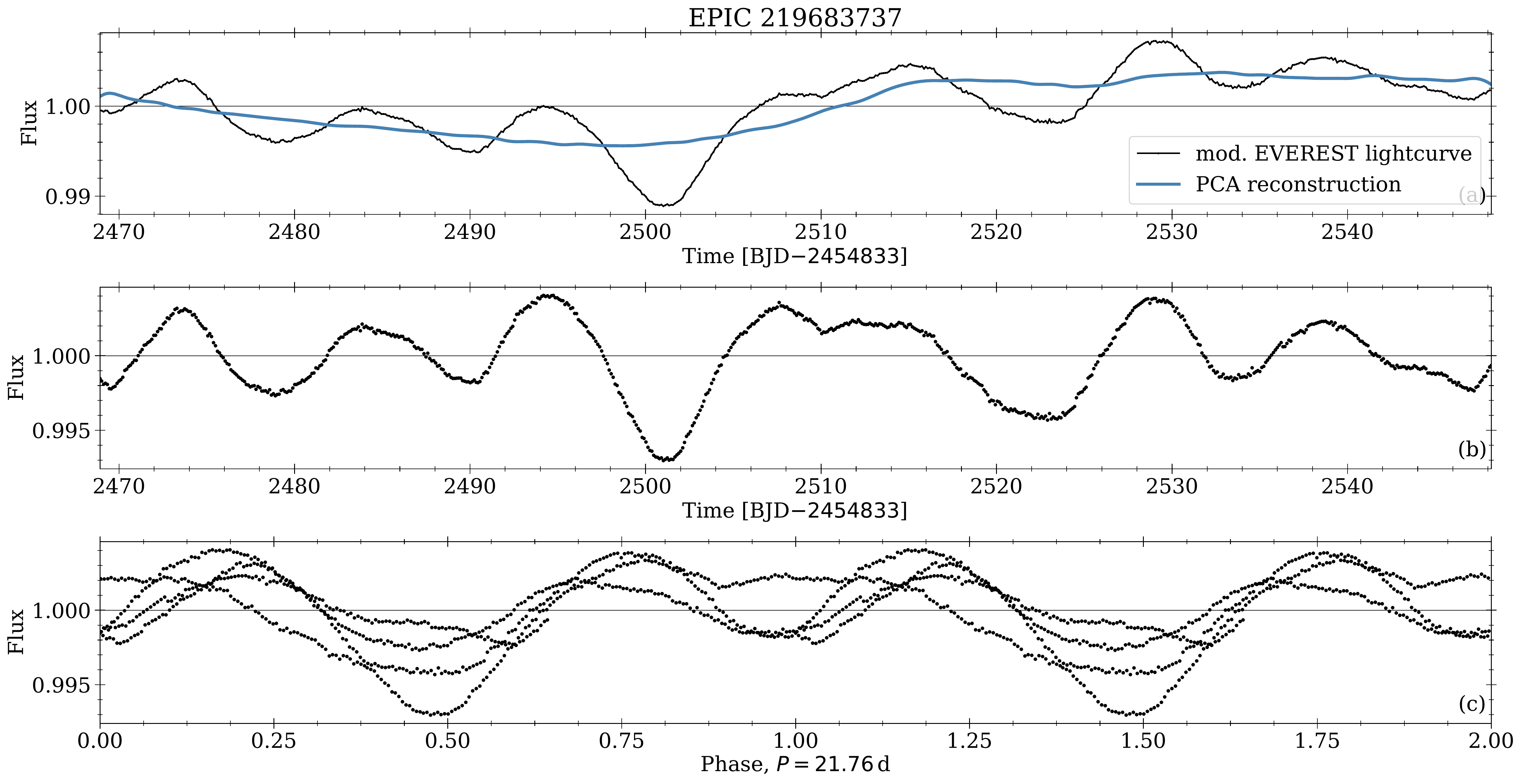} 
    \caption{Same as Fig.\,\ref{fig_lc_1}, but for EPIC 219683737 \label{fig_lc_27}} 
\end{figure*}  

\begin{figure*} 
    \centering 
    \includegraphics[width=\linewidth]{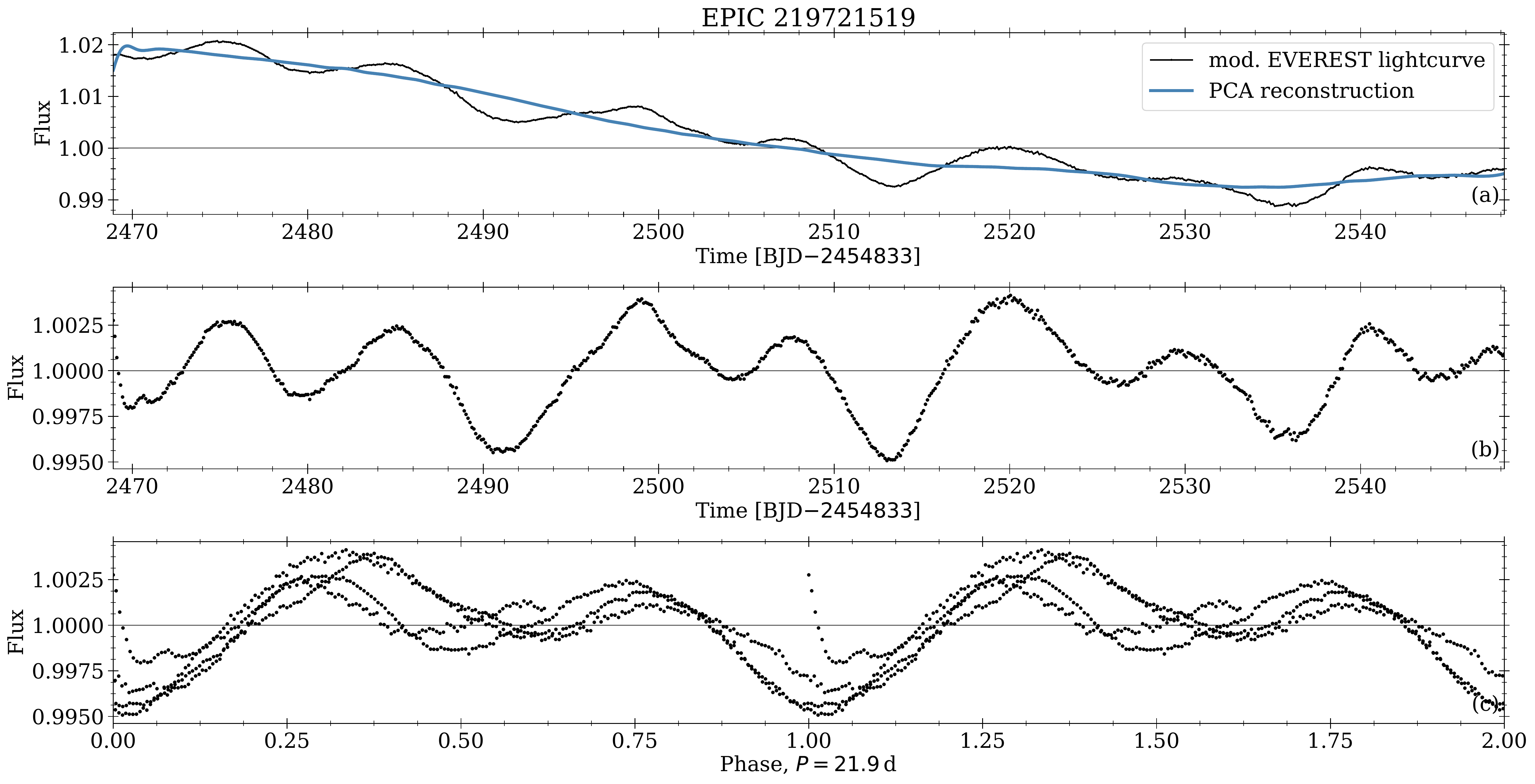} 
    \caption{Same as Fig.\,\ref{fig_lc_1}, but for EPIC 219721519 \label{fig_lc_28}} 
\end{figure*}

\begin{figure*} 
    \centering 
    \includegraphics[width=\linewidth]{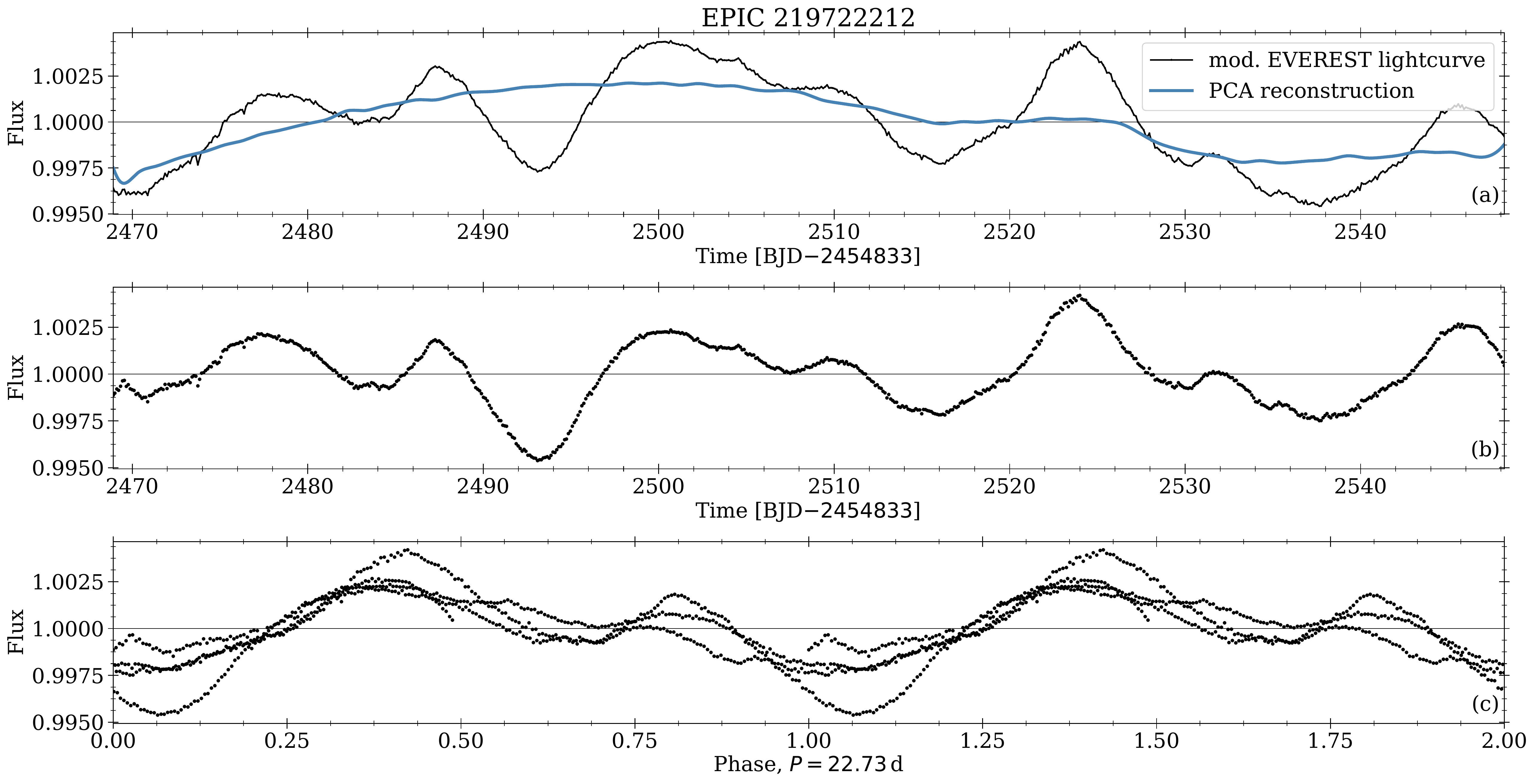} 
    \caption{Same as Fig.\,\ref{fig_lc_1}, but for EPIC 219722212 \label{fig_lc_29}} 
\end{figure*}  

\begin{figure*} 
    \centering 
    \includegraphics[width=\linewidth]{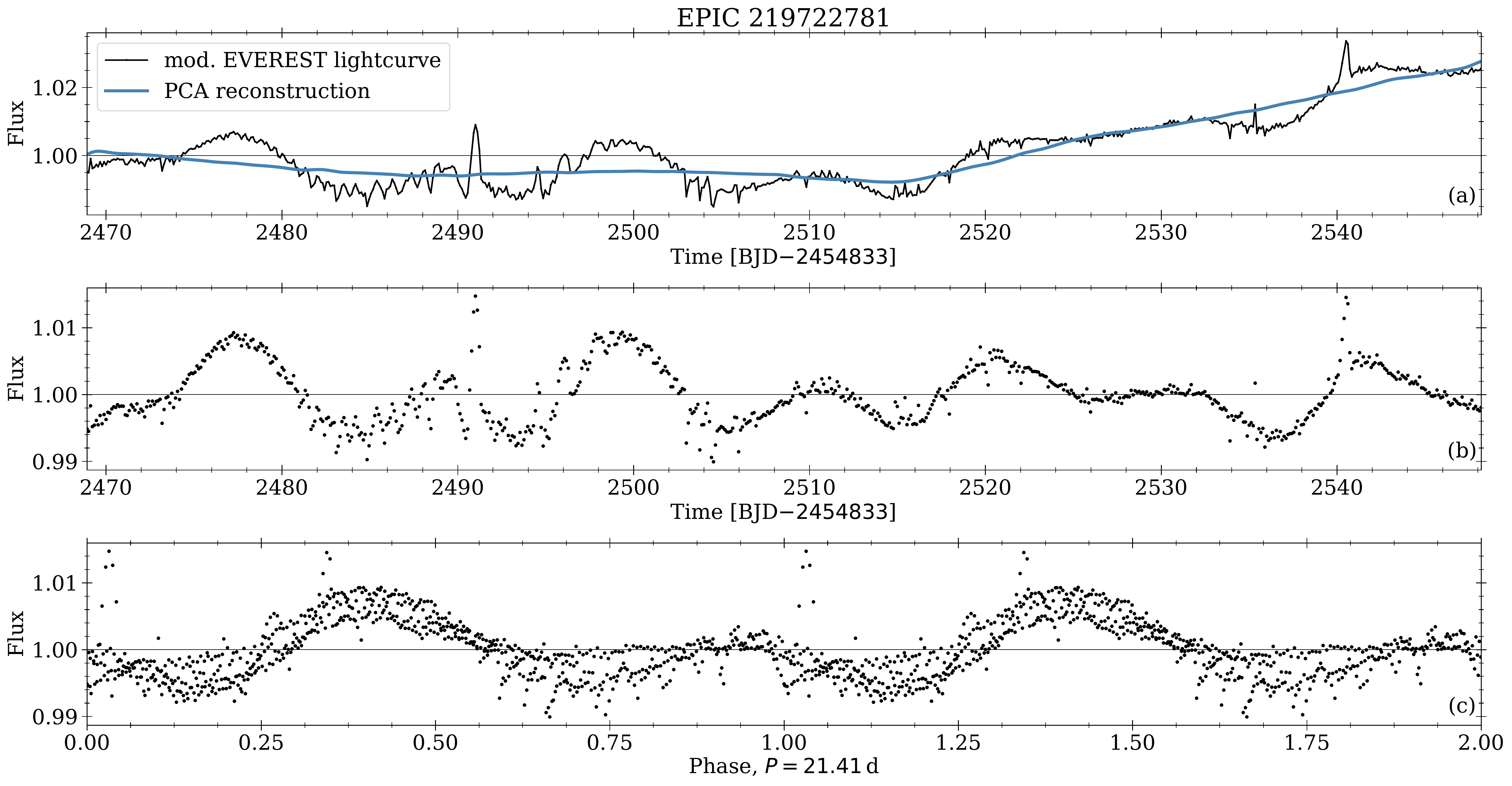} 
    \caption{Same as Fig.\,\ref{fig_lc_1}, but for EPIC 219722781 \label{fig_lc_30}} 
\end{figure*}

\begin{figure*} 
    \centering 
    \includegraphics[width=\linewidth]{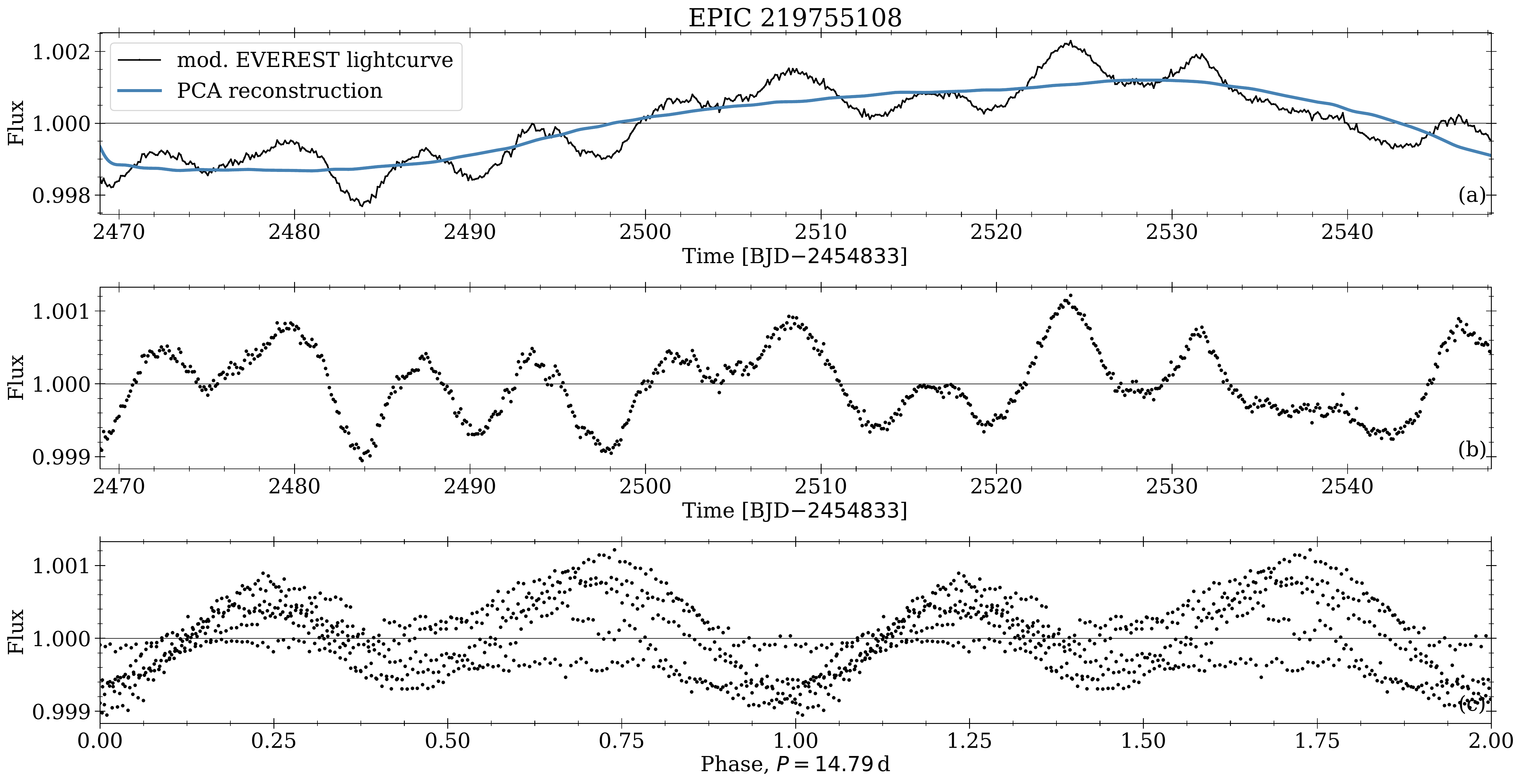} 
    \caption{Same as Fig.\,\ref{fig_lc_1}, but for EPIC 219755108 \label{fig_lc_31}} 
\end{figure*}  

\begin{figure*} 
    \centering 
    \includegraphics[width=\linewidth]{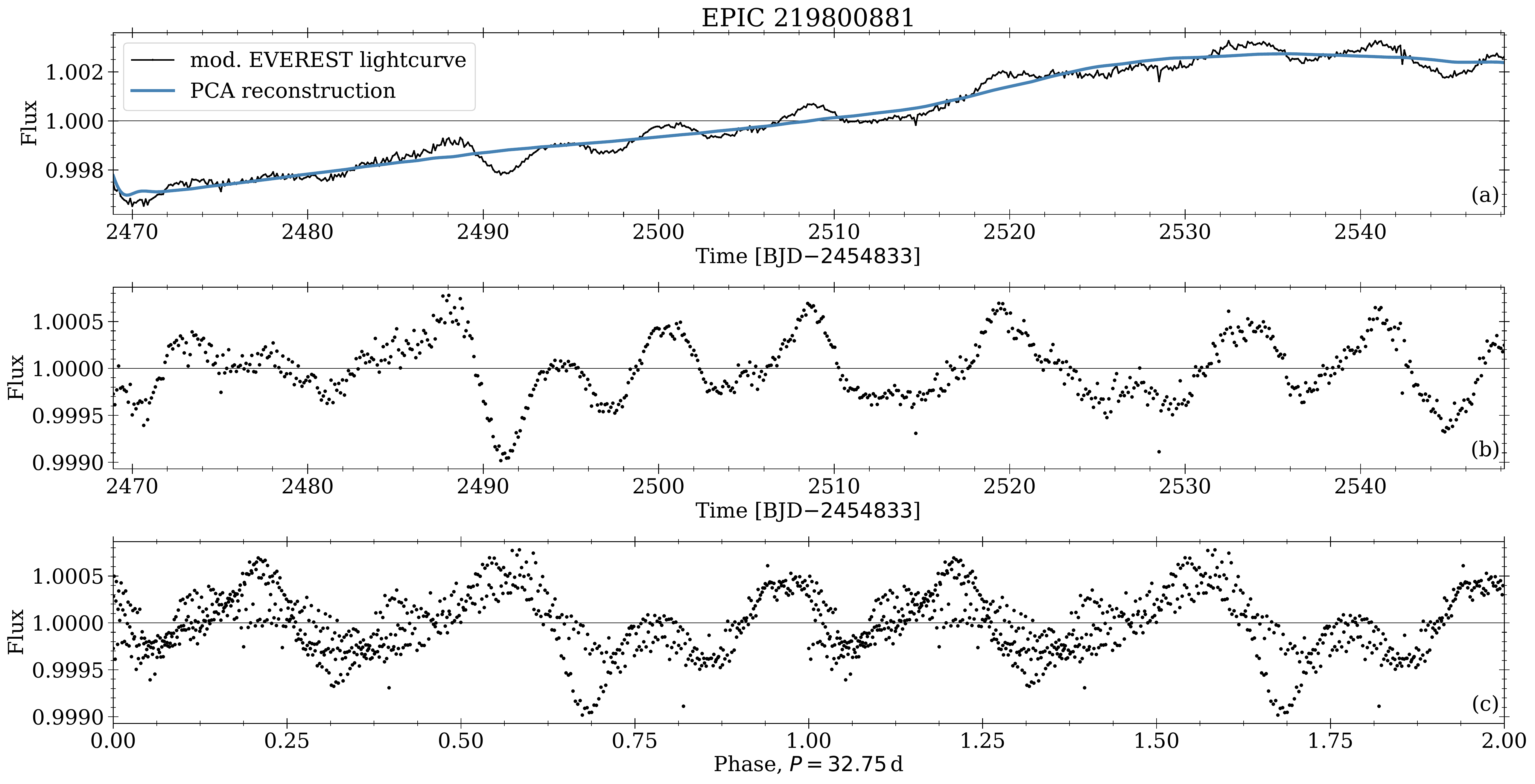} 
    \caption{Same as Fig.\,\ref{fig_lc_1}, but for EPIC 219800881 \label{fig_lc_32}} 
\end{figure*}

\end{document}